\documentclass[aps,prd,preprint,floatfix,nofootinbib,longbibliography,a4paper]{revtex4-1}
\pdfoutput=1
\usepackage{color}
\usepackage{graphicx}
\usepackage{textcomp}
\usepackage{subfigure}
\usepackage{feynmp}
\usepackage{amsmath,amssymb,array}
\usepackage{enumerate}
\usepackage{hhline}
\newcommand{\mathsym}[1]{{}} 
\def\lsim{\:\raisebox{-1.1ex}{$\stackrel{\textstyle<}{\sim}$}\:}
\def\gsim{\:\raisebox{-1.1ex}{$\stackrel{\textstyle>}{\sim}$}\:}

\DeclareSymbolFont{matha}{OML}{txmi}{m}{it}
\DeclareMathSymbol{\varv}{\mathord}{matha}{118}
\newcommand{\beqa}{\begin{eqnarray}}
\newcommand{\eeqa}{\end{eqnarray}}
\newcommand{\be}{\begin{equation}}
\newcommand{\ee}{\end{equation}}
\newcommand{\ba}{\begin{array}} 
\newcommand{\ea}{\end{array}}

\begin{document} 
\vspace*{1cm}
\title{Pseudo-Dirac Higgsino dark matter in GUT scale supersymmetry}
\bigskip
\author{V Suryanarayana Mummidi}
\email{suryam@iisermohali.ac.in}
\author{Ketan M. Patel}
\email{ketan@iisermohali.ac.in}
\affiliation{Indian Institute of Science Education and Research Mohali, Knowledge City, Sector  81, S A S Nagar, Manauli 140306, India.\\}

\bigskip
\begin{abstract}
We investigate a scenario in which supersymmetry is broken at a scale $M_S \geq 10^{14}$ GeV leaving  only a pair of Higgs doublets, their superpartners (Higgsinos) and a gauge singlet fermion (singlino) besides the standard model fermions and gauge bosons at low energy. The Higgsino-singlino mixing induces a small splitting between the masses of the electrically neutral components of Higgsinos which otherwise remain almost degenerate in GUT scale supersymmetry. The lightest combination of them provides a viable thermal dark matter if the Higgsino mass scale is close to $1$ TeV. The small mass splitting induced by the singlino turns the neutral components of Higgsinos into pseudo-Dirac fermions which successfully evade the constraints from the direct detection experiments if the singlino mass is $\lesssim 10^8$ GeV. We analyse the constraints on the effective framework, arising from the stability of electroweak vacuum, observed mass and couplings of the Higgs, and the limits on the masses of the other scalars, by matching it with the next-to-minimal supersymmetric standard model at $M_S$. It is found that the presence of singlino at an intermediate scale significantly improves the stability of electroweak vacuum and allows a stable or metastable vacuum for almost all the values of $\tan\beta$ while the observed Higgs mass together with the limit on the charged Higgs mass favours $\tan\beta \lesssim 3$.
\end{abstract} 

\maketitle

\section{Introduction}
\label{sec:introduction}
If supersymmetry (SUSY) does not stabilize the electroweak scale then the scale of its breaking is not restricted to stay in the vicinity of the weak scale. Most of the other attractive features of  low energy supersymmetry, such as precision gauge coupling unification and particle candidate(s) of weakly interacting dark matter (DM), could also be achieved in scenarios like split-supersymmetry \cite{Giudice:2004tc,ArkaniHamed:2004fb} in which only some of the superpartners are required to be close to the weak scale. The essential role played by SUSY in superstring theories \cite{Green:1987sp}, which are potential candidates for the unification of all the fundamental forces, remains the same as long as it is broken at any scale below the string scale. In fact, it is typically expected that the SUSY breaking scale in such theories is close to the string scale. Similarly, in a class of models based on supersymmetric grand unified theories (GUT) in higher spacetime dimensions (see for example, \cite{Asaka:2002my,Kim:2002im,Kitano:2003cn,Asaka:2003iy,Kobayashi:2004ud,Feruglio:2014jla,Feruglio:2015iua,Buchmuller:2015jna,Buchmuller:2017vho,Buchmuller:2017vut}), the breaking of SUSY and unified gauge symmetry are often administered by a common mechanism which gives rise to the breaking of SUSY at the GUT scale. 

In the post Higgs discovery era, scenarios of high scale SUSY are constrained by the measured Higgs mass and stability of the electroweak vacuum. For example, it is known that the standard model (SM) cannot be matched with its simplest supersymmetric counterpart - the minimal supersymmetric standard model (MSSM) - if the SUSY breaking scale is above $10^{10}$ GeV because of the constraints arising from the stability of electroweak vacuum \cite{Giudice:2011cg,EliasMiro:2011aa,Draper:2013oza,Ellis:2017erg}. The scale can be raised up to the GUT scale if SM is replaced by the two-Higgs-doublet model (THDM) as a low energy effective theory \cite{Bagnaschi:2015pwa}. The presence of additional Higgs doublet helps in achieving a stable electroweak vacuum in this case. The stability can further be improved if there exists strongly coupled right handed neutrinos below the GUT scale \cite{Mummidi:2018nph}. An effective theory involving the THDM and a pair of Higgsinos at the weak scale as remnants of GUT scale supersymmetry has also been studied in \cite{Lee:2015uza,Bagnaschi:2015pwa} and it is shown compatible with the vacuum stability constraints \cite{Bagnaschi:2015pwa}. While the weak scale Higgsinos improve the convergence of gauge couplings, they can also provide a potentially viable candidate of DM given an unbroken $R$-parity in the underlying supersymmetric theory. However, the second possibility is disfavoured by the null results from the DM direct detection experiments.

\begin{figure}[t!]
\centering
\subfigure{\includegraphics[width=0.22\textwidth]{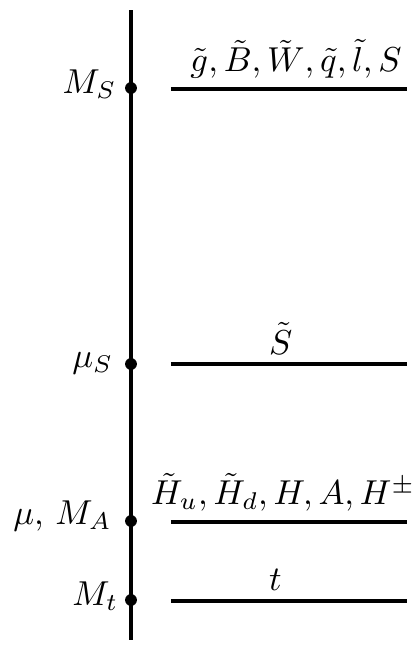}}
\caption{Schematic representation of hierarchy among the various mass scales in the framework.}
\label{fig1}
\end{figure}

The pure Dirac Higgsino DM is already ruled out because of its large elastic scattering cross section with nucleons \cite{Servant:2002hb}. The mixing of Higgsinons with bino, wino and/or the radiative corrections from supersymmetric particles can induce splitting between the masses of neutral components of Higgsinos making them Majorana fermions and hence the above constraints can be evaded. However, the required amount of mass splitting is obtained only if the mass scale of the other supersymmetric particles is $\lsim 10^8$ GeV \cite{Nagata:2014wma}. In this paper, we discuss a possibility in which  pseudo-Dirac Higgsino DM can be reconciled with GUT scale supersymmetry. We consider an effective theory consists of THDM and a pair of Higgsinos augmented by a singlet fermion, namely the singlino $\tilde{S}$. The effective theory is matched with the next-to-minimal supersymmetric standard model (NMSSM) at the SUSY breaking scale $M_S$ which is assumed close to the GUT scale. The assumed hierarchy among the different scales and corresponding mass scales of particles are depicted in Fig. \ref{fig1}. 
We find that the Higgsino-singlino-Higgs Yukawa interaction can induce required mass splitting between the neutral components of Higgsinos if the singlino mass scale is $\lsim 10^8$ GeV. The same interaction also improves the stability of electroweak vacuum significantly. It is shown that viable Higgsino DM with stable or metastable electroweak vacuum and scalar spectrum consistent with the current experimental observations can be obtained within the framework of GUT scale supersymmetry.

The paper is organized as follows. In the next section, we outline the effective framework and discuss its matching with NMSSM at the SUSY breaking scale. We provide details of various phenomenological constraints applicable on the underlying framework in section \ref{sec:constraints}. The details of numerical analysis are described in section \ref{sec:numerical} followed by the results and discussion in section \ref{sec:results} and summary in section \ref{sec:summary}. Some relevant technical details are provided in three Appendices.

\section{Framework}
\label{sec:framework}
We consider an effective theory below the SUSY breaking scale described by the following renormalizable Lagrangian:
\be \label{L}
{\cal L} = {\cal L}_{\rm THDM} + {\cal L}_{\tilde{H}} + {\cal L}_{\tilde{S}} + {\cal L}_{\tilde{H}-\tilde{S}}\,.\ee
The first term denotes the Lagrangian of the most general two-Higgs-doublet model which contains a pair of weak doublet scalars $H_{1,2}$ with hypercharge $Y=1/2$. The scalar potential and Yukawa interactions terms in ${\cal L}_{\rm THDM}$ can be written as
\beqa \label{V_THDM}
V &=& m_1^2\, H_1^\dagger H_1\, +\, m_2^2\, H_2^\dagger H_2\, - \, \left[m_{12}^2\, H_1^\dagger H_2\, +\, {\rm h.c.} \right] \nonumber \\
& + & \frac{\lambda_1}{2}\, (H_1^\dagger H_1)^2\, +\, \frac{\lambda_2}{2}\, (H_2^\dagger H_2)^2\, +\, \lambda_3\, (H_1^\dagger H_1)  (H_2^\dagger H_2)\, +\, \lambda_4\, (H_1^\dagger H_2)  (H_2^\dagger H_1) \nonumber \\
& + & \left[\frac{\lambda_5}{2}\, (H_1^\dagger H_2)^2\, +\, \lambda_6\, (H_1^\dagger H_1) (H_1^\dagger H_2)\, + \, \lambda_7\, (H_1^\dagger H_2) (H_2^\dagger H_2)\, +\, {\rm h.c.}   \right]\,,
\eeqa
and
\beqa \label{LY_THDM}
-{\cal L}_Y &=& \overline{Q}_L^i \left( Y_d^{ij} H_1+\tilde{Y}_d^{ij} H_2\right) d_R^j\,+\, \overline{Q}_L^i \left( \tilde{Y}_u^{ij} H^c_1+\, Y_u^{ij} H^c_2\right) u_R^j\nonumber \\
& + &\,  \overline{L}_L^i \left( Y_e^{ij} H_1+\tilde{Y}_e^{ij} H_2\right) e_R^j\,+\, {\rm h.c.}\,,
\eeqa
respectively. Here  $i,j=1,2,3$ denote three generations of SM fermions and $H^c_{1,2} = i \sigma^2 H^*_{1,2}$.

The second term in eq. (\ref{L}) represents free Lagrangian of a pair of fermions, namely $\tilde{H}_u$ and $\tilde{H}_{d}$, which are $SU(2)_L$ doublets with $Y=1/2$ and $-1/2$, respectively.
\be \label{L_H}
{\cal L}_{\tilde{H}} = {\cal L}_{\rm kin.} - \left[ \mu \left(\tilde{H}_u \cdot \tilde{H}_d \right) + {\rm h.c.}\right]\,,
\ee
where $\left(\tilde{H}_u \cdot \tilde{H}_d \right) = \epsilon^{\alpha \beta} (\tilde{H}_u)_\alpha (\tilde{H}_d)_\beta$ with $\alpha,\beta=1,2$ as $SU(2)_L$ indices and $\epsilon^{12}=-\epsilon^{21} = 1$. The fields $\tilde{H}_{u,d}$ have the same gauge quantum numbers as those of Higgsinos in the MSSM \cite{Martin:1997ns} and their explicit forms are:
\be \label{HuHd}
\tilde{H}_u = \left( \ba{c} \tilde{H}^+_u \\ \tilde{H}^0_u \ea \right)\,,~~~\tilde{H}_d = \left( \ba{c} \tilde{H}^0_d \\ \tilde{H}^-_d \ea \right)\,.\ee
Similarly, ${\cal L}_{\tilde{S}}$ contains a Majorana mass term of the singlino $\tilde{S}$:
\be \label{L_S}
{\cal L}_{\tilde{S}} = {\cal L}_{\rm kin.} - \left[  \frac{\mu_{S}}{2} \tilde{S} \tilde{S} + {\rm h.c.}\right] \,.
\ee
The last term in eq. (\ref{L}) includes gauge invariant renormalizable Yukawa interactions involving the fermions $\tilde{H}_u$, $\tilde{H}_d$, $\tilde{S}$ and scalars $H_{1,2}$. It is given as:
\be \label{L_HS}
- {\cal L}_{\tilde{H}-\tilde{S}} =  y_1\, \tilde{S} \left( H_1^\dagger \tilde{H}_u\right) + y_2\, \tilde{S} \left(H_2 \cdot \tilde{H}_d\right) + y_3\, \tilde{S} \left( H_2^\dagger \tilde{H}_u\right) + y_4\, \tilde{S} \left(H_1 \cdot \tilde{H}_d\right) + {\rm h.c.}\,. \ee

The free Higgsino Lagrangian ${\cal L}_{\tilde{H}}$ possesses a global $U(1)$ symmetry under which $\tilde{H}_u$ and $\tilde{H}_d$ have equal and opposite charges. However, this symmetry is broken either by the Yukawa interactions in eq. (\ref{L_HS}) or by the Majorana mass term in eq. (\ref{L_S}) given $H_{1,2}$ remain uncharged under this symmetry. As a result, the Higgsinos become pseudo-Dirac fermions in this framework. It can also be noticed that ${\cal L}$ possesses a $Z_2$ symmetry under which the fields $\tilde{H}_{u,d}$ and $\tilde{S}$ are odd while all the other fields are even. Origin of this symmetry in the effective theory can be attributed to the presence of unbroken $R$-parity in the underlying UV supersymmetric theory.

\subsection{Matching with NMSSM at $M_S$}
A minimal supersymmetric setup which can provide UV completion of an effective theory, described by ${\cal L}$ in eq. (\ref{L}), is the well-known NMSSM. We therefore match the effective theory with NMSSM at the SUSY breaking scale and obtain constraints on various parameters. The most general superpotential of NMSSM is given as \cite{Ellwanger:2009dp}
\be \label{W_NMSSM}
{\cal W} = {\cal W}_{\rm MSSM} + \lambda\, \hat{S} \left(\hat{H}_u \cdot \hat{H}_d\right) + c\, \hat{S} + \frac{\mu_S}{2} \hat{S}^2 + \frac{\kappa}{3} \hat{S}^3\,,
\ee
where ${\cal W}_{\rm MSSM} $ is the standard MSSM superpotential (see for example \cite{Martin:1997ns}) and $\hat{S}$ is chiral superfield with singlino $\tilde{S}$ and a complex scalar $S$ as its submultiplets. The corresponding soft SUSY breaking sector can be parametrized in terms of the following potential.
\be \label{soft}
V_{\rm soft} = V^{\rm MSSM}_{\rm soft} + m_S^2\, |S|^2 + \left[ \lambda A_\lambda\, S \left(H_u \cdot H_d\right) +  c_S\, S + \frac{b_S}{2} S^2 + \frac{1}{3} \kappa A_\kappa S^3 + {\rm h.c.} \right]\,
\ee
The trilinear scalar couplings in $V_{\rm soft}$ are assumed to be proportional to the corresponding Yukawa couplings in the superpotential. The term linear in $S$ is known to generate potentially dangerous quadratic divergences in supergravity \cite{Bagger:1993ji} and therefore the coupling $c_S$ needs to be suppressed. Typically, this is achieved by introducing a symmetry which forbids such term. In the most popular versions of the NMSSM, the same symmetry is often utilized to solve the so-called $\mu$ problem, see \cite{Ellwanger:2009dp,Kim:1983dt} for examples.

In our setup, we assume
\be \label{Assumption}
c = c_S =  \kappa = A_\kappa = 0\ee
for brevity and consider the remaining parameters $\mu$, $\mu_S$, $\lambda$, $A_\lambda$ and $b_S$ as real and positive. A phenomenologically consistent pseudo-Dirac Higgsino DM requires both $\mu$ and $\mu_S$ well below the scale $M_S$. We also assume $b_S \sim {\cal O}(\mu_S^2)$ and $m_S \gsim M_S$ which lead to $b_S \ll M_S^2$. We then compute the complete scalar potential of the theory from $V_{\rm soft}$ and the $D$ and $F$ terms of superpotential ${\cal W}$. After integrating out the singlet scalar $S$ from the theory and keeping only the leading order terms in $b_S/M_S^2$, the resulting effective potential is matched with the THDM potential given in eq. (\ref{V_THDM}) using the identities $H_2 = H_u$ and $H_1 = -i \sigma_2 H_d^*$. This implies the following tree level matching conditions at the SUSY breaking scale $M_S$:
\beqa \label{lm_MGUT}
\lambda_1 &=& \lambda_2 \simeq \frac{1}{4}\left(g_2^2+  g_Y^2 \right) - \frac{2 \lambda^2 \mu^2}{\tilde{m}_S^2}\left( 1-\frac{b_S}{\tilde{m}_S^2}\right)\,, \nonumber \\
\lambda_3 &\simeq & \frac{1}{4}\left(g_2^2 - g_Y^2 \right) - \frac{2 \lambda^2 \mu^2}{\tilde{m}_S^2}\left( 1-\frac{b_S}{\tilde{m}_S^2} \right)\,, \nonumber \\
\lambda_4 &\simeq & -\frac{1}{2} g_2^2 + \lambda^2 \left(1- \frac{A_\lambda^2 + \mu_S^2}{\tilde{m}_S^2} \left( 1- \frac{b_S}{\tilde{m}_S^2} \frac{2 A_\lambda \mu_S}{A_\lambda^2 + \mu_S^2} \right)\right)\,, \nonumber \\
\lambda_5 & \simeq & -\frac{\lambda^2 A_\lambda \mu_S}{\tilde{m}_S^2}\left( 1 - \frac{b_S}{\tilde{m}_S^2} \frac{A_\lambda^2 + \mu_S^2}{A_\lambda \mu_S}\right)\,, \nonumber \\
\lambda_6 & = & \lambda_7 \simeq -\frac{\mu \lambda^2 (A_\lambda + \mu_S)}{\tilde{m}_S^2}\left( 1-\frac{b_S}{\tilde{m}_S^2}\right)\,. \eeqa
where $\tilde{m}_S \approx \sqrt{m_S^2 + \mu_S^2}$ is physical mass of $S$. The parameters $m_1$ and $m_2$ in eq. (\ref{V_THDM}) are generated by the soft masses of $H_d$ and $H_u$ respectively,  while $m_{12}$ arises from the $b_\mu$ term in $V_{\rm soft}^{\rm MSSM}$ in eq. (\ref{soft}). Further, the matching between the Yukawa interactions in ${\cal W}$ and those in eq. (\ref{LY_THDM},\ref{L_HS}) leads to
\be \label{yi_MGUT}
y_1 =y_2 = \lambda\,, ~~ y_3=y_4=0\,, \ee
\be \label{Yuk_MGUT}
\tilde{Y}_d^{ij}=\tilde{Y}_u^{ij}=\tilde{Y}_e^{ij}=0\,, \ee
at $M_S$.

It can be noticed from eq. (\ref{lm_MGUT}) that the precise values of $\lambda_i$ at $M_S$ still depend on the details of SUSY breaking sector despite of the simplification achieved through ansatz in eq. (\ref{Assumption}). With $\tilde{m}_S \gsim M_S$, the couplings $\lambda_5$ and $\lambda_{6,7}$ are found to be suppressed by a factor of at least $\mu_S/M_S$ and $\mu/M_S$, respectively. Further suppression could be obtained if $|A_\lambda| \ll M_S$. One finds that $\lambda_{6,7}$ and $m_{12}$ have vanishing values in the limit $\mu,b_\mu \to 0$. Together with conditions in eqs. (\ref{yi_MGUT},\ref{Yuk_MGUT}), this implies $Z_2$ symmetry of ${\cal L}$ in eq. (\ref{L}) under which the fields $H_1$, $\tilde{H}_d$, $\tilde{S}$, $d_R^i$ and $e_R^i$ are odd. The symmetry keeps the parameters which vanish at $M_S$ under control and prevents them from taking large values through renormalization group evolution (RGE) effects. In the effective theory, the $Z_2$ symmetry is softly broken by the terms proportional to $\mu$ and $m_{12}$. Our assumption of real $\mu$, $\mu_S$, $\lambda$, $A_\lambda$, $b_\mu$ and $b_S$ implies real values for couplings in eq. (\ref{V_THDM})\footnote{Small imaginary values for these couplings can get induced through RGE because of the presence of CP violation in the SM. However, this is not a new source of CP violation and therefore we neglect such effects.}. The effective theory below $M_S$, obtained from the NMSSM with the aforementioned conditions, resembles the well-known CP conserving type II version of THDM \cite{Branco:2011iw}. The effective theory, therefore, does not give rise to low energy flavour or CP violating effects additional to those already anticipated in type-II THDM.

\subsection{Higgsino dark matter}
One of the main aims of this study is to show the existence of viable pseudo-Dirac Higgsino dark matter. We therefore discuss the Higgsino mass spectrum in detail. As already emphasized, the Higgsinos and their interactions in the effective theory below $M_S$ are well described by the terms in eqs. (\ref{L_H},\ref{L_S},\ref{L_HS}) with $y_{3,4} = 0$. Further integrating out the singlino from the low energy spectrum, the effective Higgsino Lagrangian at  scale below $\mu_S$ is given by
\be \label{L_Higgsino_mass}
{\cal L}_{\tilde{H}}^{\rm eff.} =  {\cal L}_{\tilde{H}} + \left[ \frac{c_1}{ 2 \mu_S} \left( H_1^\dagger \tilde{H}_u\right)^2 + \frac{c_2}{2 \mu_S} \left( H_2 \cdot \tilde{H}_d\right)^2 + \frac{d}{\mu_S} \left( H_1^\dagger \tilde{H}_u\right) \left( H_2 \cdot \tilde{H}_d\right) + {\rm h.c.} \right] \,,  \ee
with the following matching conditions at $\mu_S$:
\be \label{bc_cd}
c_i (\mu_S) = y_i^2(\mu_S)\,,~~d(\mu_S) = y_1(\mu_S)\, y_2(\mu_S)\,.
\ee
The electroweak symmetry breaking then contributes into the masses for the neutral and charged components of $\tilde{H}_{u,d}$. 

In the basis, $N = \left( \tilde{H}_d^0, \tilde{H}_u^0 \right)^T$, the mass term for neutral components can be written as
\be \label{neutralino_mass}
-{\cal L}_N^{\rm mass} = \frac{1}{2} N^T\, M_{N}\, N\, + {\rm h.c.}\,, 
\ee
with 
\be \label{MN}
M_N= \left( \ba{ccc} - \frac{c_2\, v_2^2}{2 \mu_S} & & -\mu + \frac{d\,  v_1 v_2}{2 \mu_S} \\ & \\
-\mu +\frac{d\,  v_1 v_2}{2 \mu_S} & & - \frac{c_1\,  v_1^2}{2 \mu_S} \ea \right)\,,\ee
where $v_1$ and $v_2$ are vacuum expectation values (VEVs) of the neutral components of $H_1$ and $H_2$ such that $\langle H_i \rangle \equiv \left(0, v_i/\sqrt{2}\right)^T$ and $\sqrt{v_1^2+v_2^2} \equiv v = 246\, {\rm GeV}$. Identifying linear combinations of $\tilde{H}_d^0$, $\tilde{H}_u^0$ as $\tilde{\chi}^0_{1,2}$ such that 
\be \label{basis_change}
\left( \ba{c} \tilde{\chi}^0_1 \\ \tilde{\chi}^0_2 \ea \right) = U_N^\dagger \left( \ba{c} \tilde{H}_d^0 \\ \tilde{H}_u^0 \ea \right) \ee
and 
\be \label{diagonalization}
U_N^T\, M_N\, U_N\, =\, {\rm Diag.}\left( m_{\tilde{\chi}^0_1},\, m_{\tilde{\chi}^0_2}\right)\,, \ee
one obtains the following expressions for the tree level neutralino masses with a convention $m_{\tilde{\chi}^0_1} < m_{\tilde{\chi}^0_2}$:
\beqa \label{m_neutralino}
m_{\tilde{\chi}^0_1} &\simeq & \mu - \frac{d\, v_1 v_2}{2 \mu_S} - \frac{c_1\, v_1^2 + c_2\, v_2^2}{4 \mu_S}\,, \nonumber \\
m_{\tilde{\chi}^0_2} &\simeq & \mu - \frac{d\, v_1 v_2}{2 \mu_S} + \frac{c_1\, v_1^2 + c_2\, v_2^2}{4 \mu_S} \,.
\eeqa
As it can be seen from eqs. (\ref{yi_MGUT},\ref{bc_cd}), $c_{1,2}$ are real and positive for real values of $\lambda$. The unitary matrix representing neutralino mixing is  
\be \label{U}
U_N=\left( \ba{cc} \cos\theta_N & i \sin\theta_N \\ -\sin\theta_N & i \cos\theta_N \ea \right)\,,~~{\rm with}~~ \theta_N \simeq  \frac{\pi}{4}+{\cal O}\left( \frac{v^2}{\mu\, \mu_S}\right)\,.\ee
Further, it is convenient to define the splitting between the masses of neutralinos as
\be \label{Dm0}
\Delta m_0 \equiv m_{\tilde{\chi}^0_2} - m_{\tilde{\chi}^0_1} = \frac{c_1\, v_1^2 + c_2\, v_2^2}{ 2 \mu_S}\,. \ee

The charged components of $\tilde{H}_{u,d}$ can be combined to form a Dirac fermion $\tilde{\chi}^+ = \left(\tilde{H}_u^+,\,  (\tilde{H}_d^-)^\dagger \right)^T $ with mass term
\be \label{chargino}
-{\cal L}_C^{\rm mass} = \mu\, \tilde{H}_u^+ \tilde{H}_d^- + {\rm h.c.} = m_{\tilde{\chi}^\pm}\, \overline{\tilde{\chi}^+} \tilde{\chi}^+ + {\rm h.c.}\,,\ee
where $m_{\tilde{\chi}^\pm} = \mu$ is mass of chargino at the tree level. As it can be seen from eq. (\ref{m_neutralino}), the contributions induced by dim-5 operators generates splitting between the masses of chargino and neutralino. In addition, it is known that loop corrections induced by the SM electroweak bosons make the chargino heavier than the neutralino \cite{Cirelli:2005uq}. The resulting mass splitting between the chargino and the lightest neutralino can be written as
\be \label{Dm+}
\Delta m_\pm \equiv m_{\tilde{\chi}^\pm} - m_{\tilde{\chi}^0_1} =  \frac{d\, v_1 v_2}{2 \mu_S} + \frac{c_1\, v_1^2 + c_2\, v_2^2}{4 \mu_S}\, + \, \Delta m_\pm^{\rm rad}\,, \ee
where $\Delta m_\pm^{\rm rad}$ is radiatively induced mass splitting. For $\mu \approx 1$ TeV, one obtains $\Delta m_\pm^{\rm rad} \approx 341$ MeV at one loop \cite{Cirelli:2005uq}. The contribution from the first two terms in eq. (\ref{Dm+}) remains positive for real $\lambda$.

\section{Phenomenological Constraints}
\label{sec:constraints}

We now discuss the set of constraints which are imposed on the effective framework described in the previous section.
\subsection{Dark matter}
The nature of the lightest neutralino $\tilde{\chi}^0_1$ is almost pure Higgsino like for $\mu_S \gg \mu$ in this framework.  Assuming that it makes all of the DM produced thermally in the Early Universe, its relic abundance is estimated in \cite{Cirelli:2005uq,Hisano:2006nn,Cirelli:2007xd} including the non-perturbative Sommerfeld corrections to DM annihilation cross sections. It is found that the observed relic abundance is obtained\footnote{This result is obtained for the SM. Since the coupling of DM with new scalars of THDM is suppressed by ${\cal O}(v/\mu_S)$ as discussed in Appendix \ref{app:dm2}, we expect that the same result holds in THDM.} if the mass of DM particle is close to $1$ TeV. This implies also $\mu \approx 1$ TeV in the present framework.

Experiments based on the direct detection of DM are known to put stringent constraints on pure Dirac Higgsino DM. If $\Delta m_0 =0$ then $\tilde{\chi}^0_1$ and $\tilde{\chi}^0_2$ can be paired to form a Dirac fermion, namely $\tilde{\chi}$. As $\tilde{\chi}$ has vector coupling with $Z$ boson, the elastic scattering with nucleon $N$ (such as $\tilde{\chi}+N \to \tilde{\chi}+N$) can proceed  through the exchange of the $Z$ boson. The scattering cross section, which is unambiguously determined for a given mass of Higgsinos, turns out too large and therefore this case is disfavoured by the non-observation of any statistically significant signal in the direct detection experiments \cite{Servant:2002hb}. Nevertheless, this constraint can be evaded if $\Delta m_0 > 0$. In this case, $\tilde{\chi}^0_1$ and $\tilde{\chi}^0_2$ are Majorana fermions and hence they do not have vector coupling with $Z$ boson.

The inelastic scattering, $\tilde{\chi}^0_1+N \to \tilde{\chi}^0_2+N$, is still subject to the constraints from the direct detection experiments for Majorana $\tilde{\chi}^0_{1,2}$, if $\Delta m_0$ is very small. Such processes arise through the t-channel exchange of $Z$ boson. Considering this, a lower limit on $\Delta m_0$ has been derived in \cite{Nagata:2014wma} using the then available data from XENON 10 and XENON 100 experiments. We update this analysis for the latest available data, including those from XENON 1T. The details are described in Appendix \ref{app:dm1}. We find that the present observations from the direct detection experiments lead to a lower bound on neutralino mass splitting, $\Delta m_0 \ge 200$ keV, at $90 \%$ confidence level. In the present framework, this bound translates into an upper limit on the singlino mass scale $\mu_S$. Using eq. (\ref{m_neutralino}) and assuming $c_i$ as ${\cal O}(1)$ numbers, we find
\be \label{bound_muS}
\mu_S \approx \frac{c_1\, v_2^2 + c_2\, v_1^2}{2\, \Delta m_0}\, \lsim\, 10^{8} ~{\rm GeV}\,. 
\ee
A similar bound on gaugino mass scale was obtained earlier in the case in which the mixing of Higgsino with bino or wino was responsible for mass splitting \cite{Fox:2014moa,Nagata:2014wma}. Higgsino mass splitting can also be induced radiatively through stop-top loop if the stop mixing angle is nonzero \cite{Giudice:1995np}. This however also requires the stop masses $\lsim 10^{8}$ GeV for $\Delta m_0 > 200 $ keV. In our framework, all the super-partners can have GUT scale masses while the presence of singlino, with $\mu_S \lsim 10^{8}$ GeV, can induce the required $\Delta m_0$.

The spin-independent and spin-dependent elastic scattering processes, like $\tilde{\chi}^0_1+N \to \tilde{\chi}^0_1+N$, can also occur in the underlying framework through the exchange of THDM scalars or $Z$ boson, respectively. We find that the scattering cross sections of the first type of interactions are suppressed due to $\mu_S \gg \mu$ while those of the latter are negligible because of pseudo-Dirac nature of Higgsino DM. After determining the constraints on the scalar spectrum and couplings, we estimate these cross sections and show that the obtained results are in agreement with the current experimental limits. This is described in detail in Appendix \ref{app:dm2}.

 The Higgsino DM can give rise to indirect signatures through their pair annihilation into $W^+W^-$ at tree level and $ZZ$, $\gamma \gamma$, $Z\gamma$ at loop level, which subsequently leads to the production of gamma rays and anti-protons. The latest constraints on almost pure Higgsino DM from the indirect searches are reviewed in \cite{Krall:2017xij,Kowalska:2018toh}. The constraints from the observations of gamma-ray by Fermi-LAT  \cite{Ackermann:2015zua} exclude Higgsino DM with mass $\le$ 330 GeV while HESS \cite{Abdallah:2016ygi} observations do not put any such limit. The strongest indirect search constraints on Higgsino DM arise from the latest AMS-02 results \cite{Aguilar:2016kjl} on anti-protons which put a conservative limit $m_{\tilde{\chi}^0_1} \ge 500$ GeV. Nevertheless, the almost pure Higgsino DM with mass $\approx 1$ TeV considered in our framework remains unconstrained from the current indirect detection experiments.

\subsection{Vacuum stability and perturbativity}
\label{subsec:vacuum}
The matching conditions obtained in eq. (\ref{lm_MGUT}) determine the values of quartic couplings in terms of the gauge couplings and several of NMSSM parameters. With the assumed hierarchies in the scales, i.e. $\mu<\mu_S \ll \tilde{m}_S \sim M_S$, the couplings $\lambda_5$ and $\lambda_{6,7}$ are  suppressed by factors of ${\cal O}(\mu_S/M_S)$ and ${\cal O}(\mu/M_S)$, respectively. They are even more suppressed if $A_\lambda \ll M_S$ and/or $b_S \approx M_S^2$. Further, the approximate $Z_2$ symmetry of the effective theory forbids them from taking large values through running effects. Therefore, the contributions of the terms involving the couplings $\lambda_{5,6,7}$ in the scalar potential remain negligible at all the scales below $M_S$. The remaining quartic couplings must satisfy the following conditions, at the intermediate scales between $M_S$ and $M_t$, for the absolute stability of electroweak vacuum \cite{Gunion:2002zf}
\beqa \label{stability_cond}
\lambda_1 & > & 0\,, \nonumber\\
\lambda_2 & > & 0\,, \nonumber\\
\lambda_3 + \sqrt{\lambda_1 \lambda_2} & > & 0\,, \nonumber\\ 
\lambda_4 +\lambda_3 + \sqrt{\lambda_1 \lambda_2}  & > & 0\,.
\eeqa

In a more conservative approach, it is assumed that the scalar potential in eq. (\ref{V_THDM}) may have multiple minima and the electroweak vacuum can decay into a more stable minimum through quantum tunnelling. However, the lifetime of electroweak vacuum is required to be greater than the current age of the Universe in order to make it phenomenologically viable. Such a minimum of potential is termed as metastable vacuum and its lifetime, in case of single scalar field, is estimated in \cite{Isidori:2001bm} including the quantum effects. This was generalized in \cite{Bagnaschi:2015pwa} for THDM scalar potential after mapping the underlying potential into single filed potential using the first three of the conditions given in eq. (\ref{stability_cond}). The requirement of metastable vacuum replaces the last condition in eq. (\ref{stability_cond}) with \cite{Bagnaschi:2015pwa}
\be \label{meta_cond}
\frac{4 \sqrt{\lambda_1 \lambda_2}\, \left( \lambda_4 +\lambda_3 + \sqrt{\lambda_1 \lambda_2} \right)}{\lambda_1 + \lambda_2 + 2 \sqrt{\lambda_1 \lambda_2}} \gsim -\frac{2.82}{41.1 + \log_{10}\left( \frac{Q}{\rm GeV}\right)}\,,
\ee
where $Q$ is the renormalization scale.

The first three of the conditions listed in eq. (\ref{stability_cond}) are found to be satisfied as a consequence of the high scale boundary conditions given in eq. (\ref{lm_MGUT}). Note that with  $\mu<\mu_S \ll \tilde{m}_S \sim M_S$, the couplings $\lambda_{1,2}$ are positive while $|\lambda_3| \ll \sqrt{\lambda_1 \lambda_2}$ \cite{Bagnaschi:2015pwa,Mummidi:2018nph}. Hence, it is the last condition in eq. (\ref{stability_cond}) or eq. (\ref{meta_cond}) which determine the stability or metastability of scalar potential, respectively. The same has been the case for THDM matched with MSSM at the GUT scale \cite{Bagnaschi:2015pwa,Mummidi:2018nph} in which the running of $\lambda_4$ dominantly decides the stability of vacuum. However, two important differences arise in the present framework. Firstly, the matching with NMSSM modifies boundary condition for $\lambda_4$ as can be seen from eq. (\ref{lm_MGUT}). Unlike in the case of MSSM, $\lambda_4$ can be made positive at $M_S$ with appropriately chosen values of $\lambda$ and $A_\lambda$. Secondly, even if the tree level enhancement in $\lambda_4$ is absent (for example, if $A_\lambda \approx M_S$), the contribution from singlino-Higgsino loop modifies the running of $\lambda_4$ and drives its value toward the positive side while running from $M_S$ down to the $M_t$ as it can be seen from appropriate RG equations given in Appendix \ref{app:RGE}.

\subsection{Scalar spectrum}
The matching of the effective theory with the NMSSM determines the values of all the quartic couplings which in turn provides useful correlations among the masses of THDM scalars. In order to check the viability of such correlations, we evaluate the mass spectrum of these scalars and impose various experimental constraints on the remaining free parameters of the potential. We closely follow the procedure and notations of \cite{Mummidi:2018nph} which is briefly outlined in the following. 

The VEVs of the neutral components of $H_1$ and $H_2$ break the electroweak symmetry giving rise to five physical Higgs bosons: two CP even and electrically neutral ($h$, $H$), two CP even and charged ($H^\pm$) and a CP odd and neutral ($A$). At the minimum, the parameters $m_1$, $m_2$ and $m_{12}$ in eq. (\ref{V_THDM}) can be replaced by appropriate functions of $M_A$, $\tan\beta$, $v$ and quartic couplings \cite{Gunion:2002zf}. Here, $M_A$ represents the physical mass of pseudo-scalar Higgs in $\overline{\rm MS}$ renormalization scheme while $\tan\beta \equiv v_2/v_1$. This replacement allows us to express the masses of CP even neutral and charged Higgs bosons in terms of the unknown parameters $M_A$, $\tan\beta$ and known parameters $v$ and $\lambda_i$. The physical CP even neutral Higgs bosons can be obtained as linear combinations of neutral components in $H_1$ and $H_2$ such that $H = \cos\alpha\, H_1 + \sin\alpha\, H_2$ and $h = -\sin\alpha\, H_1 + \cos\alpha\, H_2$. The state $h$ is identified with the observed SM like Higgs and $H$ is assumed to be heavier than $h$. For a consistent matching between the theoretically predicted mass of $h$ and that of the observed Higgs, we covert the running mass into pole mass, namely $M_h$, using the method described in \cite{Mummidi:2018nph}. The mixing angle $\alpha$ and the masses of heavy CP even Higgs ($M_H$) and charged Higgs ($M_{H^\pm}$) are also determined in terms of $M_A$, $\tan\beta$, $v$ and $\lambda_i$. Their expressions are given in \cite{Mummidi:2018nph} with all the  necessary details.

We consider the following set of constraints on the masses of Higgs bosons and the mixing angle $\alpha$.
\beqa \label{cons_lowscale}
M_h & = & (125 \pm 3)\, {\rm GeV}\,, \nonumber \\
|\cos(\beta - \alpha)| & \leq & 0.055\,, \nonumber \\
M_{H^\pm} & \geq & 580\, {\rm GeV}\,. \eeqa
For $M_h$, the experimentally measured value from \cite{Aad:2015zhl} is considered. In order to account for  theoretical uncertainties in estimating the Higgs mass, we allow a deviation of $\pm 3$ GeV from the measured mean value. Further, the couplings of $h$ with $W^\pm$  and $Z$ bosons are proportional to $\sin^2(\beta -\alpha)$ and therefore they are constrained from the observed signal strengths of $h \to W^+ W^-$ and $h \to Z Z$. This, through the results of the latest global fit performed in \cite{Chowdhury:2017aav}, implies that the deviation from the so-called alignment limit, i.e. $\beta-\alpha = \pi/2$, cannot be greater than 0.055 in the case of THDM of type II which gives rise to the second constraint in eq. (\ref{cons_lowscale}). The quoted lower bound on the mass of charged Higgs boson arises form the observed branching ratio of $b \to s + \gamma$ at $95\%$ confidence level \cite{Misiak:2017bgg}. Note that this bound is applicable for almost all values of $\tan\beta$ in THDM of type II. 

Since the masses of different scalars are correlated in the framework under consideration, we find that the lower limit on $M_{H^\pm}$ translates into lower limits on $M_A$ and $M_H$, thus making all these scalars heavier than $580$ GeV. Such heavy scalars already satisfy the direct search bounds and limits arising from the flavour observables such as $B_s \to \mu^+ \mu^-$, see \cite{Chowdhury:2017aav} for example and references therein. We also observe that for $M_A \ge 580$ GeV, the masses of the scalars $H$, $A$ and $H^\pm$ remain approximately degenerate in this framework. It is found that such a heavy and degenerate spectrum of scalars always satisfies the constraints imposed by the electroweak precision observables \cite{Mummidi:2018nph,Broggio:2014mna}.

\section{Numerical analysis}
\label{sec:numerical}
The viability of the proposed framework with respect to the various constraints, discussed in the previous section, is investigated by solving two-loop RGE equations numerically and implementing 1-loop corrected matching conditions. The one and two loop beta functions for the underlying framework are generated using publicly available package SARAH \cite{Staub:2013tta} and are listed in Appendix \ref{app:RGE}. We obtain the values of gauge and Yukawa couplings at the top quark pole mass scale $M_t$ from the experimental values of gauge couplings and fermion masses measured at the different scales. The procedure of this with relevant details is described in our previous work \cite{Mummidi:2018nph}. The obtained values of gauge couplings and fermion masses at the scale $M_t$ are listed in Table \ref{tab:inputs}. The values of Yukawa couplings at $M_t$ are extracted from the fermion masses as described in \cite{Mummidi:2018nph}. For $M_t$, we use the latest PDG average $M_t = 173.1 \pm 0.9$ GeV \cite{Tanabashi:2018oca}.
\begin{table}[!ht]
\begin{center}
\begin{tabular}{|cc|cc|cc|cc|}
    \hline
    Parameter & Value & Parameter & Value & Parameter & Value & Parameter & Value\\
    \hline
    $g_1$ & 0.46315 & $m_u$ & 1.21 MeV & $m_d$ & 2.58 MeV & $m_e$ & 0.499 MeV \\
    $g_2$ & 0.65403 & $m_c$ & 0.61 GeV & $m_s$ &  52.75  MeV & $m_{\mu}$&0.104 GeV\\
    $g_3$ & 1.1631 & $m_t$ & 163.35 GeV & $m_b$ & 2.72 GeV & $m_{\tau}$&1.759 GeV \\
    \hline
\end{tabular}
\caption{The SM parameters at renormalization scale $M_t = 173.1$ GeV in ${\overline{\rm MS}}$ scheme. See for details, Appendix C of \cite{Mummidi:2018nph}.}
\label{tab:inputs}
\end{center}
\end{table}

Following a similar procedure as described in \cite{Mummidi:2018nph}, we first evolve the gauge and Yukawa couplings from $M_t$ to $M_S$ using one loop RGE equations as the quartic couplings do not contribute in their running at this level. We then obtain the quartic couplings at $M_S$ using the matching conditions given in eq. (\ref{lm_MGUT}). The tree level matching of $\lambda_i$ should be corrected by one loop threshold corrections which are explicitly given in \cite{Haber:1993an,Lee:2015uza}. However, such corrections depend on the exact details of the SUSY spectrum and hence require explicit details of SUSY breaking. For simplicity, we assume all the squarks, sleptons and gauginos to be degenerate with mass $\sim M_S$ and vanishing trilinear parameters. Although such a choice of SUSY spectrum is very specific, it is naturally realized in the models of SUSY breaking based on flux compactification, see for example \cite{Buchmuller:2015jna}. With this choice of SUSY spectrum  and from the fact that $\mu \ll M_S$, the one loop threshold corrections in Yukawa couplings are found  to be suppressed by the degeneracy in the masses of the superpartners or by the smallness of $\mu/M_S$ as it can be seen from the relevant expressions given in \cite{Lee:2015uza}. The threshold corrections in quartic couplings are also suppressed by either vanishing trilinear couplings or $(\mu/M_S)^2$ and hence they are negligible in the present framework.

After obtaining the values of quartic couplings at $M_S$ as described in the above, all the gauge, Yukawa and quartic couplings are evolved down to $\mu_S$ using full two loop RGE equations. At $\mu_S$, we obtain the coefficients $c_i$ and $d$ using the conditions given in eq. (\ref{bc_cd}). After integrating out the singlino at this scale, the gauge, Yukawa and quartic couplings are evolved from $\mu_S$ to $M_t$ using appropriate two loop RGE equations. All these couplings are again evolved from $M_t$ to $M_S$ and back to $M_t$ iteratively until the couplings converge to their input values supplied at $M_t$. The stability and metastability of the potential is checked at  the intermediate scales using the conditions given in eq. (\ref{stability_cond}) and eq. (\ref{meta_cond}), respectively. After the convergence is achieved, we calculate the masses of scalars and the Higgs mixing angle $\alpha$ as function of input parameters $M_A$ and $\tan \beta$ using the obtained values of quartic couplings at $M_t$. We also evolve the effective operators given in eq. (\ref{L_Higgsino_mass}) from $\mu_S$ to $\mu$ using the one loop beta functions:
\beqa \label{reno_operators}
\beta_{c_1} & = & \frac{c_1}{16 \pi^2}  \left(6\, {\rm Tr}\left( Y_d^\dagger Y_d\right)+2\, {\rm Tr}\left( Y_e^\dagger Y_e\right) + 2 \lambda_1 -3 g_2^2\right)\,, \nonumber \\
\beta_{c_2} & = & \frac{c_2}{16 \pi^2}  \left(6\, {\rm Tr}\left( Y_u^\dagger Y_u\right)+ 2 \lambda_2 -3 g_2^2\right)\,, \nonumber \\
\beta_{d} & = & \frac{d}{16 \pi^2} \left(6\, {\rm Tr}\left( Y_u^\dagger Y_u\right)+6\, {\rm Tr}\left( Y_d^\dagger Y_d\right)+2\, {\rm Tr}\left( Y_e^\dagger Y_e\right)+2 \lambda_3 -3 g_Y^2 - 6 g_2^2\right)\,,\eeqa
where $\beta_X = dX/d(\ln Q)$. We have derived the above equations following a procedure similar to the one given in \cite{Antusch:2001vn,Antusch:2001ck} in the context of neutrinos. The neutralino  mass spectrum is then obtained by substituting the values of $c_i$ and $d$ at the scale $\mu$ in eq. (\ref{MN}). 

We set $\mu = 1\, {\rm TeV}$, as required by the observed DM abundance, and evaluate the vacuum stability constraints on the values of $\tan\beta$ and $\lambda$, for $M_S = 2 \times 10^{14}$ GeV and two sample values of $\mu_S$, simultaneously checking their viability to produce large enough $\Delta m_0$. Since $A_\lambda$ has direct implication to the boundary value of $\lambda_4$, we perform this analysis for $A_\lambda =0$ and $A_\lambda = M_S$. We repeat all these cases for $M_S = 2 \times 10^{16}$ GeV as well. Once the consistent values of $\lambda$ are found, we select some benchmark points and evaluate the low energy spectrum and constraints on $M_A$ and $\tan \beta$. The results of the numerical analysis are discussed in the next section.

\section{Results and Discussions}
\label{sec:results}
\begin{figure}[t]
\centering
\subfigure{\includegraphics[width=0.43\textwidth]{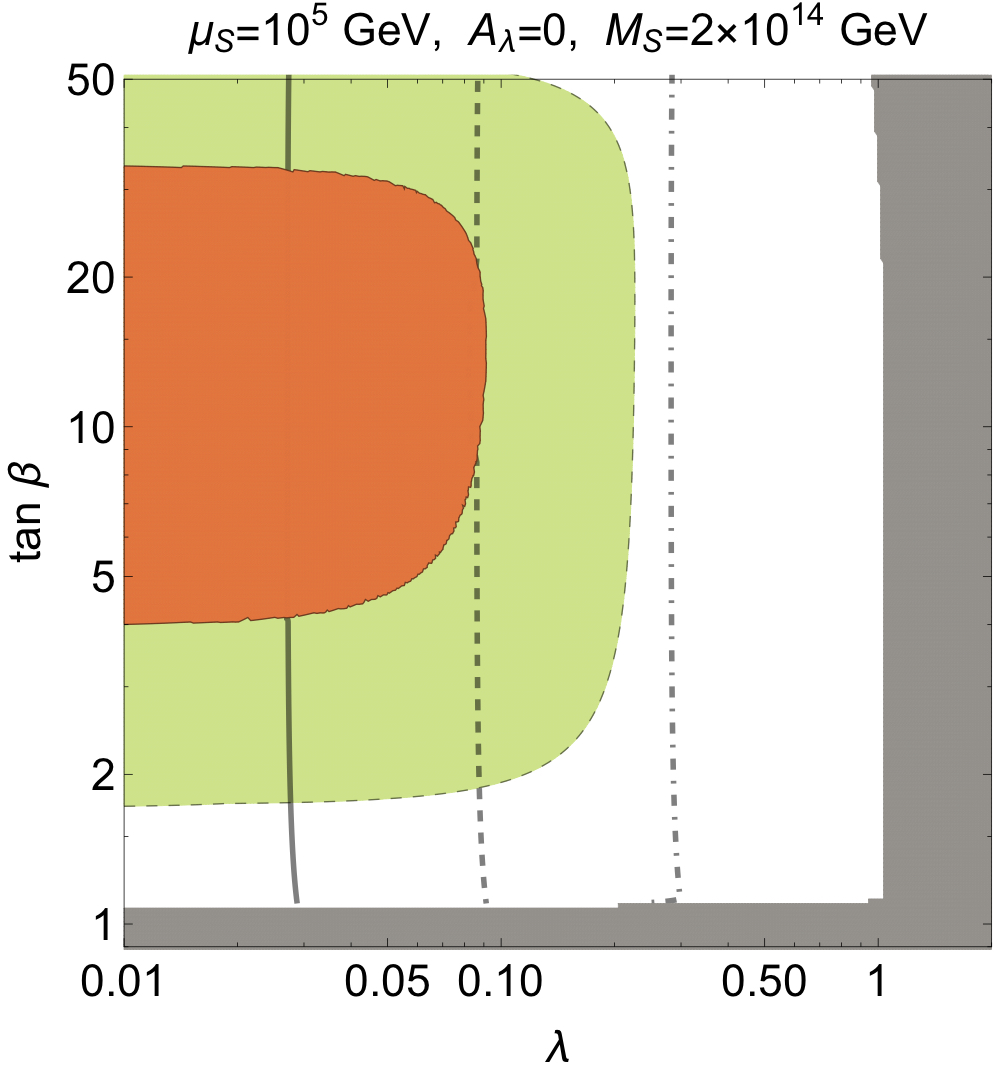}} \hspace*{0.5cm}
\subfigure{\includegraphics[width=0.43\textwidth]{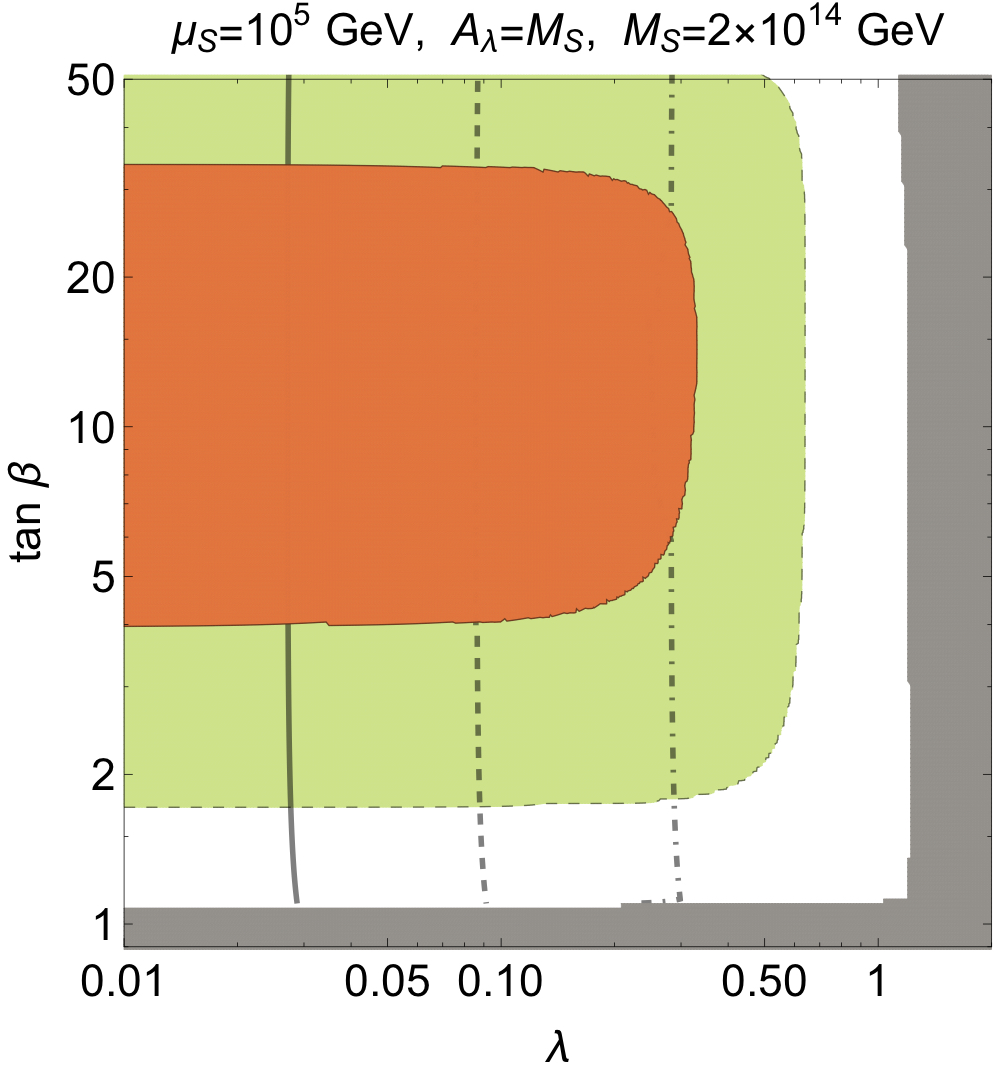}}\\
\subfigure{\includegraphics[width=0.43\textwidth]{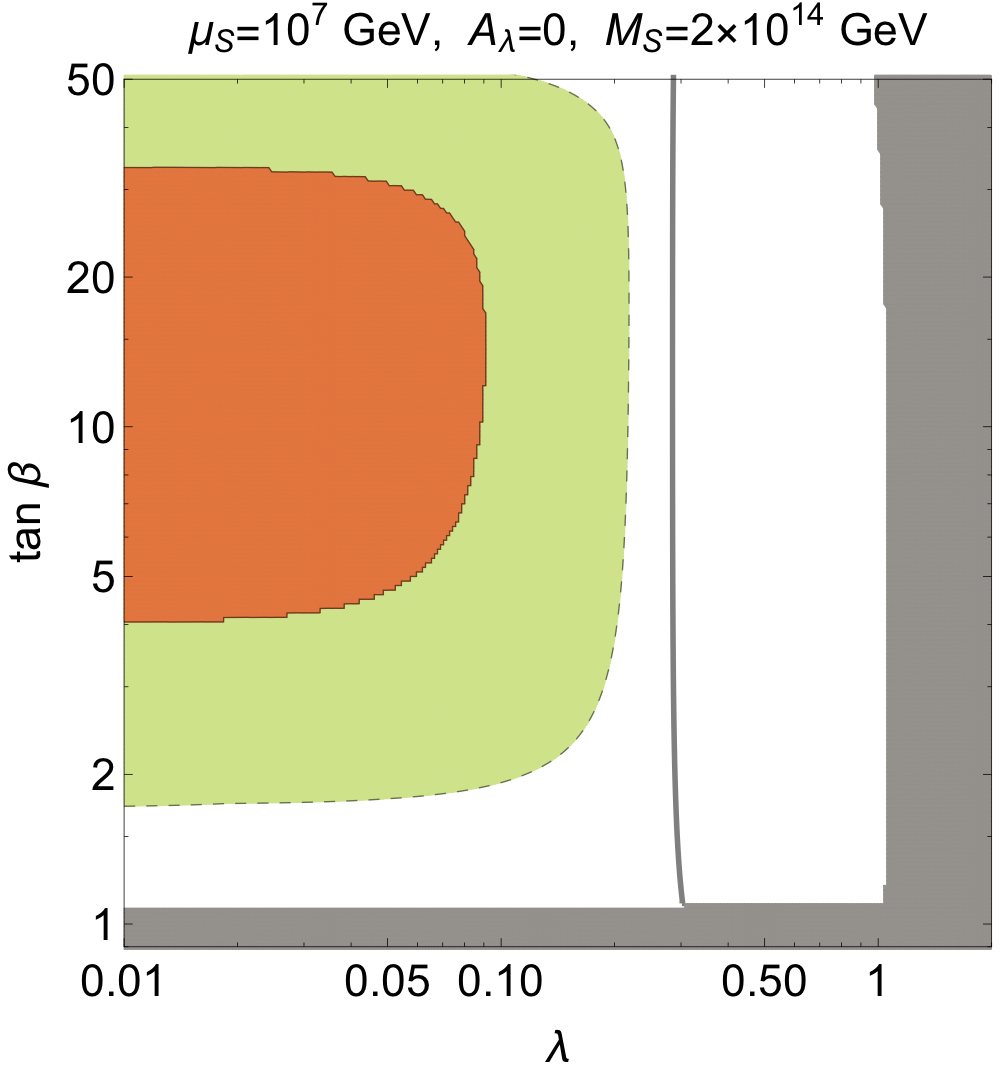}} \hspace*{0.5cm}
\subfigure{\includegraphics[width=0.43\textwidth]{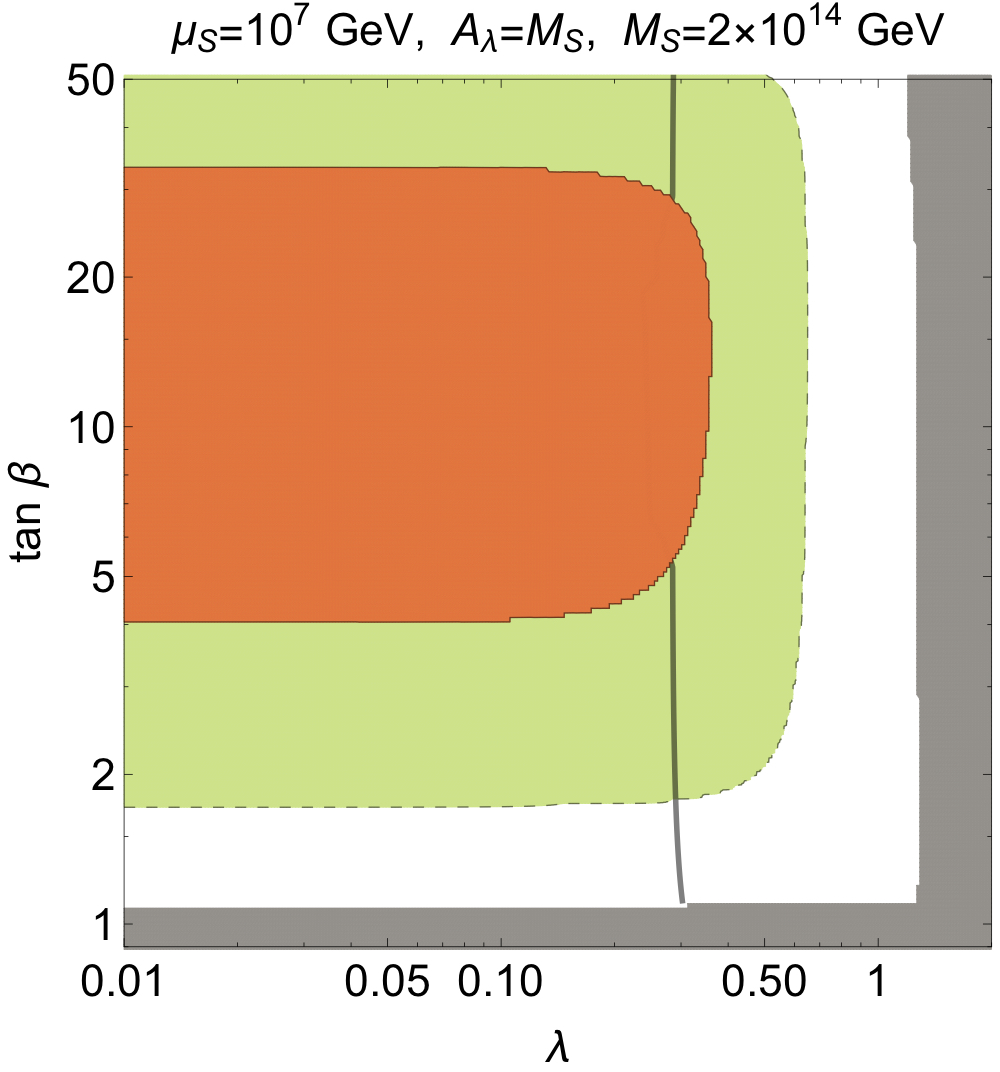}}\\
\caption{The regions corresponding to absolute stability (unshaded), metastability (green) and instability (orange) of the electroweak vacuum in $\tan\beta$-$\lambda$ plane for different values of $\mu_S$ and $A_\lambda$, and for $\mu=1$ TeV, $M_t = 173.1$ GeV and $M_S = 2 \times 10^{14}$ GeV. The region shaded in grey is disfavoured by the non-perturbativity of at least one or more couplings. The vertical black contour lines correspond to $\Delta m_0 = 200$ keV (solid), 2 MeV (dashed) and 20 MeV (dot-dashed).}
\label{fig:stability_14}
\end{figure}

\begin{figure}[t]
\centering
\subfigure{\includegraphics[width=0.43\textwidth]{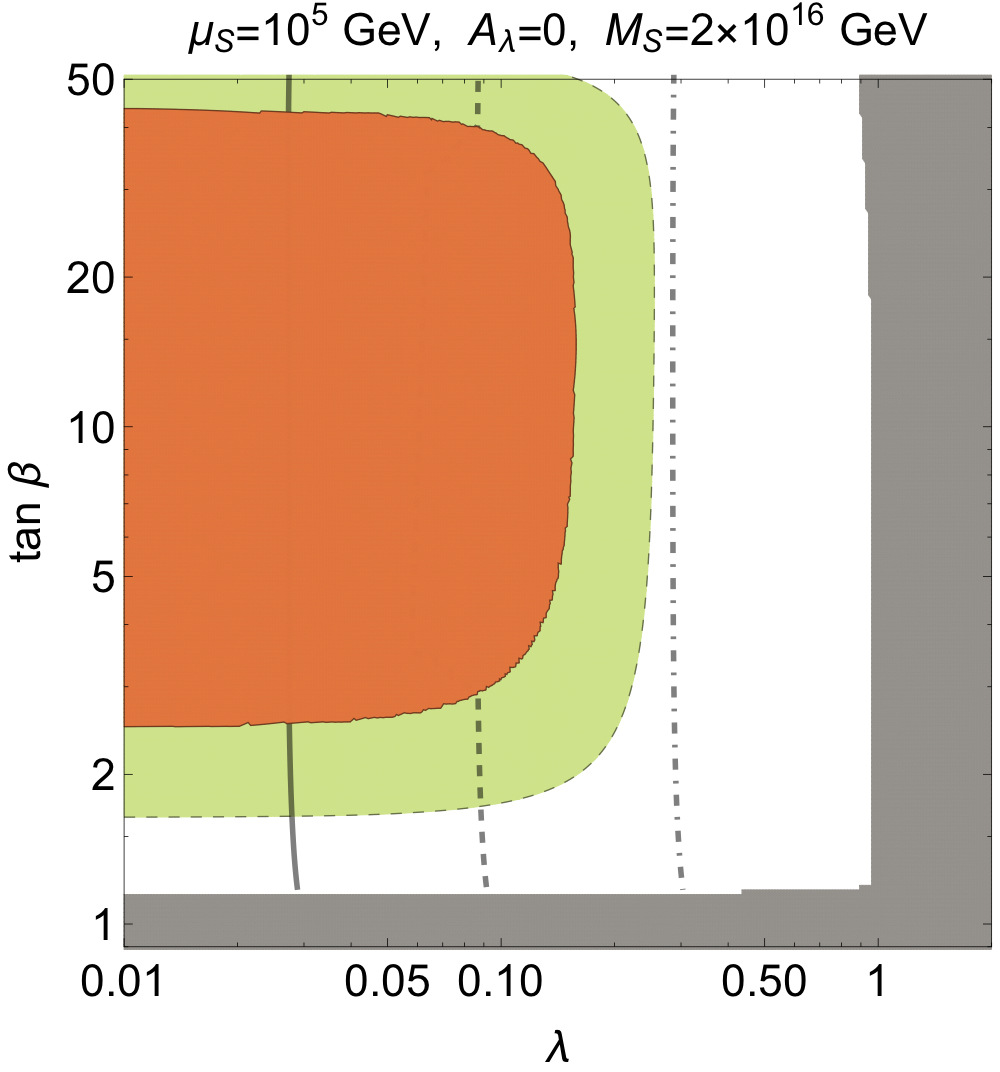}} \hspace*{0.5cm}
\subfigure{\includegraphics[width=0.43\textwidth]{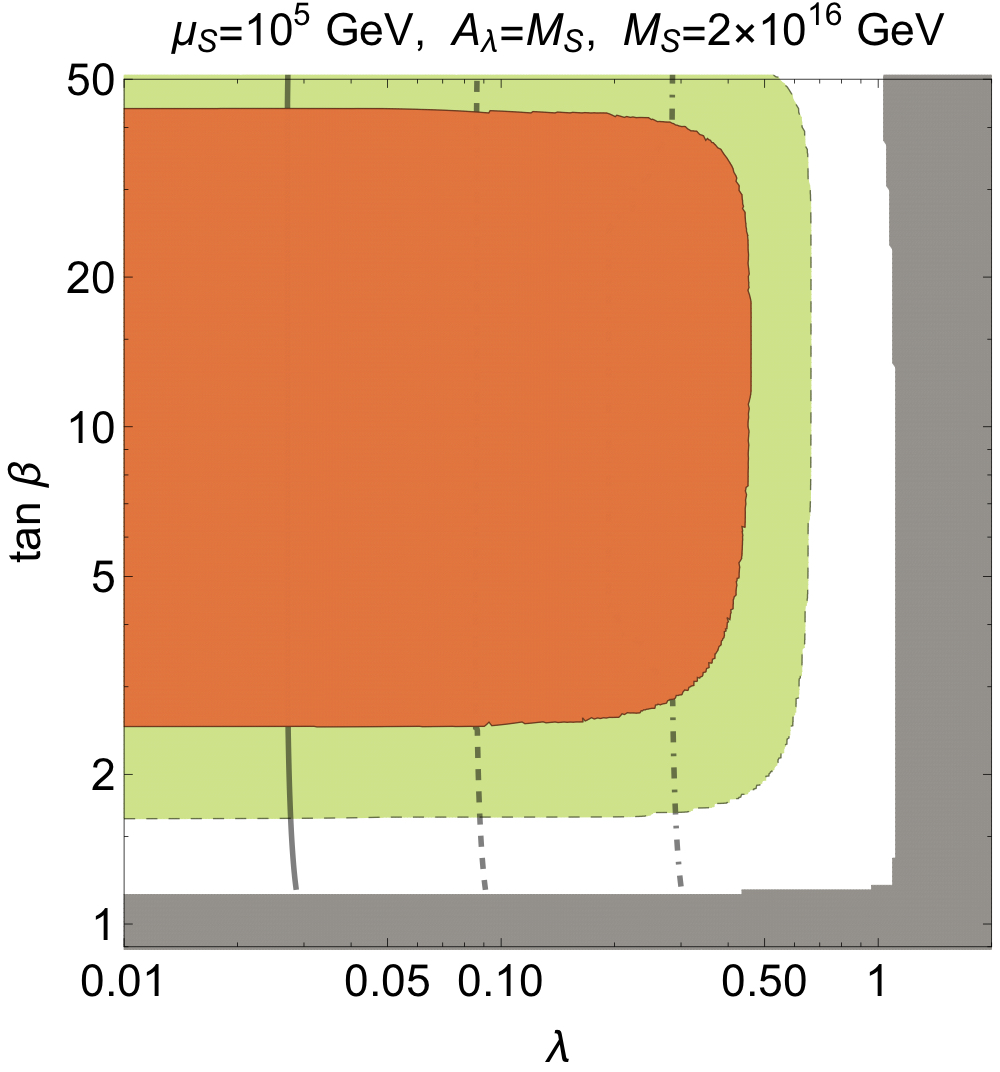}}\\
\subfigure{\includegraphics[width=0.43\textwidth]{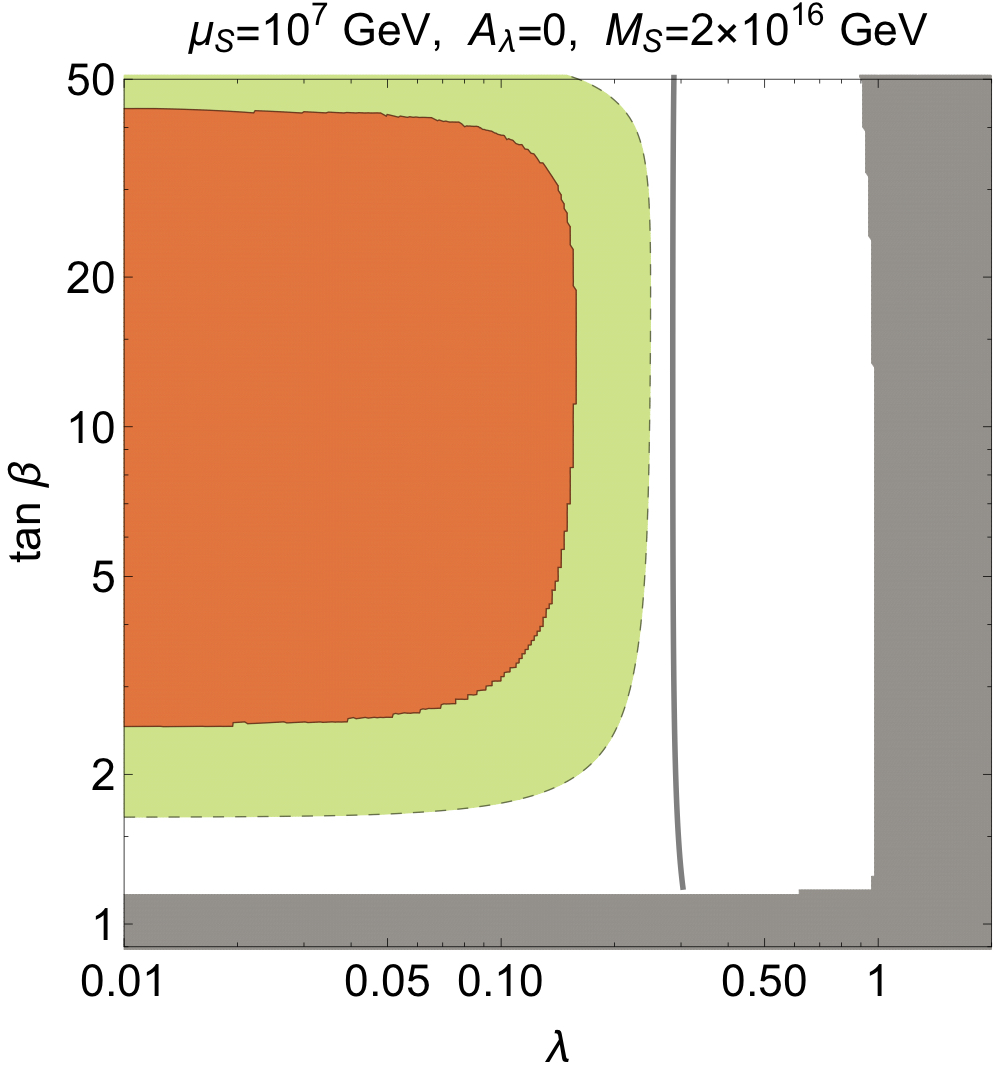}} \hspace*{0.5cm}
\subfigure{\includegraphics[width=0.43\textwidth]{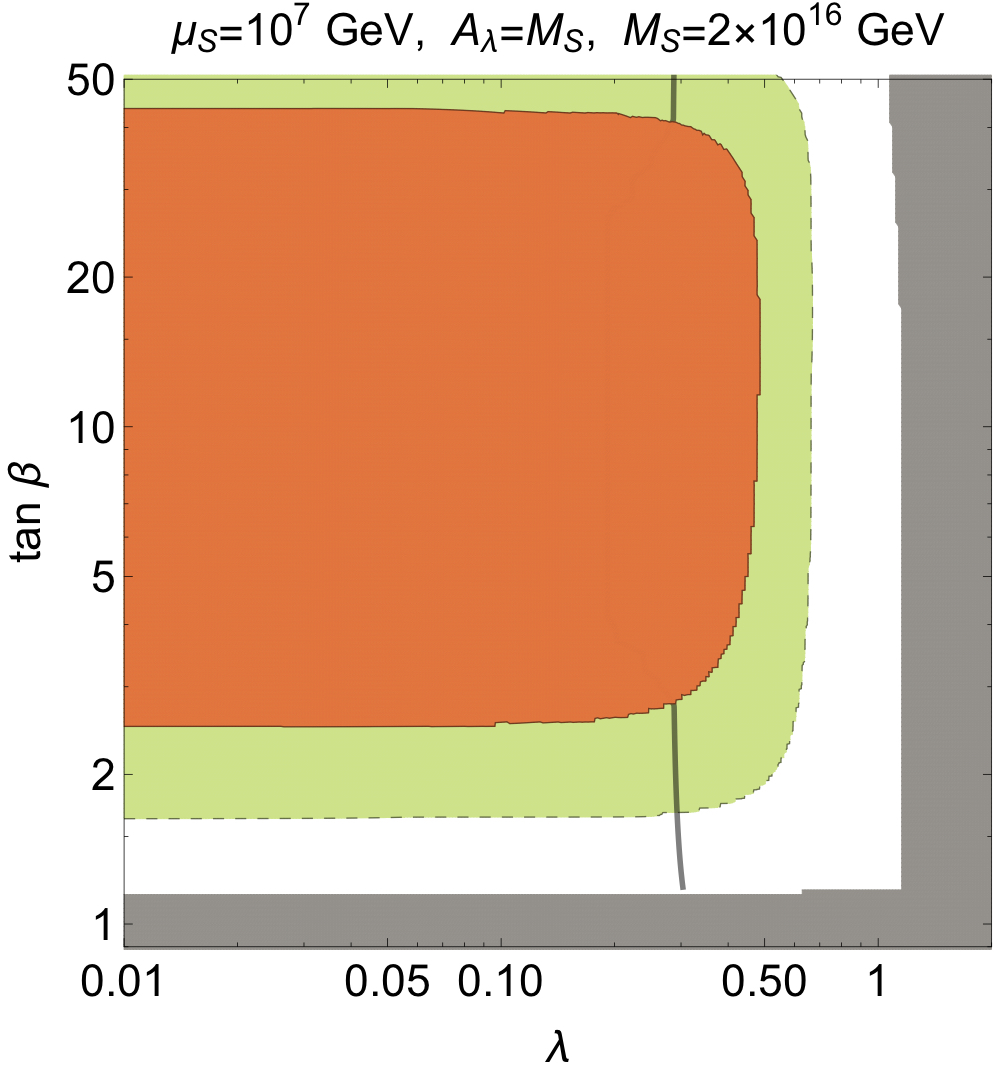}}\\
\caption{Same as in Fig. \ref{fig:stability_14} but for $M_S = 2 \times 10^{16}$ GeV.}
\label{fig:stability_16}
\end{figure}

The constraints on $\tan\beta$ and $\lambda$ arising from the vacuum stability, perturbativity and neutralino mass difference are displayed in Figs. \ref{fig:stability_14} and \ref{fig:stability_16} for two example values of $M_S$.
In all these cases, the perturbativity of the effective theory disfavours $\lambda \gsim 1$ and $\tan\beta \lsim 1.2$. The latter leads to non-perturbative values for the top quark Yukawa coupling. We find that the bound from direct detection experiment, $\Delta m_0 > 200$ keV, implies $\lambda \ge 0.03$ ($ 0.3$) for $\mu_S=10^5$ ($10^7$) GeV for all the values of $\tan\beta$. From the obtained results, we find an approximate relation:
\be \label{lambda_bound}
\frac{\lambda^2}{\mu_S} \approx 2.72 \left( \frac{\Delta m_0}{v^2}\right)\,,\ee
which is a direct consequence of eq. (\ref{bound_muS}). In this case, the perturbativity constraint alone  implies a robust upper bound, $\mu_S \le 10^8$ GeV, for a viable pseudo-Dirac Higgsino DM within this framework. 

The presence of Higgsino-singlino-Higgs Yukawa interaction also has interesting implications on the stability of the electroweak vacuum. It is known that in the case of pure THDM matched with MSSM at the GUT scale, only a narrow range in $\tan\beta \in [1.1,1.8]$ is allowed by the absolute stability of the electroweak vacuum \cite{Bagnaschi:2015pwa,Mummidi:2018nph}. This conclusion is changed in the present framework for large enough values of $\lambda$ as it can be seen from the stability, metastability and instability contours displayed in Figs. \ref{fig:stability_14} and \ref{fig:stability_16}. For example, the electroweak vacuum remains stable or metastable for all values of $\tan\beta$ if $\lambda \ge 0.1$ ($\lambda  \ge 0.3$) for $A_\lambda =0$ $(A_\lambda =M_S)$  and $M_S = 2 \times 10^{14}$ GeV. As discussed in section \ref{subsec:vacuum}, large $\lambda$ enhances the stability of potential in two ways. For $A_\lambda = 0$, $\lambda$ directly modifies the boundary value of $\lambda_4$, making it positive for $\lambda^2 > g_2^2/2$, and hence rescues the electroweak vacuum from instability. This direct effect becomes feeble if $A_\lambda \approx M_S$. However, the contribution of $\lambda$ in the running of $\lambda_4$ is still able to improve the stability of the vacuum. This indirect effect requires relatively larger values of $\lambda$ in order to achieve metastability or absolute stability in comparison to the case with $A_\lambda = 0$. We find that change in $\mu_S$, within the range $10^5$-$10^8$ GeV, has negligible effects on the results of vacuum stability. From Figs. \ref{fig:stability_14} and \ref{fig:stability_16}, it can also be observed that the instability regions grow when $M_S$ is increased because of the relatively longer running.

\begin{figure}[t]
\centering
\subfigure{\includegraphics[width=0.43\textwidth]{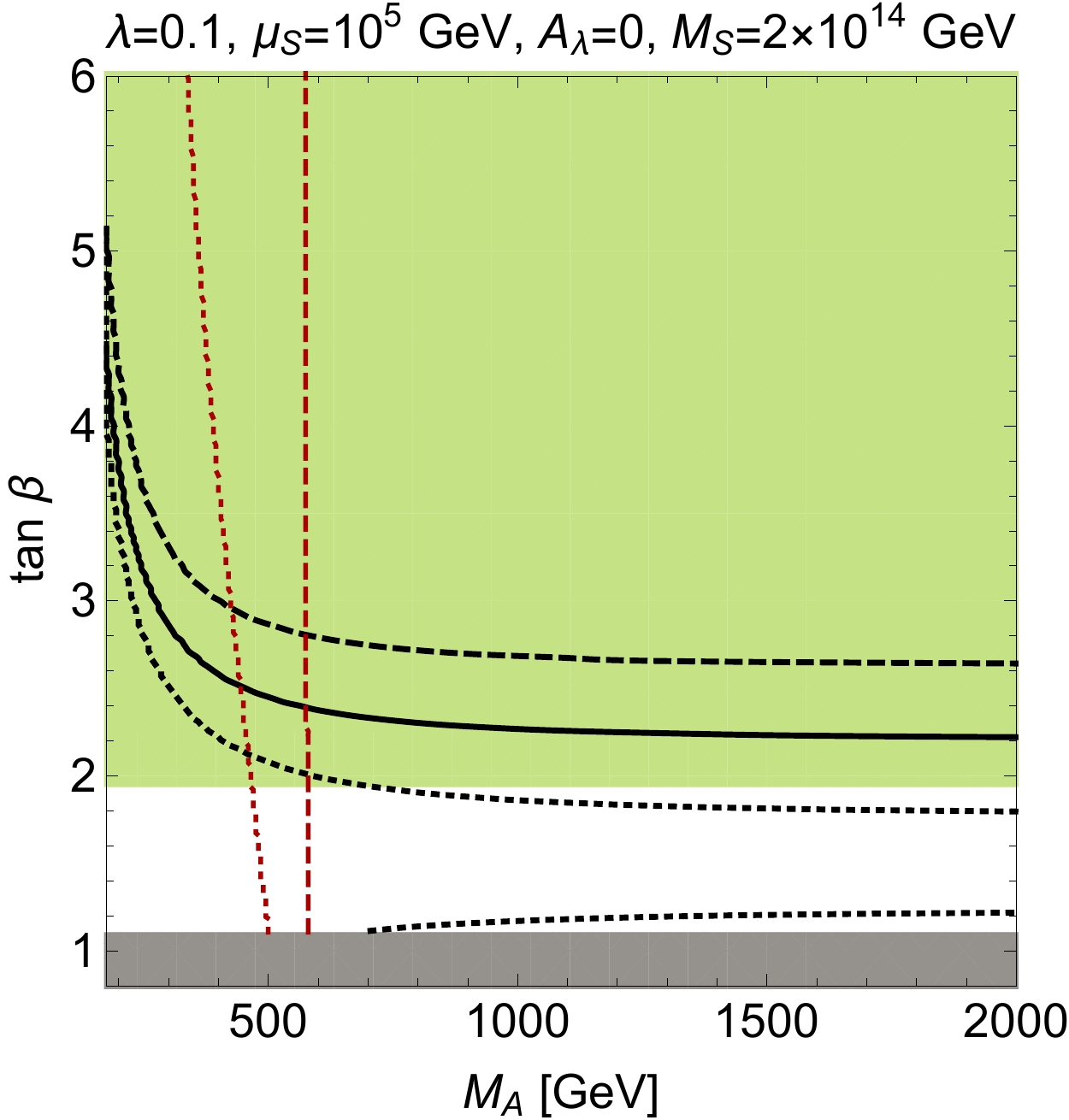}} \hspace*{0.5cm}
\subfigure{\includegraphics[width=0.43\textwidth]{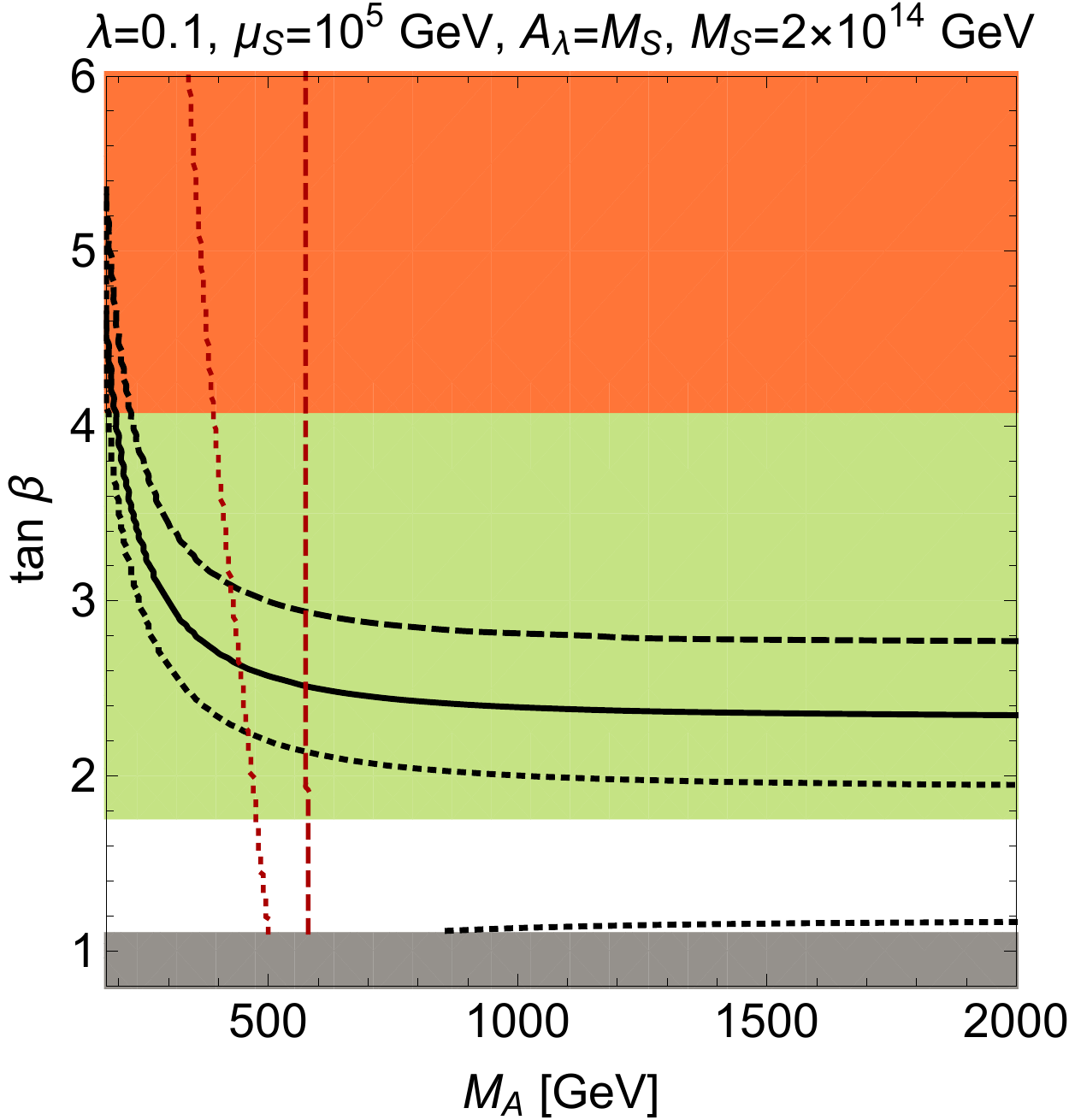}}\\
\subfigure{\includegraphics[width=0.43\textwidth]{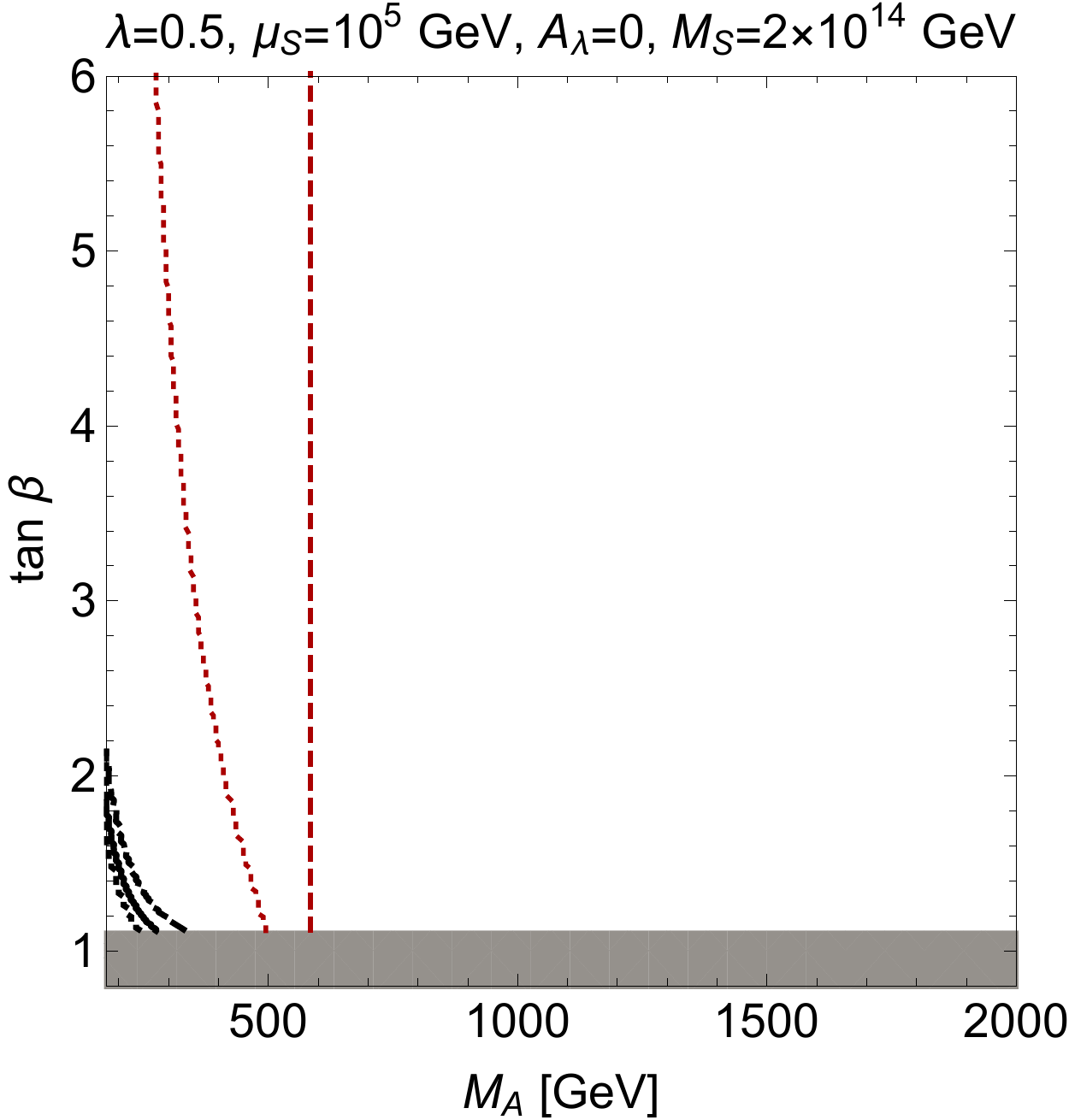}} \hspace*{0.5cm}
\subfigure{\includegraphics[width=0.43\textwidth]{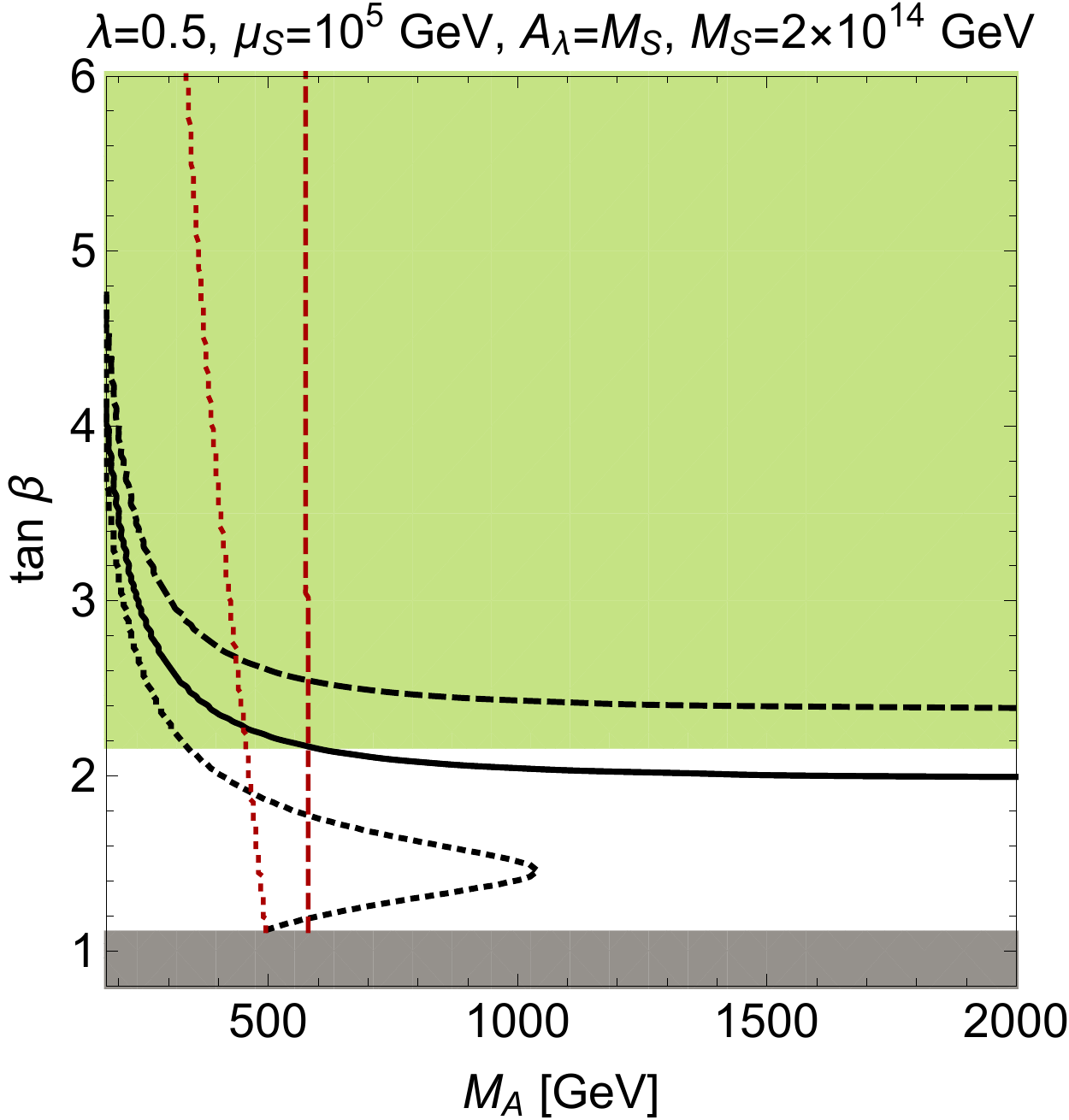}}\\
\caption{Contours of $M_h=125$ GeV (solid black), $122$ GeV (dotted black) and $128$ GeV (dashed black) on $\tan\beta$-$M_A$ plane for $M_t = 173.1$ GeV, $\mu=1$ TeV,  $\mu_S = 10^5$ GeV, $M_S = 2 \times 10^{14}$ GeV, and for different values of $\lambda$ and $A_\lambda$. The grey region is disfavoured by perturbativity while the unshaded, green and orange regions correspond to absolute stability, metastability and instability of electroweak vacuum, respectively. The region on the left side of the dashed and dotted red contours is disfavoured by the constraint $M_{H^\pm} \gsim 580$ GeV and $|\cos(\beta-\alpha)| \lsim 0.055$, respectively.}
\label{fig:Higgs_mus5} 
\end{figure}

\begin{figure}[t]
\centering
\subfigure{\includegraphics[width=0.43\textwidth]{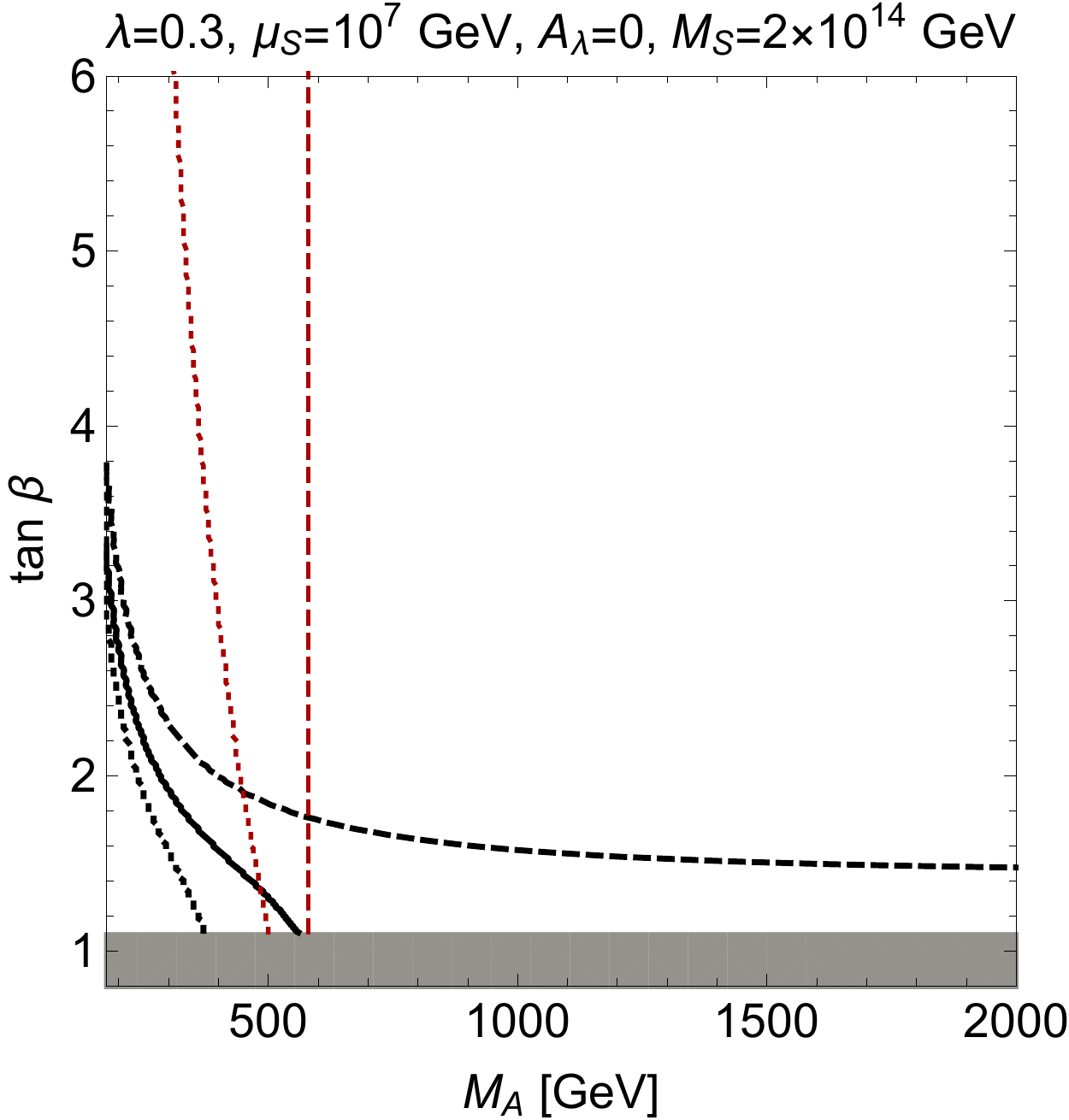}} \hspace*{0.5cm}
\subfigure{\includegraphics[width=0.43\textwidth]{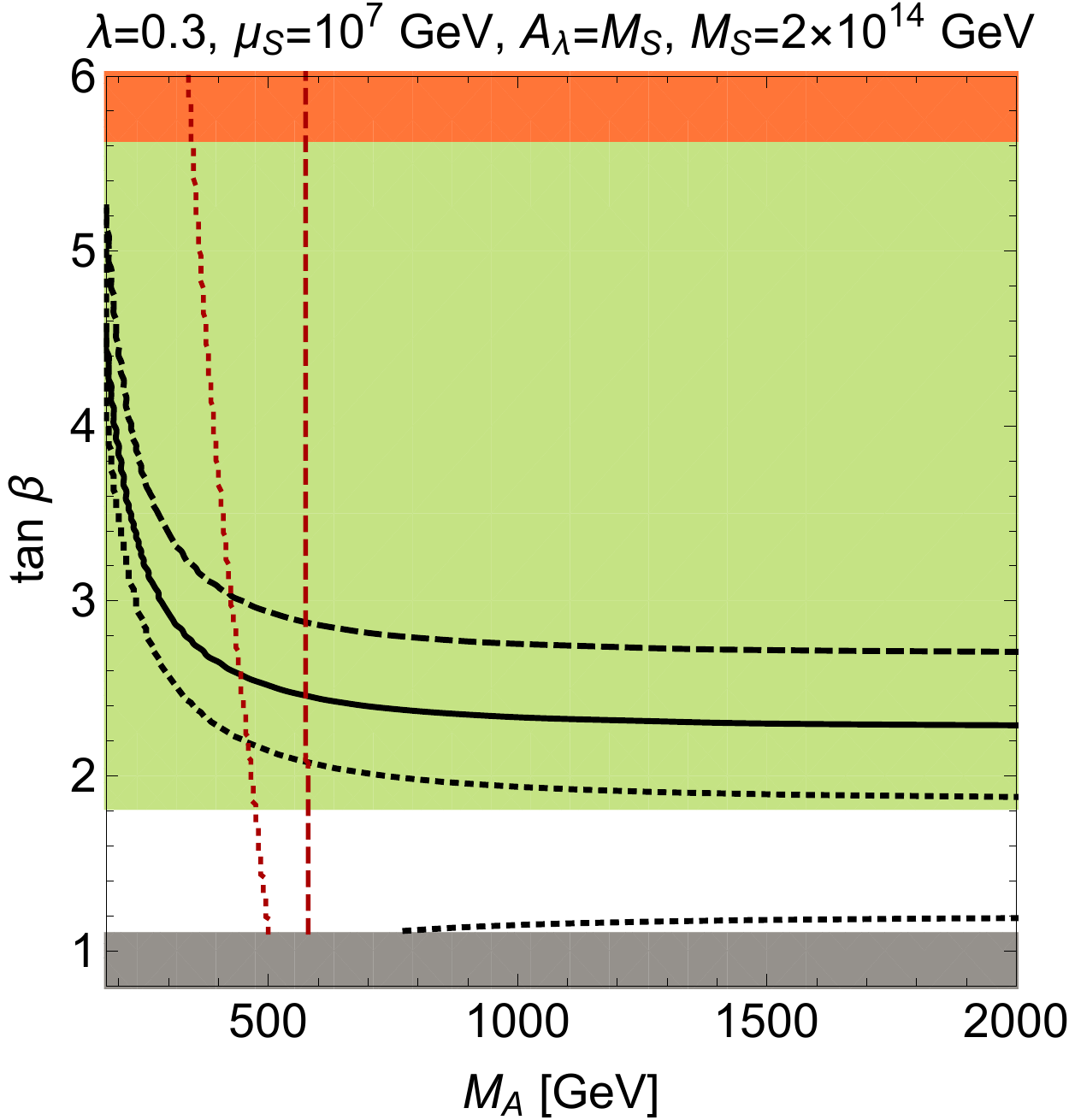}}\\
\subfigure{\includegraphics[width=0.43\textwidth]{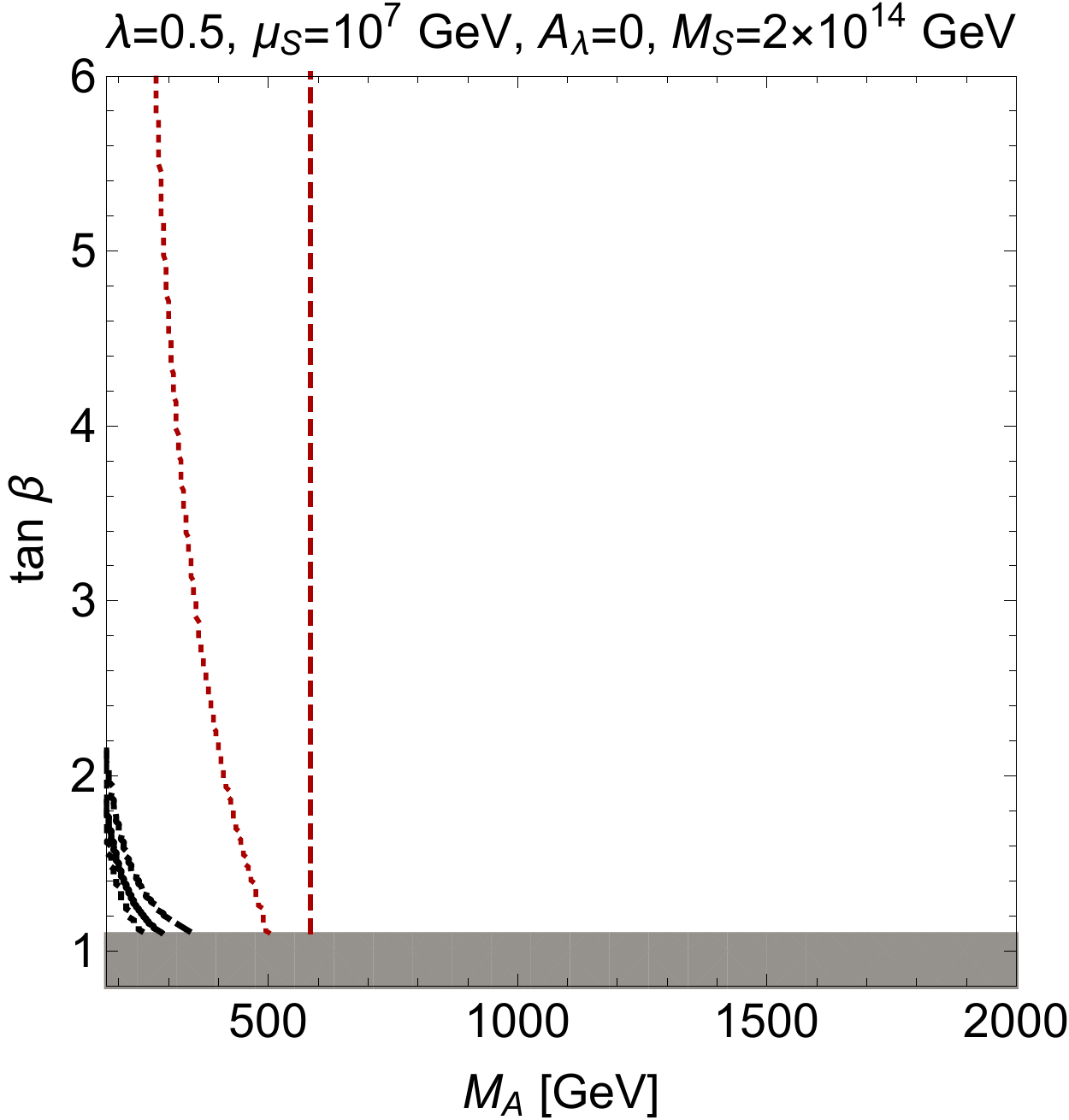}} \hspace*{0.5cm}
\subfigure{\includegraphics[width=0.43\textwidth]{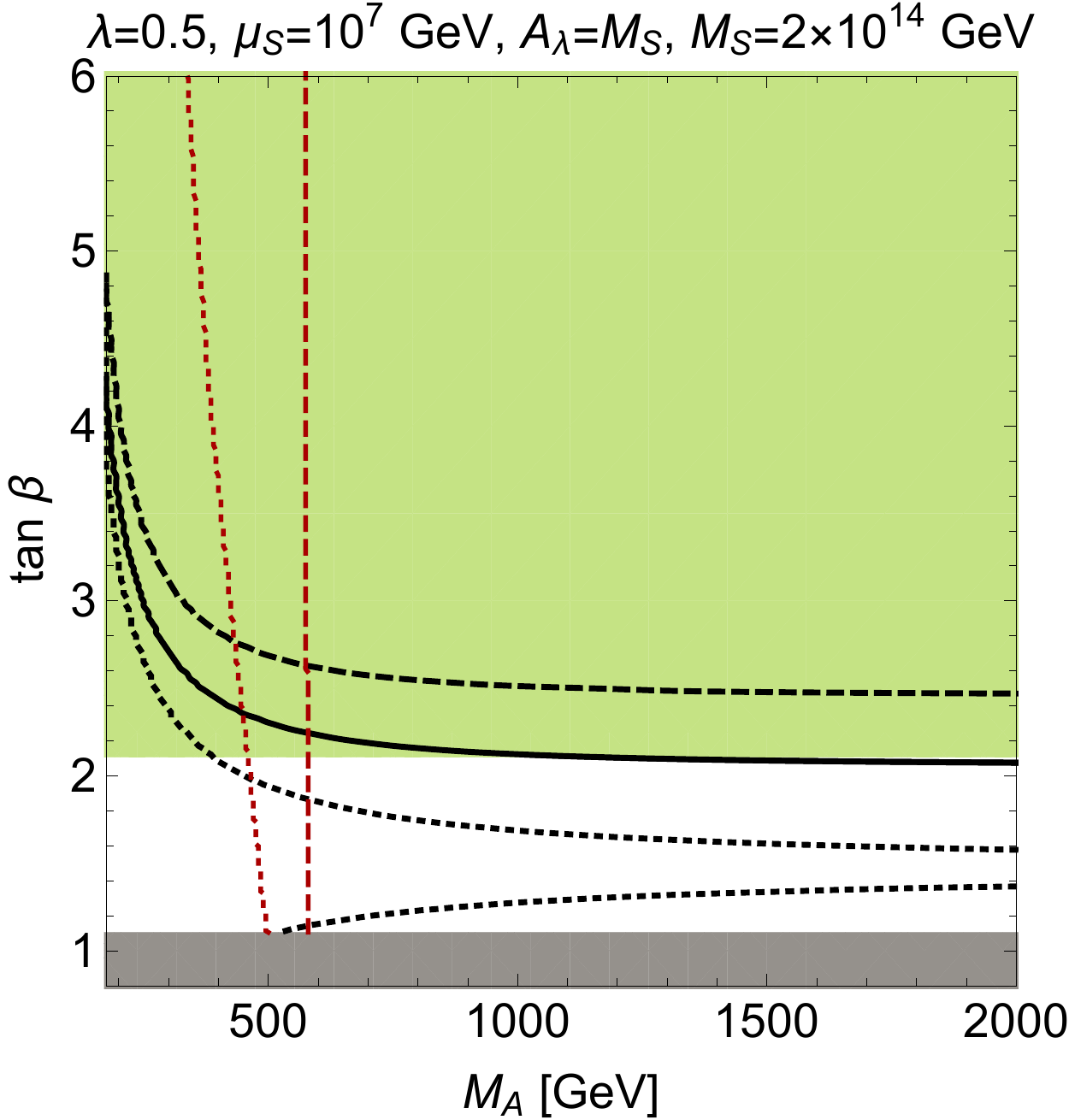}}\\
\caption{Same as in Fig. \ref{fig:Higgs_mus5} but for $\mu_S =10^7$ GeV.}
\label{fig:Higgs_mus7}
\end{figure}

We now discuss the consequences of the low energy constraints, listed in eq. (\ref{cons_lowscale}), on the parameters of the underlying framework. For this, we select two benchmark values of $\lambda$,  allowed by $\Delta m_0 \ge 200$ keV, from each of the panels in Fig. \ref{fig:stability_14} and obtain the scalar spectrum at $M_t$ as function of parameters $M_A$ and $\tan\beta$. The results are displayed in Figs. \ref{fig:Higgs_mus5} and \ref{fig:Higgs_mus7}. 
As it can be seen, $\lambda \ge 0.5$ for $A_\lambda = 0$ is disfavoured by the low energy constraints. In this case, large value of $\lambda$ results into relatively heavier SM like Higgs through $\lambda_4$ \cite{Mummidi:2018nph}. The observed range in $M_h$ requires either $\tan\beta < 1$ or $M_A < 400$ GeV. The first is ruled out by perturbativity while the later is disfavoured by the lower bound on $M_{H^\pm}$ as well as by the upper bound on $|\cos(\beta-\alpha)|$, as it can be seen from the lower left panels in Figs. \ref{fig:Higgs_mus5} and \ref{fig:Higgs_mus7}. For the cases with $A_\lambda = M_S$, there is no such restrictions on the values of $\lambda$ and it is always possible to obtain 125 GeV Higgs for some values of $\tan\beta$ and $M_A$ which are allowed by stable/metastable vacuum and other low energy constraints. As it can be seen from Figs. \ref{fig:Higgs_mus5} and \ref{fig:Higgs_mus7}, the observed Higgs mass typically requires $\tan\beta \le 3$ in all the viable cases.

We have used the top quark pole mass, $M_t=173.1$ GeV, in our analysis so far. The results, in particular, the constraint arising from the stability of vacuum and prediction for $M_h$, are sensitive to the precise value of $M_t$. In order to quantify these effects, we repeat our analysis corresponding to the results displayed in Figs. \ref{fig:Higgs_mus5} and \ref{fig:Higgs_mus7} for $M_t = 172.2$ GeV and $174$ GeV which are top quark pole masses at $-1\sigma$ and $+1 \sigma$ from the mean value, respectively. The obtained results are shown in Appendix \ref{app:top_mass}.

The discovery prospects of light Higgsinos at colliders, with the other SUSY particles much heavier, have been widely studied, see for example \cite{Berggren:2013vfa,Schwaller:2013baa,Low:2014cba,Nagata:2014wma,Han:2014kaa,Bobrovskyi:2011jj,Xiang:2016ndq,Mahbubani:2017gjh,Kowalska:2018toh}. In the absence of coloured superpartners, the Higgsinos are produced through only weak interactions and therefore their production rate is relatively smaller. Further, the collider searches are known to be highly insensitive if $\Delta m_0 < 5$ GeV \cite{Schwaller:2013baa,Han:2014kaa}. The decay of chargino into neutralino and charged pion is also difficult to detect if $\Delta m_\pm \gsim 300$ MeV \cite{Nagata:2014wma,Thomas:1998wy}. Therefore, the present framework seems to remain unconstrained from the current collider searches of Higgsinos. It is more feasible to probe almost-degenerate Higgsinos at future lepton colliders with sufficient centre-of-mass energy through tagging of the photon produced in association with a pair of neutralinos or charginos \cite{Berggren:2013vfa}. The future electron-proton colliders may also provide a potential platform for the searches of almost-degenerate Higgsinos \cite{Curtin:2017bxr}. A dedicated analysis in these directions, for the ranges of masses and couplings of Higgsinos obtained in the underlying framework, would be useful.

Although the underlying framework is shown to accommodate TeV scale pseudo-Dirac Higgsino DM consistent with GUT scale SUSY breaking, it is not complete in all aspects. One of the pertinent issues with the GUT scale SUSY breaking is that it does not lead to precision unification of the SM gauge couplings.  The convergence of the couplings, at one-loop, requires \cite{Giveon:1991zm,Patel:2011eh}
\be \label{gcu}
{\cal R} \equiv \frac{b_2 - b_3}{b_1-b_2} = \frac{5}{8}\, \frac{\sin^2\theta_W - \alpha/\alpha_S}{3/8-\sin^2\theta_W} = 0.716 \pm 0.005\,,\ee
where $b_i$ for $i=1,2,3$ are effective beta function coefficients of $U(1)$, $SU(2)_L$ and $SU(3)_C$ gauge couplings, respectively (see \cite{Patel:2011eh} for more details). One obtains ${\cal R} \approx 0.528$ and $0.556$ for the SM and THDM, respectively. The THDM together with a pair of weak scale Higgsinos lead to ${\cal R} = 0.673$, bringing it closer to the value as required in eq. (\ref{gcu}). Therefore it improves the convergence of the gauge couplings significantly. However, the scale of unification turns out to be of the ${\cal O}(10^{14})$ GeV which is one or two orders of magnitude lower than what is naively expected from the latest experimental limit on the proton lifetime \cite{Miura:2016krn}. In this case, achieving the precise unification and/or satisfying constraints from the proton lifetime require new effects at the scale $M_S$ or below. This may change the derived results depending on the exact nature of these new effects. Further, if the type-I seesaw mechanism is invoked for non-vanishing neutrino masses then the presence of singlet neutrinos may also modify the vacuum stability constraints if they are strongly coupled \cite{Mummidi:2018nph}. All these cases require dedicated analysis similar to the one presented in this paper in order to obtain a precise limit on the parameters of the underlying framework.

\section{Summary}
\label{sec:summary}
There exists a class of models in which supersymmetry is broken at the GUT scale giving rise to an effective theory which contains the SM with an additional Higgs doublet and a pair of Higgsinos. If $R$-parity is unbroken, the lightest of the neutral components of Higgsinos can be dark matter candidate. The observed thermal relic abundance can be achieved if its mass is $\sim 1$ TeV. However, in the absence of any other new physics below the GUT scale, the neutral components of Higgsinos are almost degenerate with very tiny mass splitting between them, $\Delta m_0 \approx {\cal O}(0.1)$ eV. Such an almost pure Dirac-like Higgsino DM is already disfavoured by the direct detection experiments. 

The problem can be circumvented minimally by introducing a gauge singlet Majorana fermion, namely the singlino, which is odd under $R$-parity. The Higgsino-singlino-Higgs Yukawa interaction gives rise to the Higgsino-singlino mixing through electroweak symmetry breaking, which induces large enough $\Delta m_0$ to evade the direct detection constraints. We show that the same Yukawa interaction also improves the stability of the electroweak vacuum. In order to quantify these effects, we match the effective theory with NMSSM which minimally accommodates the singlino with MSSM. We derive matching conditions between effective theory parameters and those of NMSSM and perform two loop RGE analysis with one loop threshold corrections in order to check the viability of effective theory with respect to the various phenomenological constraints.

We find that viable pseudo-Dirac Higgsino DM puts an upper bound on the singlino mass scale, $\mu_S \lsim 10^8$ GeV. The ${\cal O}(1)$ Yukawa coupling between Higgsino, singlino and Higgs makes the electroweak vacuum metastable or stable at all the scale up to $M_S$ for all the values of $\tan\beta$ allowed by perturbativity constraints. However, the observed Higgs mass together with constraints on the charged Higgs mass and Higgs to gauge boson couplings strongly disfavour $\tan\beta \gsim 3$ in almost all the cases discussed in this paper. It is shown that the viable pseudo-Dirac Higgsino DM, consistent with other phenomenological constraints, can be obtained within the underlying framework of GUT scale superymmetry for $\mu_S \in [10^5, 10^7]$ GeV, $\mu \approx 1$ TeV, $\tan\beta \in [1.2, 3]$ and with ${\cal O}(1)$ Higgsino-singlino-Higgs Yukawa coupling.

\section*{Acknowledgements}
This work is supported by Early Career Research Award (ECR/2017/000353) and by a research grant under INSPIRE Faculty Award (DST/INSPIRE/04/2015/000508) from the Department of Science and Technology, Government of India.

\appendix 
\section{Constraints from the direct detection dark matter searches}
\label{app:dm}

\subsection{Inelastic scattering}
\label{app:dm1}
A dark matter particle $\chi$ of mass $m_{\chi}$, streaming with velocity distribution $f(\varv)$ in our galaxy can  interact with target nucleus $X$ of mass $m_T$. This interaction can be recorded by observing the recoil energy, $E_R$, of the nucleus in the direct detection experiments, such as \cite{Akerib:2016vxi,Angle:2009xb,Aprile:2012nq,Aprile:2018dbl}. The differential rate of such events is given by \cite{Savage:2006qr}
\be \label{rate}
\frac{dR}{dE_R}=\frac{N_T\,m_T}{2\,m_{\chi}\,\mu_T^2}\,\rho_{\chi}\sigma\,\int_{\varv_{\rm min}}^{\varv_{\rm max}}\frac{f(\varv)}{\varv} \,d^3\varv\,,
\ee
where $\mu_T = m_\chi m_T/(m_\chi + m_T)$ is the reduced mass of the nucleus-DM system, $N_T$ is number of target nuclei in the detector, $\rho_\chi=0.3$ GeV/cm$^3$ is the local DM density in our galaxy. $\varv_{\rm min}$ is the minimum velocity of DM that can trigger a nuclear recoil of energy $E_R$. $\varv_{\rm max}$ is the maximum velocity of DM in the galaxy which is equal to $\varv_{\rm esc}$, i.e. the minimum velocity required for DM particle in order to escape from the galaxy. $\sigma$ is the scattering cross section of DM particle with nucleus which is determined by the underlying particle physics framework. The integral in eq. (\ref{rate}) is estimated in \cite{Savage:2006qr} using Maxwellian distribution for velocities. The obtained result depends on $\varv_{\rm obs}$ in addition to $\varv_{\rm min}$ and $\varv_{\rm esc}$ where  $\varv_{\rm obs}$ takes into account for the relative motion between the earth and the rest frame of DM. We use the result of \cite{Savage:2006qr} for estimation of $\varv_{\rm obs}$.

In the case of elastic scattering with a nucleus, i.e. $\chi+X \rightarrow \chi+X$, the minimum velocity of DM particle required to produce nuclear recoil with energy $E_R$ is given by
\be \label{vmin}
\varv_{\rm min}=\frac{1}{\sqrt{2\,m_T\,E_R}} \left(\frac{m_T\,E_R}{\mu_T}\right)\,,
\ee
If $\chi$ is a pure Higgsino DM with $m_\chi \sim 1$ TeV, the number of events estimated using the above formulae turns out to be much larger than the total observed events in the experiments such as  Xenon 10 and Xenon 100. Therefore, the case of pure Dirac Higgsino DM is disfavoured.

In the case of pseudo-Dirac Higgsinos discussed in the paper, we have Majorana neutralinos $\tilde{\chi}^0_1$ and $\tilde{\chi}^0_2$ with mass difference $\Delta m_0$ as given by eq. (\ref{Dm0}). In this case $\tilde{\chi}^0_1$ can scatter inelastically off the nucleus giving rise to a process: $\tilde{\chi}^0_1 +X \rightarrow \tilde{\chi}^0_2+X$. If $\Delta m_0$ is small enough, the nuclear recoil can also be observed in direct detection experiments. The value of $\varv_{\rm min}$ that can give rise to nuclear recoil with energy $E_R$ is then given by
\be \label{vmin_in}
\varv_{\rm min}=\frac{1}{\sqrt{2\,m_T\,E_R}} \left(\frac{m_T\,E_R}{\mu_T}+\Delta m_0 \right)\,,
\ee
and the scattering cross section $\sigma$ in eq. (\ref{rate}) is replaced by inelastic cross section, $\sigma^{\rm Inelastic}\equiv \sigma(\tilde{\chi}^0_1 +X \rightarrow \tilde{\chi}^0_2+X)$. In the framework under consideration, the dominant contribution to $\sigma^{\rm Inelastic}$ arises from the exchange of $Z$ bosons as the couplings between $\tilde{\chi}^0_{1,2}$ and THDM scalars are suppressed by ${\cal O}(v/\mu_S)$, see eq. (\ref{L_Higgsino_mass}). Using eq. (\ref{basis_change}) and kinetic term of Higgsinos in eq. (\ref{L_Higgsino_mass}), one obtain the following effective interaction between the $\tilde{\chi}^0_{1,2}$ and quarks after integrating out the $Z$ boson:
\be \label{L-in}
{\cal L}^{\rm Inelastic} = - i \sin 2\theta\, \frac{G_F}{\sqrt{2}} (T_3^q - 2 Q_q \sin^2\theta_W)\, \overline{\tilde{\chi}^0_2} \gamma^\mu \tilde{\chi}^0_1\, \overline{q} \gamma_\mu q\,,
\ee
where $T_3^q$ ($Q_q$) is $1/2$ ($2/3$) for $q=u,c,t$ and $-1/2$ ($-1/3$) for $q=d,s,b$. $G_F$ is Fermi constant and $\theta_W$ is weak mixing angle. We have $\sin 2 \theta \approx 1$ for pseudo-Dirac Higgsinos in our framework. The effective spin-independent inelastic cross section between the neutralinos and proton or neutron are estimated by summing over the valance quark contributions using the above results. This gives the following result for the cross section between the neutralinos and nucleus 
\be \label{sigma_inelastic}
\sigma^{\rm Inelastic}=\frac{G_F^2\,\mu_T^2}{8\,\pi}\,\left[N_n-(1-4\,\sin^2{\theta_W})\,N_p\right]^2\,F(E_R)^2\,,
\ee
where $N_n$ ($N_p$) is number of neutrons (protons) in a given nucleus and $F(E_R)$ is the nuclear form factor which parametrizes the distributions of protons and neutrons in the nucleus \cite{Duda:2006uk}.

\begin{figure}[t]
\centering
\subfigure{\includegraphics[width=0.5\textwidth]{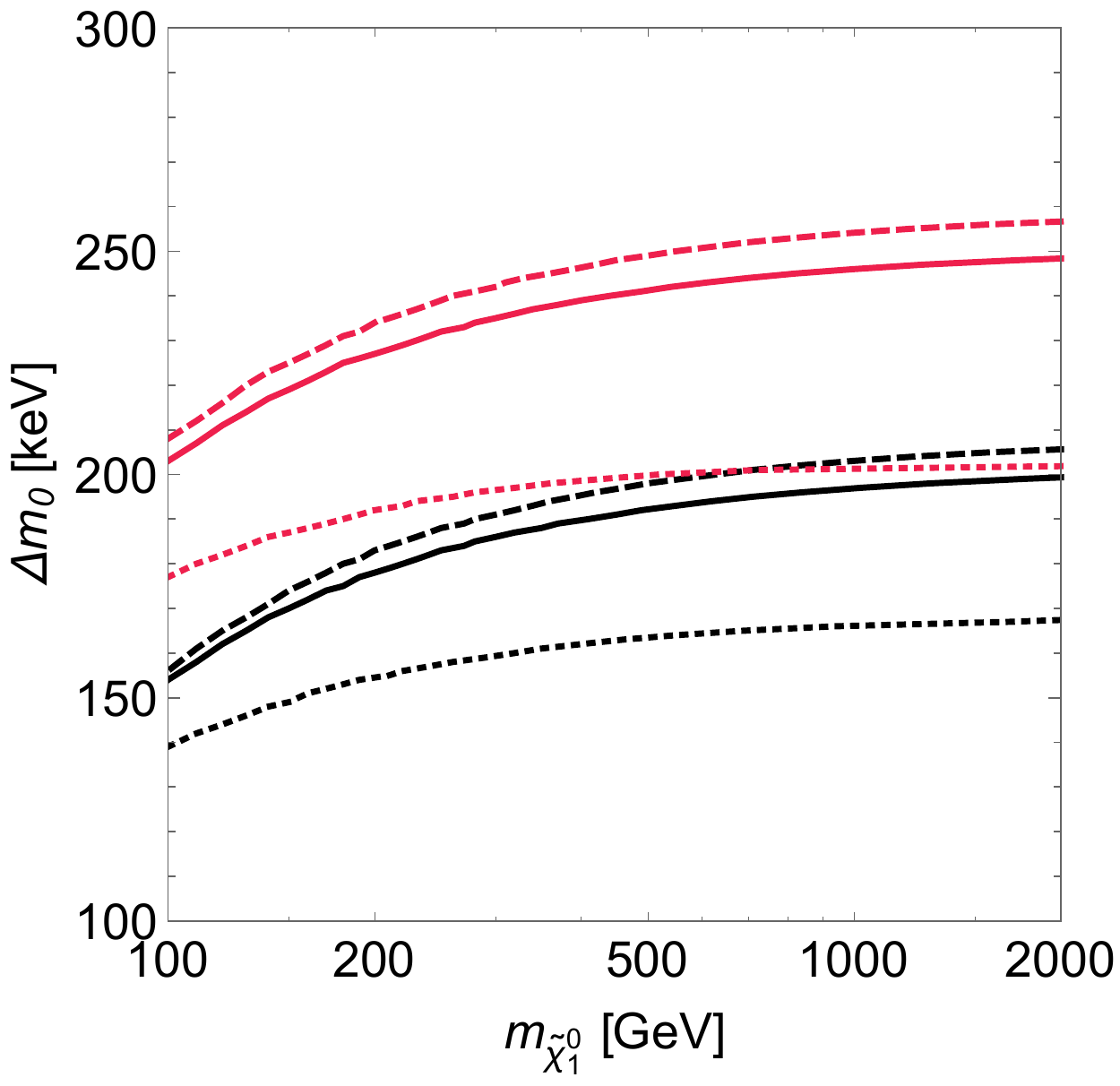}}
\caption{Constraint on $\Delta m_0$ for given $m_{\tilde{\chi}^0_1}$. The regions below solid, dashed and dotted lines are disfavoured at $90\%$ confidence level by the Xenon 1T, Xenon 100 and Xenon 10 data, respectively. The upper red lines correspond to $\varv_{\rm esc.} = 650$ km/sec while the lower black lines are for $\varv_{\rm esc.} = 500$ km/sec.}
\label{fig:Xenon}
\end{figure}

We estimate the total number of events $R$ using eqs. (\ref{rate},\ref{vmin_in},\ref{sigma_inelastic}) for given range of recoil energy $E_R$. The results are then compared with the data collected by the experiments: Xenon 10 with 58.6 live days of data and a target mass of 5.4 Kg  \cite{Angle:2009xb}, Xenon 100 with 224.6 live days of data and a target mass of 34 kg  \cite{Aprile:2012nq} , and Xenon 1T  with 278.8 live days of data with target mass of 1300 Kg \cite{Aprile:2018dbl}. The values of parameters, $m_{\tilde{\chi}^0_1}$ and $\Delta m_0$, are excluded with 90\% confidence level if theoretically estimated number of events are greater than the observed number of events in accordance with poisson statistics.  The results are displayed in Fig. \ref{fig:Xenon}.
We find a conservative limit $\Delta m_0 \geq 200$ keV for $m_{\tilde{\chi}^0_1} \approx 1$ TeV in order to evade the constraints from direct detection experiments on pseudo-Dirac Higgsino DM.

\subsection{Elastic scattering}
\label{app:dm2}
In this section we discuss the elastic scattering of pseudo-Dirac Higgsino DM in the underlying framework. The relavant terms can be read from the effective Lagrangian given in eq. (\ref{L_Higgsino_mass}) and are given by
\beqa \label{elastic_L}
{\cal L}_{\tilde{H}}^{\rm eff.} &\supset & -\frac{g_Z}{4}\,\cos 2\theta\,\overline{\tilde{\chi}^0_1}\gamma^\mu \gamma^5 \tilde{\chi}^0_1\, Z_\mu + \frac{v}{2\,\mu_S}\left(g_{h}\,h+g_H\,H\right) \overline{\tilde{\chi}^0_1}\tilde{\chi}^0_1\,, \eeqa
where 
\beqa \label{g_i_DM}
g_Z &=& \sqrt{g_Y^2 + g_2^2}\,, \nonumber \\
g_h &=& -c_1\, \sin\alpha \cos\beta + c_2\,\cos\alpha \sin\beta + d \, \cos(\alpha+\beta)\,, \nonumber \\
g_H &=&c_1\, \cos\alpha \cos\beta + c_2\, \sin\alpha \sin\beta + d \, \sin(\alpha+\beta)\,. \nonumber \eeqa
The first term in eq. (\ref{elastic_L}) gives rise to spin-dependent (SD) contribution while the remaining terms induce spin-independent (SI) contributions to the elastic scattering. Upon integrating out the $Z$ boson and THDM scalars, one obtains the following relevant effective operators:
\beqa \label{eff_elastic}
{\cal L}^{\rm Elastic}_{\rm SI} & =& \sum_q \frac{m_q}{4 \mu_S} \left(\frac{g_h b_h^q}{M_h^2}+\frac{g_H b_H^q}{M_H^2}  \right)\, \overline{\tilde{\chi}^0_1}\tilde{\chi}^0_1 \bar{q}q\,, \nonumber \\
{\cal L}^{\rm Elastic}_{\rm SD} & =& \sum_q \cos2\theta \frac{G_F}{\sqrt{2}} T^q_3\,  \overline{\tilde{\chi}^0_1}\gamma^\mu \gamma^5 \tilde{\chi}^0_1\, \bar{q}\gamma_\mu \gamma^5q\,,
\eeqa
where $b^q_h = - b^q_H= - \sin\alpha/\sin\beta$ for $q=u,c,t$ and $b^q_h = b^q_H= \cos\alpha/\cos\beta$ for $q=d,s,b$.  The SD cross section is proportional to $\cos2\theta$ hence it is negligible in our framework. 

Using the first in eq. (\ref{eff_elastic}), we compute SI cross section of DM with a single proton which is given by
\be \label{Sigma_SI}
\sigma^{\rm SI}_p=\frac{4}{\pi}\mu_p^2 f_p^2\,, \ee
where
\be \label{fp}
f_p=\sum_{\phi=h,H}\frac{m_p g_\phi}{8\mu_S M_\phi^2}\left(\sum_{q=u,d,s}b_\phi^qf_{Tq}^{(p)}+\frac{2}{27}f_{TG}^{(p)}\sum_{q=c,b,t}b_\phi^q\right)\, \ee
and $m_p$ is mass of proton. The factors $f_{Tu}^{(p)}=0.019$, $f_{Td}^{(p)}=0.029$, $f_{Ts}^{(p)}=0.009$ and $f_{TG}^{(p)}=1-\sum_{u,d,s}f_{Tq}^{(p)}$ are obtained from the lattice computation  \cite{Oksuzian:2012rzb}. We estimate $\sigma^{\rm SI}_p$ for the values of parameters favoured by our results, i.e. $\tan\beta = 1.5$, $\beta-\alpha = \pi/2$, $M_h = 125$ GeV, $M_H = 1$ TeV and $\mu_S=10^5$ GeV.We find $\sigma^{\rm SI}_p \approx 10^{-57}$ cm$^2$ which is much smaller than the sensitivity ($\gsim 10^{-44}$ cm$^2$)  of current generation experiments \cite{Akerib:2016vxi}. Typically, the neutrino background, with cross section $\sim 10^{-48}$ cm$^2$, dominates over the DM signal \cite{Billard:2013qya}. Hence, it remains challenging to constraint the pseudo-Higgsino DM discussed in this framework from their elastic scattering signatures.

\section{Renormalization group equations}
\label{app:RGE}
We list 2-loop renormalization group equations for various couplings appearing in our framework which are obtained using publicly available package SARAH \cite{Staub:2013tta}. The couplings evolve according to the following equation:
\begin{eqnarray} \label{RG}
Q\,\frac{d C}{dQ} &=& \frac{1}{16\,\pi^2}\,\left(\beta_{C}^{(1,0)}+\Theta(Q-\mu)\, \beta_{C}^{(1,1)}+\Theta(Q-\mu_S)\, \beta_{C}^{(1,2)}\right)\\ \nonumber
&+&\Big(\frac{1}{16\,\pi^2}\Big)^2\,\left(\beta_{C}^{(2,0)}+\Theta(Q-\mu)\, \beta_{C}^{(2,1)}+\Theta(Q-\mu_S)\, \beta_{C}^{(2,2)}\right)\,,
\end{eqnarray}
where $C$ represents gauge, Yukawa or quartic couplings and $Q$ is the renormalization scale. The step function $\Theta(Q-Q_0) = 1$ for $Q>Q_0$ and it vanishes otherwise. The one and two-loop beta functions for the different couplings are as the following.

\subsection{Gauge couplings}
{\allowdisplaybreaks  \begin{align} 
	\beta_{g_1}^{(1,0)} & =  
	\frac{21}{5}\, g_{1}^{3},\nonumber \\ 
		\beta_{g_1}^{(1,1)} & =  
	\frac{2}{5}\, g_{1}^{3}, \nonumber \\ 
	\beta_{g_1}^{(1,2)} & =  0, \nonumber \\ 
	\beta_{g_1}^{(2,0)} & =  
	\frac{1}{50} \,g_{1}^{3}\, \Big(180\, g_{2}^{2}  + 208\, g_{1}^{2}  + 440\, g_{3}^{2}  -75 \,\mbox{Tr}\Big({Y_{e}^{\dagger}Y_e  }\Big) \Big) \nonumber\\
	& -25 \,\mbox{Tr}\Big({ Y_{d}^{\dagger}Y_d }\Big)  -85\, \mbox{Tr}\Big({Y_{u}^{\dagger} Y_u  }\Big) \Big),\nonumber \\ 
		\beta_{g_1}^{(21)} & = 	\frac{1}{50} \,g_{1}^{3}\, \Big( 9 g_1^2+45 g_2^2\Big),\nonumber \\ 
			\beta_{g_1}^{(22)} & = -\frac{15}{50} \,g_{1}^{3}\, \Big(y_1^2+y_2^2+y_3^2+y_4^2\Big),\nonumber \\ 
	\beta_{g_2}^{(1,0)} & =  
	-3\, g_{2}^{3} ,\nonumber \\ 
	\beta_{g_2}^{(1,1)} & =  \frac{2}{3}\, g_{2}^{3}, \nonumber \\ 
	\beta_{g_2}^{(1,2)} & = 0,\nonumber \\ 
	\beta_{g_2}^{(2,0)} & =  
	\frac{1}{10}\, g_{2}^{3}\, \Big(80\, g_{2}^{2}+120\, g_{3}^{2}  + 12\, g_{1}^{2}  -5\, \mbox{Tr}\Big({Y_e  Y_{e}^{\dagger}}\Big)  \Big)-15 \,\mbox{Tr}\Big({Y_d  Y_{d}^{\dagger}}\Big) -15\, \mbox{Tr}\Big({Y_u  Y_{u}^{\dagger}}\Big)     \Big),\nonumber \\ 
		\beta_{g_2}^{(2,1)} & =  
	 g_{2}^{3}\, \Big(\frac{3}{10}g_1^2+\frac{49}{6}g_2^2\Big),\nonumber \\ 
	\beta_{g_2}^{(2,2)} & =  
	-\frac{1}{2}\, g_{2}^{3}\, \Big(y_1^2+y_2^2+y_3^2+y_4^2\Big),\nonumber \\ 
	\beta_{g_3}^{(1,0)} & =  
	-7\, g_{3}^{3},\nonumber \\ 
		\beta_{g_3}^{(1,1)} & =  0,\nonumber \\ 
		\beta_{g_3}^{(1,2)} & =  0,\nonumber \\ 
		\beta_{g_3}^{(2,1)} & =  0,\nonumber \\ 
		\beta_{g_3}^{(2,2)} & = 0,\nonumber \\ 
	\beta_{g_3}^{(2,0)} & =  
	-\frac{1}{10}\, g_{3}^{3}\, \Big(-11\, g_{1}^{2}  + 260\, g_{3}^{2}  -45\, g_{2}^{2}+ 20\, \mbox{Tr}\Big({Y_d  Y_{d}^{\dagger}}\Big)  + 20\, \mbox{Tr}\Big({Y_u  Y_{u}^{\dagger}}\Big)   \Big).\nonumber 
	\end{align}} 
\subsection{Yukawa couplings}
{\allowdisplaybreaks  \begin{align} 
	\beta_{Y_u}^{(1,0)} & =  
	Y_u\, \Big(-8\, g_{3}^{2}  -\frac{17}{20}\, g_{1}^{2}  -\frac{9}{4}\, g_{2}^{2} + 3\, \mbox{Tr}\Big({Y_{u}^{\dagger}  Y_{u}}\Big)\Big)+\frac{1}{2} \,\Big(3\,{Y_u  Y_{u}^{\dagger}  Y_u}  + {Y_d  Y_{d}^{\dagger}  Y_u  }\Big), \nonumber \\ 
		\beta_{Y_u}^{(1,1)} & =0,\nonumber \\ 
		\beta_{Y_u}^{(1,2)} & =Y_u\Big( |y_2|^2 + |y_3|^2\Big),\nonumber \\ 
\beta_{Y_u}^{(2,0)} & =  Y_u \,\Big(\frac{1267}{600} \,g_{1}^{4} -\frac{9}{20}\, g_{1}^{2} g_{2}^{2} -\frac{21}{4}\, g_{2}^{4} +\frac{19}{15}\, g_{1}^{2} g_{3}^{2} +9\, g_{2}^{2} g_{3}^{2} -108\, g_{3}^{4} +\frac{3}{2}\, \lambda_{2}^{2} +\lambda_{3}^{2}\nonumber\\
	&+\lambda_3 \lambda_4 +\lambda_{4}^{2}+\frac{1}{8}\, \Big(160\, g_{3}^{2}  + 17\, g_{1}^{2}  + 45 \,g_{2}^{2} \Big)\,\mbox{Tr}\Big({Y_u  Y_{u}^{\dagger}}\Big) +\frac{3}{8} \,\Big(5\, g_{2}^{2}  \Big)\nonumber\\
	&-\frac{9}{4} \,\mbox{Tr}\Big({Y_d  Y_{d}^{\dagger}  Y_u  Y_{u}^{\dagger}}\Big)\Big) -\frac{27}{4}\, \mbox{Tr}\Big({Y_u  Y_{u}^{\dagger}  Y_u  Y_{u}^{\dagger}}\Big) \Big) \Big)\nonumber\\ 
	&+{Y_d  Y_{d}^{\dagger}  Y_u}\, \Big(-2\, \lambda_3  + 2\, \lambda_4  + \frac{16}{3} \,g_{3}^{2}  + \frac{33}{16}\, g_{2}^{2}  -\frac{3}{4}\, \mbox{Tr}\Big({Y_e  Y_{e}^{\dagger}}\Big)  -\frac{41}{240}\, g_{1}^{2}  -\frac{9}{4}\, \mbox{Tr}\Big({Y_d  Y_{d}^{\dagger}}\Big) \Big)\nonumber \\ 
	&+\frac{1}{80}\,{Y_u  Y_{u}^{\dagger}  Y_u} \Big(1280\, g_{3}^{2}  \Big)  + 223\, g_{1}^{2}  -540\, \mbox{Tr}\Big({Y_u  Y_{u}^{\dagger}}\Big)  + 675\, g_{2}^{2}  -480\, \lambda_2 \Big)\nonumber\\
	&-\frac{1}{4} \, \Big(-6\, {Y_u  Y_{u}^{\dagger}  Y_u  Y_{u}^{\dagger}  Y_u}  + {Y_d  Y_{d}^{\dagger}  Y_d  Y_{d}^{\dagger}  Y_u} + {Y_u  Y_{u}^{\dagger}  Y_d  Y_{d}^{\dagger}  Y_u}\Big), \nonumber \\ 
	\beta_{Y_u}^{(2,1)} & =Y_u\Big[\frac{38}{200}g_1^4+\frac{1}{2}g_2^4\Big],\nonumber \\ 
	\beta_{Y_u}^{(2,2)}&=Y_u\Big(-\frac{9}{4} |y_2|^4 -\frac{9}{4} |y_3|^4 +\frac{1}{8} y_2^* \Big(-12 y_2 |y_1|^2  + 3 y_2 \Big(5 g_{2}^{2}  -6 |y_4|^2  + g_{1}^{2}\Big) -8 \Big(5 y_2 y_3  + y_1 y_4 \Big)y_3^* \Big)\nonumber\\
	&- y_2 y_3 y_1^* y_4^* +\frac{3}{8} |y_3|^2 \Big(-4 y_4 y_4^*  + 5 g_{2}^{2}  -6 y_1 y_1^*  + g_{1}^{2}\Big)-\Big(\frac{3}{4}y_1^2+\frac{3}{4}y_4^2\Big)Y_u  Y_{d}^{\dagger}  Y_d\nonumber\\
	&-\Big(\frac{9}{4} y_2^2+\frac{9}{4} y_3^2\Big) Y_u  Y_{u}^{\dagger}  Y_u,\nonumber \\ 
	\beta_{Y_d}^{(1,0)} & =  Y_d\,\big(-8\, g_{3}^{2}  -\frac{1}{4}\, g_{1}^{2}  -\frac{9}{4}\, g_{2}^{2} +3\, \mbox{Tr}\Big({Y_d  Y_{d}^{\dagger}}\Big)  + \mbox{Tr}\Big({Y_e  Y_{e}^{\dagger}}\Big)\Big)+\frac{1}{2} \,\Big(3\,{Y_d  Y_{d}^{\dagger}  Y_d}  + {Y_u  Y_{u}^{\dagger}  Y_d  }\Big), \nonumber \\ 
		\beta_{Y_d}^{(1,1)} & =0,\nonumber \\ 
	\beta_{Y_d}^{(1,2)} & =Y_d (y_1^2+y_4^2),\nonumber \\ 
	\beta_{Y_d}^{(2,0)} & =  
	Y_d \,\Big(-\frac{113}{600}\, g_{1}^{4} -\frac{27}{20}\, g_{1}^{2} g_{2}^{2} -\frac{21}{4}\, g_{2}^{4} +\frac{31}{15}\, g_{1}^{2} g_{3}^{2} +9\, g_{2}^{2} g_{3}^{2} -108\, g_{3}^{4} +\frac{3}{2}\, \lambda_{1}^{2} +\lambda_{3}^{2}\nonumber\\
	&+\lambda_3 \lambda_4 +\lambda_{4}^{2}+\frac{5}{8}\, \Big(32\, g_{3}^{2}  + 9\, g_{2}^{2}  + g_{1}^{2}\Big)\mbox{Tr}\Big({Y_d  Y_{d}^{\dagger}}\Big) +\frac{15}{8}\, \Big(g_{1}^{2} + g_{2}^{2}\Big)\mbox{Tr}\Big({Y_e  Y_{e}^{\dagger}}\Big)\nonumber\\
	&-\frac{27}{4}\, \mbox{Tr}\Big({Y_d  Y_{d}^{\dagger}  Y_d  Y_{d}^{\dagger}}\Big)-\frac{9}{4}\, \mbox{Tr}\Big({Y_{d}^{\dagger} Y_u Y_{u}^{\dagger} Y_d }\Big) -\frac{9}{4}\, \mbox{Tr}\Big({Y_e  Y_{e}^{\dagger}  Y_e Y_{e}^{\dagger} }\Big) \Big) \Big)\nonumber\\
	&+\frac{1}{80}\, {Y_d  Y_{d}^{\dagger}  Y_d}\, \Big(1280\, g_{3}^{2}  -180\, \mbox{Tr}\Big({Y_e  Y_{e}^{\dagger}}\Big)  + 187\, g_{1}^{2}  -540\, \mbox{Tr}\Big({Y_d  Y_{d}^{\dagger}}\Big)  + 675\, g_{2}^{2}  -480\, \lambda_1 \Big)\nonumber \\ 
	&+\frac{1}{240}\,{Y_u  Y_{u}^{\dagger} Y_d  } \,\Big(1280\, g_{3}^{2} \Big)  -480\, \lambda_3  + 480\, \lambda_4  + 495\, g_{2}^{2}  -53 \,g_{1}^{2} \nonumber\\ 
	&-540\, \mbox{Tr}\Big({Y_u  Y_{u}^{\dagger}}\Big) \Big)
	+\frac{1}{4} \, \Big(6\, {Y_d  Y_{d}^{\dagger}  Y_d  Y_{d}^{\dagger}  Y_d}  - {Y_d Y_{d}^{\dagger} Y_u  Y_{u}^{\dagger} Y_d  }  - { Y_u Y_{u}^{\dagger} Y_u Y_{u}^{\dagger} Y_d } \Big),\nonumber \\ 
	\beta_{Y_d}^{(2,1)} & =-Y_d \Big[\frac{4}{600}g_1^4-\frac{1}{2}g_2^4\Big],\nonumber \\ 
	\beta_{Y_d}^{(2,2)}&=Y_d\Big(+\frac{3}{8} g_{1}^{2} |y_4|^2 +\frac{15}{8} g_{2}^{2} |y_4|^2 -\frac{9}{4} |y_1|^4 -\frac{9}{4} |y_4|^4 -\frac{3}{2} y_4 |y_3|^2 y_4^* -\frac{1}{4} y_4 y_2^* \Big(4 y_1 y_3^*  + 9 y_2 y_4^* \Big)\nonumber \\ 
	&+\frac{1}{8} y_1^* \Big(-12 y_1 |y_2|^2  + 15 g_{2}^{2} y_1  -18 y_1 |y_3|^2  + 3 g_{1}^{2} y_1  -40 y_1 |y_4|^2  -8 y_2 y_3 y_4^* \Big)\Big)\nonumber\\
	&-\frac{3}{4}(y_2^2+y_3^2) Y_d  Y_{u}^{\dagger}  Y_u-\frac{9}{4} (y_1^2+y_4^2) Y_d  Y_{d}^{\dagger}  Y_d,\nonumber \\ 
    \beta_{Y_e}^{(1,0)} & =  
	\frac{1}{4} \, \Big(6\, {Y_e  Y_{e}^{\dagger}  Y_e} \Big) + Y_e\, \Big(3\, \mbox{Tr}\Big({Y_d  Y_{d}^{\dagger}}\Big)  + \mbox{Tr}\Big({Y_e  Y_{e}^{\dagger}}\Big)  -\frac{9}{4} \Big(g_{1}^{2} + g_{2}^{2}\Big)\Big),\nonumber \\ 
		\beta_{Y_e}^{(1,1)} & =0,\nonumber \\ 
	\beta_{Y_e}^{(1,2)} & =Y_e (y_1^2+y_4^2),\nonumber \\ 
	\beta_{Y_e}^{(2,0)} & =  Y_e\, \Big(\frac{1449}{200}\, g_{1}^{4} +\frac{27}{20}\, g_{1}^{2} g_{2}^{2} -\frac{21}{4}\, g_{2}^{4} +\frac{3}{2}\, \lambda_{1}^{2} +\lambda_{3}^{2}+\lambda_3 \lambda_4 +\lambda_{4}^{2}\nonumber \\ 
	&+\frac{5}{8} \,\Big(32\, g_{3}^{2}  + 9 \,g_{2}^{2}  + g_{1}^{2}\Big)\mbox{Tr}\Big({Y_d  Y_{d}^{\dagger}}\Big) 
	+\frac{15}{8} \,\Big(g_{1}^{2} + g_{2}^{2}\Big)\,\mbox{Tr}\Big({Y_e  Y_{e}^{\dagger}}\Big) -\frac{27}{4}\, \mbox{Tr}\Big({Y_d  Y_{d}^{\dagger}  Y_d  Y_{d}^{\dagger}}\Big) \nonumber\\
	&-\frac{9}{4}\, \mbox{Tr}\Big({Y_{d}^{\dagger} Y_u Y_{u}^{\dagger}  Y_d  }\Big) -\frac{9}{4}\, \mbox{Tr}\Big({Y_e  Y_{e}^{\dagger}  Y_e  Y_{e}^{\dagger}}\Big) \Big) \Big)\nonumber\\
	&+\frac{3}{80} \,{Y_e  Y_{e}^{\dagger}  Y_e} \Big(129\, g_{1}^{2}  -180\, \mbox{Tr}\Big({Y_d  Y_{d}^{\dagger}}\Big)  + 225\, g_{2}^{2}  -160\,\lambda_1  -60\, \mbox{Tr}\Big({Y_e  Y_{e}^{\dagger}}\Big) \Big),\nonumber \\ 
	\beta_{Y_e}^{(2,1)} & =Y_e \Big[\frac{132}{200}g_1^4+\frac{1}{2}g_2^4\Big],\nonumber \\ 
	\beta_{Y_e}^{(2,2)}&=\frac{1}{200}Y_e\Big(+75 g_{1}^{2} |y_4|^2 +375 g_{2}^{2} |y_4|^2 -450 |y_1|^4 -450 |y_4|^4 -300 y_4 |y_3|^2 y_4^* \nonumber\\
	&-50 y_4 y_2^* \Big(4 y_1 y_3^*  + 9 y_2 y_4^* \Big)+25 y_1^* \Big(-12 y_1 |y_2|^2  + 15 g_{2}^{2} y_1  -18 y_1 |y_3|^2  + 3 g_{1}^{2} y_1  \nonumber\\
	&-40 y_1 |y_4|^2  -8 y_2 y_3 y_4^* \Big) \Big)-\frac{9}{4} (y_1^2+y_4^2) Y_e  Y_{e}^{\dagger}  Y_e,\nonumber \\ 
\beta_{y_4}^{(1,0)} & = 0,\nonumber\\
\beta_{y_4}^{(1,1)} & = 0,\nonumber\\
\beta_{y_4}^{(1,2)} & =  
+\Big(3 y_2 y_3  + 4 y_1 y_4 \Big)y_1^* \nonumber \\ 
&+\frac{1}{20} y_4 \Big(20 \mbox{Tr}\Big({Y_e  Y_{e}^{\dagger}}\Big)  + 20 |y_3|^2  -45 g_{2}^{2}  + 50 |y_2|^2  + 50 |y_4|^2  + 60 \mbox{Tr}\Big({Y_d  Y_{d}^{\dagger}}\Big)  -9 g_{1}^{2} \Big),\nonumber\\ 
\beta_{y_4}^{(2,0)} & = 0,\nonumber\\
\beta_{y_4}^{(2,1)} & = 0,\nonumber\\
\beta_{y_4}^{(2,2)} & =  
-\frac{1}{4} y_1 \Big(29 y_2 y_3  + 36 y_1 y_4 \Big)y_{1}^{*2} +\frac{1}{40} y_1^* \Big(27 g_{1}^{2} y_2 y_3 +135 g_{2}^{2} y_2 y_3 -80 \lambda_3 y_2 y_3 -160 \lambda_4 y_2 y_3 \nonumber\\
&+6 g_{1}^{2} y_1 y_4 +390 g_{2}^{2} y_1 y_4 -240 \lambda_1 y_1 y_4 -10 \Big(18 y_2 y_3  + 25 y_1 y_4 \Big)|y_2|^2 -10 \Big(29 y_2 y_3  + 43 y_1 y_4 \Big)|y_3|^2 \nonumber\\
&-530 y_2 y_3 |y_4|^2 -600 y_1 y_{4}^{2} y_4^*
-240 y_2 y_3 \mbox{Tr}\Big({Y_d  Y_{d}^{\dagger}}\Big) -420 y_1 y_4 \mbox{Tr}\Big({Y_d  Y_{d}^{\dagger}}\Big) -80 y_2 y_3 \mbox{Tr}\Big({Y_e  Y_{e}^{\dagger}}\Big)\nonumber \\ 
& -140 y_1 y_4 \mbox{Tr}\Big({Y_e  Y_{e}^{\dagger}}\Big) \Big)
-\frac{1}{400} y_4 \Big(-258 g_{1}^{4} +540 g_{1}^{2} g_{2}^{2} +1900 g_{2}^{4} -600 \lambda_{1}^{2} -400 \lambda_{3}^{2} -400 \lambda_3 \lambda_4\nonumber\\
& -400 \lambda_{4}^{2} -1545 g_{1}^{2} |y_4|^2 -4125 g_{2}^{2} |y_4|^2 +2400 \lambda_1 |y_4|^2 +1200 |y_2|^4 +700 |y_3|^4 +1200 |y_4|^4 \nonumber\\
&-250 g_{1}^{2} \mbox{Tr}\Big({Y_d  Y_{d}^{\dagger}}\Big) -2250 g_{2}^{2} \mbox{Tr}\Big({Y_d  Y_{d}^{\dagger}}\Big) -8000 g_{3}^{2} \mbox{Tr}\Big({Y_d  Y_{d}^{\dagger}}\Big) +2700 |y_4|^2 \mbox{Tr}\Big({Y_d  Y_{d}^{\dagger}}\Big)\nonumber\\
& -750 g_{1}^{2} \mbox{Tr}\Big({Y_e  Y_{e}^{\dagger}}\Big) -750 g_{2}^{2} \mbox{Tr}\Big({Y_e  Y_{e}^{\dagger}}\Big) +900 |y_4|^2 \mbox{Tr}\Big({Y_e  Y_{e}^{\dagger}}\Big) +10 |y_3|^2 \Big(160 \lambda_3  + 180 \mbox{Tr}\Big({Y_u  Y_{u}^{\dagger}}\Big)  \nonumber\\
&+ 21 g_{1}^{2} -255 g_{2}^{2}  + 70 y_4 y_4^*  + 80 \lambda_4 \Big)+5 y_2^* \Big(140 \Big(5 y_1 y_4  + 6 y_2 y_3 \Big)y_3^*  + 3 y_2 \Big(-103 g_{1}^{2}  + 160 \lambda_3  + 160 \lambda_4  \nonumber\\
&+ 160 |y_4|^2  + 180 \mbox{Tr}\Big({Y_u  Y_{u}^{\dagger}}\Big)  -275 g_{2}^{2} \Big)\Big)+2700 \mbox{Tr}\Big({Y_d  Y_{d}^{\dagger}  Y_d  Y_{d}^{\dagger}}\Big) +900 \mbox{Tr}\Big({Y_d  Y_{u}^{\dagger}  Y_u  Y_{d}^{\dagger}}\Big)\nonumber\\
& +900 \mbox{Tr}\Big({Y_e  Y_{e}^{\dagger}  Y_e  Y_{e}^{\dagger}}\Big) \Big),\nonumber\\ 
\beta_{y_2}^{(1,0)} & = 0,\nonumber\\
\beta_{y_2}^{(1,1)} & =  0,\nonumber\\
\beta_{y_2}^{(1,2)} & =  
-\frac{9}{20} g_{1}^{2} y_2 -\frac{9}{4} g_{2}^{2} y_2 +y_2 |y_1|^2 +4 y_2 |y_3|^2 +\frac{5}{2} y_2 |y_4|^2 +\frac{5}{2} y_{2}^{2} y_2^* +3 y_1 y_4 y_3^* \nonumber \\ 
&+3 y_2 \mbox{Tr}\Big({Y_u  Y_{u}^{\dagger}}\Big),\nonumber \\ 
\beta_{y_2}^{(2,0)} & = 0,\nonumber\\
\beta_{y_2}^{(2,1)} & = 0,\nonumber\\
\beta_{y_2}^{(2,2)} & =  
+\frac{129}{200} g_{1}^{4} y_2 -\frac{27}{20} g_{1}^{2} g_{2}^{2} y_2 -\frac{19}{4} g_{2}^{4} y_2 +\frac{3}{2} \lambda_{2}^{2} y_2 +\lambda_{3}^{2} y_2 +\lambda_3 \lambda_4 y_2 +\lambda_{4}^{2} y_2 +\frac{3}{20} g_{1}^{2} y_2 |y_3|^2 \nonumber \\ 
&+\frac{39}{4} g_{2}^{2} y_2 |y_3|^2 -6 \lambda_2 y_2 |y_3|^2 +\frac{309}{80} g_{1}^{2} y_2 |y_4|^2 +\frac{165}{16} g_{2}^{2} y_2 |y_4|^2 -6 \lambda_3 y_2 |y_4|^2 \nonumber \\ 
&-6 \lambda_4 y_2 |y_4|^2 -\frac{7}{4} y_2 |y_1|^4 -9 y_2 |y_3|^4 -3 y_2 |y_4|^4 -3 y_{2}^{3} y_{2}^{*2} +\frac{27}{40} g_{1}^{2} y_1 y_4 y_3^* +\frac{27}{8} g_{2}^{2} y_1 y_4 y_3^* \nonumber \\ 
&-2 \lambda_3 y_1 y_4 y_3^* -4 \lambda_4 y_1 y_4 y_3^* -\frac{29}{4} y_1 y_3 y_4 y_{3}^{*2} -\frac{25}{4} y_2 y_4 |y_3|^2 y_4^* -\frac{9}{2} y_1 y_{4}^{2} y_3^* y_4^* \nonumber \\ 
&-\frac{27}{4} y_2 |y_4|^2 \mbox{Tr}\Big({Y_d  Y_{d}^{\dagger}}\Big) -\frac{9}{4} y_2 |y_4|^2 \mbox{Tr}\Big({Y_e  Y_{e}^{\dagger}}\Big) -\frac{1}{40} y_1^* \Big(70 y_1 y_{2}^{2} y_2^* +10 y_1 \Big(29 y_1 y_4  + 43 y_2 y_3 \Big)y_3^* \nonumber\\
&+70 y_2 \Big(5 y_2 y_3  + 6 y_1 y_4 \Big)y_4^* +y_1 y_2 \Big(160 \lambda_3  + 180 \mbox{Tr}\Big({Y_d  Y_{d}^{\dagger}}\Big)  + 21 g_{1}^{2}  -255 g_{2}^{2}  \nonumber\\
&+ 60 \mbox{Tr}\Big({Y_e  Y_{e}^{\dagger}}\Big)  + 80 \lambda_4 \Big)\Big)+\frac{17}{8} g_{1}^{2} y_2 \mbox{Tr}\Big({Y_u  Y_{u}^{\dagger}}\Big) +\frac{45}{8} g_{2}^{2} y_2 \mbox{Tr}\Big({Y_u  Y_{u}^{\dagger}}\Big) +20 g_{3}^{2} y_2 \mbox{Tr}\Big({Y_u  Y_{u}^{\dagger}}\Big)\nonumber\\
& -\frac{21}{2} y_2 |y_3|^2 \mbox{Tr}\Big({Y_u  Y_{u}^{\dagger}}\Big) -6 y_1 y_4 y_3^* \mbox{Tr}\Big({Y_u  Y_{u}^{\dagger}}\Big) -\frac{1}{80} |y_2|^2 \Big(20 \Big(53 y_1 y_4  + 60 y_2 y_3 \Big)y_3^*  \nonumber\\
&+ 3 y_2 \Big(-103 g_{1}^{2}  + 160 \lambda_2  + 160 y_4 y_4^*  + 180 \mbox{Tr}\Big({Y_u  Y_{u}^{\dagger}}\Big)  -275 g_{2}^{2} \Big)\Big)\nonumber \\ 
&-\frac{9}{4} y_2 \mbox{Tr}\Big({Y_d  Y_{u}^{\dagger}  Y_u  Y_{d}^{\dagger}}\Big) -\frac{27}{4} y_2 \mbox{Tr}\Big({Y_u  Y_{u}^{\dagger}  Y_u  Y_{u}^{\dagger}}\Big),\nonumber \\ 
\beta_{y_1}^{(1,0)} & = 0,\nonumber\\
\beta_{y_1}^{(1,1)} & = 0,\nonumber\\
\beta_{y_1}^{(1,2)} & =  
-\frac{9}{20} g_{1}^{2} y_1 -\frac{9}{4} g_{2}^{2} y_1 +y_1 |y_2|^2 +\frac{5}{2} y_1 |y_3|^2 +4 y_1 |y_4|^2 +\frac{5}{2} y_{1}^{2} y_1^* +3 y_2 y_3 y_4^* \nonumber \\ 
&+3 y_1 \mbox{Tr}\Big({Y_d  Y_{d}^{\dagger}}\Big) +y_1 \mbox{Tr}\Big({Y_e  Y_{e}^{\dagger}}\Big),\nonumber \\ 
\beta_{y_1}^{(2,0)} & = 0,\nonumber\\
\beta_{y_1}^{(2,1)} & = 0,\nonumber\\
\beta_{y_1}^{(2,2)} & =  
+\frac{129}{200} g_{1}^{4} y_1 -\frac{27}{20} g_{1}^{2} g_{2}^{2} y_1 -\frac{19}{4} g_{2}^{4} y_1 +\frac{3}{2} \lambda_{1}^{2} y_1 +\lambda_{3}^{2} y_1 +\lambda_3 \lambda_4 y_1 +\lambda_{4}^{2} y_1 +\frac{309}{80} g_{1}^{2} y_1 |y_3|^2 \nonumber \\ 
&+\frac{165}{16} g_{2}^{2} y_1 |y_3|^2 -6 \lambda_3 y_1 |y_3|^2 -6 \lambda_4 y_1 |y_3|^2 +\frac{3}{20} g_{1}^{2} y_1 |y_4|^2 +\frac{39}{4} g_{2}^{2} y_1 |y_4|^2 \nonumber \\ 
&-6 \lambda_1 y_1 |y_4|^2 -\frac{7}{4} y_1 |y_2|^4 -3 y_1 |y_3|^4 -9 y_1 |y_4|^4 -3 y_{1}^{3} y_{1}^{*2} +\frac{27}{40} g_{1}^{2} y_2 y_3 y_4^* +\frac{27}{8} g_{2}^{2} y_2 y_3 y_4^* \nonumber \\ 
&-2 \lambda_3 y_2 y_3 y_4^* -4 \lambda_4 y_2 y_3 y_4^* -\frac{25}{4} y_1 y_4 |y_3|^2 y_4^* -\frac{9}{2} y_2 y_{3}^{2} y_3^* y_4^* -\frac{29}{4} y_2 y_3 y_4 y_{4}^{*2} \nonumber \\ 
&+\frac{5}{8} g_{1}^{2} y_1 \mbox{Tr}\Big({Y_d  Y_{d}^{\dagger}}\Big) +\frac{45}{8} g_{2}^{2} y_1 \mbox{Tr}\Big({Y_d  Y_{d}^{\dagger}}\Big) +20 g_{3}^{2} y_1 \mbox{Tr}\Big({Y_d  Y_{d}^{\dagger}}\Big) -\frac{21}{2} y_1 |y_4|^2 \mbox{Tr}\Big({Y_d  Y_{d}^{\dagger}}\Big) \nonumber \\ 
&-6 y_2 y_3 y_4^* \mbox{Tr}\Big({Y_d  Y_{d}^{\dagger}}\Big) +\frac{15}{8} g_{1}^{2} y_1 \mbox{Tr}\Big({Y_e  Y_{e}^{\dagger}}\Big) +\frac{15}{8} g_{2}^{2} y_1 \mbox{Tr}\Big({Y_e  Y_{e}^{\dagger}}\Big) -\frac{7}{2} y_1 |y_4|^2 \mbox{Tr}\Big({Y_e  Y_{e}^{\dagger}}\Big) \nonumber \\ 
&-2 y_2 y_3 y_4^* \mbox{Tr}\Big({Y_e  Y_{e}^{\dagger}}\Big)+\frac{1}{80} |y_1|^2 \Big(309 g_{1}^{2} y_1 +825 g_{2}^{2} y_1-480 \lambda_1 y_1 -140 y_1 y_2 y_2^* \nonumber\\
& -480 y_1 y_3 y_3^* -1060 y_2 y_3 y_4^* -1200 y_1 y_4 y_4^* -540 y_1 \mbox{Tr}\Big({Y_d  Y_{d}^{\dagger}}\Big) -180 y_1 \mbox{Tr}\Big({Y_e  Y_{e}^{\dagger}}\Big) \Big)\nonumber\\
&-\frac{27}{4} y_1 |y_3|^2 \mbox{Tr}\Big({Y_u  Y_{u}^{\dagger}}\Big) -\frac{1}{40} y_2^* \Big(70 y_1 \Big(5 y_1 y_4  + 6 y_2 y_3 \Big)y_3^* +10 y_2 \Big(29 y_2 y_3  + 43 y_1 y_4 \Big)y_4^* \nonumber\\
&+y_1 y_2 \Big(160 \lambda_3  + 180 \mbox{Tr}\Big({Y_u  Y_{u}^{\dagger}}\Big)  + 21 g_{1}^{2}  -255 g_{2}^{2}  + 80 \lambda_4 \Big)\Big)\nonumber \\ 
&-\frac{27}{4} y_1 \mbox{Tr}\Big({Y_d  Y_{d}^{\dagger}  Y_d  Y_{d}^{\dagger}}\Big) -\frac{9}{4} y_1 \mbox{Tr}\Big({Y_d  Y_{u}^{\dagger}  Y_u  Y_{d}^{\dagger}}\Big) -\frac{9}{4} y_1 \mbox{Tr}\Big({Y_e  Y_{e}^{\dagger}  Y_e  Y_{e}^{\dagger}}\Big),\nonumber \\ 
\beta_{y_3}^{(1,0)} & =0,\nonumber\\
\beta_{y_3}^{(1,1)} & =0,\nonumber\\
\beta_{y_3}^{(1,2)} & =  
\Big(3 y_1 y_4  + 4 y_2 y_3 \Big)y_2^*  + \frac{1}{20} y_3 \Big(20 |y_4|^2  -45 g_{2}^{2}  + 50 |y_3|^2  + 60 \mbox{Tr}\Big({Y_u  Y_{u}^{\dagger}}\Big)  -9 g_{1}^{2} \Big) + \frac{5}{2} y_3 |y_1|^2,\nonumber \\ 
\beta_{y_3}^{(2,0)} & =0,\nonumber\\
\beta_{y_3}^{(2,1)} & =0,\nonumber\\
\beta_{y_3}^{(2,2)} & =  
-3 y_3 |y_1|^4 -\frac{1}{4} y_2 \Big(29 y_1 y_4  + 36 y_2 y_3 \Big)y_{2}^{*2} -\frac{1}{80} y_1^* \Big(20 y_1 \Big(18 y_1 y_4  + 25 y_2 y_3 \Big)y_2^* \nonumber\\
&+y_3 \Big(480 y_1 |y_3|^2 +140 \Big(5 y_2 y_3  + 6 y_1 y_4 \Big)y_4^* +3 y_1 \Big(-103 g_{1}^{2}  + 160 \lambda_3  + 160 \lambda_4 \nonumber\\
& + 180 \mbox{Tr}\Big({Y_d  Y_{d}^{\dagger}}\Big)  -275 g_{2}^{2}  + 60 \mbox{Tr}\Big({Y_e  Y_{e}^{\dagger}}\Big) \Big)\Big)\Big)+\frac{1}{40} y_2^* \Big(6 g_{1}^{2} y_2 y_3 +390 g_{2}^{2} y_2 y_3 \nonumber\\
&-240 \lambda_2 y_2 y_3 +27 g_{1}^{2} y_1 y_4 +135 g_{2}^{2} y_1 y_4 -80 \lambda_3 y_1 y_4 -160 \lambda_4 y_1 y_4 -10 \Big(53 y_1 y_4  \nonumber\\
&+ 60 y_2 y_3 \Big)|y_3|^2 -10 \Big(29 y_1 y_4  + 43 y_2 y_3 \Big)|y_4|^2 -420 y_2 y_3 \mbox{Tr}\Big({Y_u  Y_{u}^{\dagger}}\Big) -240 y_1 y_4 \mbox{Tr}\Big({Y_u  Y_{u}^{\dagger}}\Big) \Big)\nonumber \\ 
&-\frac{1}{400} y_3 \Big(-258 g_{1}^{4} +540 g_{1}^{2} g_{2}^{2} +1900 g_{2}^{4} -600 \lambda_{2}^{2} -400 \lambda_{3}^{2} -400 \lambda_3 \lambda_4 -400 \lambda_{4}^{2} \nonumber\\
&+1200 |y_3|^4 +700 |y_4|^4 +10 |y_4|^2 \Big(160 \lambda_3  + 180 \mbox{Tr}\Big({Y_d  Y_{d}^{\dagger}}\Big)  + 21 g_{1}^{2}  -255 g_{2}^{2}  + 60 \mbox{Tr}\Big({Y_e  Y_{e}^{\dagger}}\Big) \nonumber\\
& + 80 \lambda_4 \Big)
-5 |y_3|^2 \Big(-140 y_4 y_4^*  + 309 g_{1}^{2}  -480 \lambda_2  -540 \mbox{Tr}\Big({Y_u  Y_{u}^{\dagger}}\Big)  + 825 g_{2}^{2} \Big)-850 g_{1}^{2} \mbox{Tr}\Big({Y_u  Y_{u}^{\dagger}}\Big) \nonumber \\ 
&-2250 g_{2}^{2} \mbox{Tr}\Big({Y_u  Y_{u}^{\dagger}}\Big) -8000 g_{3}^{2} \mbox{Tr}\Big({Y_u  Y_{u}^{\dagger}}\Big) +900 \mbox{Tr}\Big({Y_d  Y_{u}^{\dagger}  Y_u  Y_{d}^{\dagger}}\Big) +2700 \mbox{Tr}\Big({Y_u  Y_{u}^{\dagger}  Y_u  Y_{u}^{\dagger}}\Big)\Big).\nonumber	
\end{align}} 
\subsection{Quartic scalar couplings}
{\allowdisplaybreaks  \begin{align} 
	\beta_{\lambda_1}^{(1,0)} & =  
	+\frac{27}{100} g_{1}^{4} +\frac{9}{10} g_{1}^{2} g_{2}^{2} +\frac{9}{4} g_{2}^{4} -\frac{9}{5} g_{1}^{2} \lambda_1 -9 g_{2}^{2} \lambda_1 +12 \lambda_{1}^{2} +4 \lambda_{3}^{2} +4 \lambda_3 \lambda_4 +2 \lambda_{4}^{2}  \nonumber \\ 
	&+12 \lambda_1 \mbox{Tr}\Big({Y_d  Y_{d}^{\dagger}}\Big) +4 \lambda_1 \mbox{Tr}\Big({Y_e  Y_{e}^{\dagger}}\Big) -12 \mbox{Tr}\Big({Y_d  Y_{d}^{\dagger}  Y_d  Y_{d}^{\dagger}}\Big) -4 \mbox{Tr}\Big({Y_e  Y_{e}^{\dagger}  Y_e  Y_{e}^{\dagger}}\Big), \nonumber\\ 
	\beta_{\lambda_1}^{(1,1)} & = 0,\nonumber\\
	\beta_{\lambda_1}^{(1,2)} & = +4 \lambda_1 |y_4|^2 -4 |y_1|^4 -4 |y_4|^4 +4 |y_1|^2 \Big(-2 y_4 y_4^*  + \lambda_1\Big),\nonumber\\
	\beta_{\lambda_1}^{(2,0)} & =  
	-\frac{153}{40} g_{1}^{6} -\frac{363}{40} g_{1}^{4} g_{2}^{2} -\frac{67}{8} g_{1}^{2} g_{2}^{4} +\frac{259}{8} g_{2}^{6} +\frac{2073}{200} g_{1}^{4} \lambda_1 +\frac{117}{20} g_{1}^{2} g_{2}^{2} \lambda_1 -\frac{11}{8} g_{2}^{4} \lambda_1\nonumber \\ 
	& +\frac{54}{5} g_{1}^{2} \lambda_{1}^{2} +54 g_{2}^{2} \lambda_{1}^{2} -78 \lambda_{1}^{3} +\frac{9}{5} g_{1}^{4} \lambda_3 +15 g_{2}^{4} \lambda_3 +\frac{24}{5} g_{1}^{2} \lambda_{3}^{2} +24 g_{2}^{2} \lambda_{3}^{2} -20 \lambda_1 \lambda_{3}^{2} \nonumber \\ 
	&-16 \lambda_{3}^{3} +\frac{9}{10} g_{1}^{4} \lambda_4 +3 g_{1}^{2} g_{2}^{2} \lambda_4 +\frac{15}{2} g_{2}^{4} \lambda_4 +\frac{24}{5} g_{1}^{2} \lambda_3 \lambda_4 +24 g_{2}^{2} \lambda_3 \lambda_4 -20 \lambda_1 \lambda_3 \lambda_4\nonumber \\ 
	& -24 \lambda_{3}^{2} \lambda_4 +\frac{12}{5} g_{1}^{2} \lambda_{4}^{2} +6 g_{2}^{2} \lambda_{4}^{2} -12 \lambda_1 \lambda_{4}^{2} -32 \lambda_3 \lambda_{4}^{2} -12 \lambda_{4}^{3} +\frac{9}{10} g_{1}^{4} \mbox{Tr}\Big({Y_d  Y_{d}^{\dagger}}\Big) \nonumber\\
	&+\frac{27}{5} g_{1}^{2} g_{2}^{2} \mbox{Tr}\Big({Y_d  Y_{d}^{\dagger}}\Big) -\frac{9}{2} g_{2}^{4} \mbox{Tr}\Big({Y_d  Y_{d}^{\dagger}}\Big) +\frac{5}{2} g_{1}^{2} \lambda_1 \mbox{Tr}\Big({Y_d  Y_{d}^{\dagger}}\Big) +\frac{45}{2} g_{2}^{2} \lambda_1 \mbox{Tr}\Big({Y_d  Y_{d}^{\dagger}}\Big) \nonumber\\
	&+80 g_{3}^{2} \lambda_1 \mbox{Tr}\Big({Y_d  Y_{d}^{\dagger}}\Big) -72 \lambda_{1}^{2} \mbox{Tr}\Big({Y_d  Y_{d}^{\dagger}}\Big) -\frac{9}{2} g_{1}^{4} \mbox{Tr}\Big({Y_e  Y_{e}^{\dagger}}\Big) +\frac{33}{5} g_{1}^{2} g_{2}^{2} \mbox{Tr}\Big({Y_e  Y_{e}^{\dagger}}\Big)\nonumber\\
	& -\frac{3}{2} g_{2}^{4} \mbox{Tr}\Big({Y_e  Y_{e}^{\dagger}}\Big) +\frac{15}{2} g_{1}^{2} \lambda_1 \mbox{Tr}\Big({Y_e  Y_{e}^{\dagger}}\Big) +\frac{15}{2} g_{2}^{2} \lambda_1 \mbox{Tr}\Big({Y_e  Y_{e}^{\dagger}}\Big)-24 \lambda_{1}^{2} \mbox{Tr}\Big({Y_e  Y_{e}^{\dagger}}\Big) \nonumber \\ 
	& -24 \lambda_{3}^{2} \mbox{Tr}\Big({Y_u  Y_{u}^{\dagger}}\Big) -24 \lambda_3 \lambda_4 \mbox{Tr}\Big({Y_u  Y_{u}^{\dagger}}\Big) -12 \lambda_{4}^{2} \mbox{Tr}\Big({Y_u  Y_{u}^{\dagger}}\Big) +\frac{8}{5} g_{1}^{2} \mbox{Tr}\Big({Y_d  Y_{d}^{\dagger}  Y_d  Y_{d}^{\dagger}}\Big) \nonumber \\ 
	&-64 g_{3}^{2} \mbox{Tr}\Big({Y_d  Y_{d}^{\dagger}  Y_d  Y_{d}^{\dagger}}\Big) -3 \lambda_1 \mbox{Tr}\Big({Y_d  Y_{d}^{\dagger}  Y_d  Y_{d}^{\dagger}}\Big) -9 \lambda_1 \mbox{Tr}\Big({Y_d  Y_{u}^{\dagger}  Y_u  Y_{d}^{\dagger}}\Big) \nonumber \\ 
	&-\frac{24}{5} g_{1}^{2} \mbox{Tr}\Big({Y_e  Y_{e}^{\dagger}  Y_e  Y_{e}^{\dagger}}\Big) - \lambda_1 \mbox{Tr}\Big({Y_e  Y_{e}^{\dagger}  Y_e  Y_{e}^{\dagger}}\Big) +60 \mbox{Tr}\Big({Y_d  Y_{d}^{\dagger}  Y_d  Y_{d}^{\dagger}  Y_d  Y_{d}^{\dagger}}\Big) 
	\nonumber \\ 
	&+12 \mbox{Tr}\Big({Y_d  Y_{u}^{\dagger}  Y_u  Y_{d}^{\dagger}  Y_d  Y_{d}^{\dagger}}\Big) 
	+20 \mbox{Tr}\Big({Y_e  Y_{e}^{\dagger}  Y_e  Y_{e}^{\dagger}  Y_e  Y_{e}^{\dagger}}\Big),\nonumber \\ 
	\beta_{\lambda_1}^{(2,2)} & =-8 \lambda_{3}^{2} |y_3|^2 -8 \lambda_3 \lambda_4 |y_3|^2 -4 \lambda_{4}^{2} |y_3|^2 -\frac{9}{50} g_{1}^{4} |y_4|^2 -\frac{3}{5} g_{1}^{2} g_{2}^{2} |y_4|^2 -\frac{3}{2} g_{2}^{4} |y_4|^2 \nonumber\\
	&+\frac{3}{2} g_{1}^{2} \lambda_1 |y_4|^2 +\frac{15}{2} g_{2}^{2} \lambda_1 |y_4|^2 -24 \lambda_{1}^{2} |y_4|^2 - \lambda_1 |y_4|^4 +8 |y_3|^2 |y_4|^4 +20 y_{1}^{3} y_{1}^{*3} \nonumber \\ 
	&+y_2^* \Big(4 y_1 y_4 \Big(2 \Big(\lambda_3 + \lambda_4\Big) + 4 |y_4|^2  - \lambda_1 \Big)y_3^*  - y_2 \Big(-20 |y_4|^4  + 4 \lambda_{4}^{2}  + 8 \lambda_{3}^{2}  + 8 \lambda_3 \lambda_4 \nonumber\\
	& + 9 \lambda_1 |y_4|^2 \Big)\Big)-6 \lambda_1 y_4 |y_3|^2 y_4^* +20 y_{4}^{3} y_{4}^{*3} +y_1 y_{1}^{*2} \Big(16 y_2 y_3 y_4^*  + 20 y_1 |y_3|^2  + 68 y_1 |y_4|^2  \nonumber\\
	&+ 8 y_1 |y_2|^2  - \lambda_1 y_1 \Big)+y_1^* \Big(-\frac{9}{50} g_{1}^{4} y_1 -\frac{3}{5} g_{1}^{2} g_{2}^{2} y_1 -\frac{3}{2} g_{2}^{4} y_1 +\frac{3}{2} g_{1}^{2} \lambda_1 y_1 +\frac{15}{2} g_{2}^{2} \lambda_1 y_1 \nonumber\\
	&-24 \lambda_{1}^{2} y_1 +12 \lambda_1 y_1 |y_4|^2 +68 y_1 |y_4|^4 +2 y_1 y_2^* \Big(14 y_2 |y_4|^2  -3 \lambda_1 y_2  + 8 y_1 y_4 y_3^* \Big)\nonumber\\
	&-4 \lambda_1 y_2 y_3 y_4^* +8 \lambda_3 y_2 y_3 y_4^* +8 \lambda_4 y_2 y_3 y_4^* +16 y_2 y_3 y_4 y_{4}^{*2} +y_1 |y_3|^2 \Big(28 y_4 y_4^*  -9 \lambda_1 \Big)\Big),\nonumber \\ 
		\beta_{\lambda_4}^{(1,0)} & =  
	+\frac{9}{5} g_{1}^{2} g_{2}^{2} -\frac{9}{5} g_{1}^{2} \lambda_4 -9 g_{2}^{2} \lambda_4 +2 \lambda_1 \lambda_4 +2 \lambda_2 \lambda_4 +8 \lambda_3 \lambda_4 +4 \lambda_{4}^{2}  \nonumber \\ 
	&+6 \lambda_4 \mbox{Tr}\Big({Y_d  Y_{d}^{\dagger}}\Big) +2 \lambda_4 \mbox{Tr}\Big({Y_e  Y_{e}^{\dagger}}\Big) +6 \lambda_4 \mbox{Tr}\Big({Y_u  Y_{u}^{\dagger}}\Big) +12 \mbox{Tr}\Big({Y_d  Y_{u}^{\dagger}  Y_u  Y_{d}^{\dagger}}\Big), \nonumber\\
	\beta_{\lambda_4}^{(1,1)} & =  0 ,\nonumber\\
	\beta_{\lambda_4}^{(1,2)} & = +2 \lambda_4 |y_3|^2 +2 \lambda_4 |y_4|^2 -2 y_{3}^{2} y_{1}^{*2} -2 y_{4}^{2} y_{2}^{*2} -2 y_{1}^{2} y_{3}^{*2} +2 y_2^* \Big(-2 y_1 y_4 y_3^*  -2 y_2 |y_4|^2  \nonumber\\
	&+ \lambda_4 y_2 \Big)-4 y_1 y_2 y_3^* y_4^* -2 y_{2}^{2} y_{4}^{*2} +2 y_1^* \Big(-2 y_1 |y_3|^2  -2 y_2 y_3 y_4^*  -2 y_3 y_4 y_2^*  + \lambda_4 y_1 \Big),\nonumber\\
	\beta_{\lambda_4}^{(2,0)} & =  
	-\frac{141}{10} g_{1}^{4} g_{2}^{2} -10 g_{1}^{2} g_{2}^{4} +27 g_{2}^{6} +3 g_{1}^{2} g_{2}^{2} \lambda_1 +3 g_{1}^{2} g_{2}^{2} \lambda_2 +\frac{6}{5} g_{1}^{2} g_{2}^{2} \lambda_3 +\frac{1533}{200} g_{1}^{4} \lambda_4 \nonumber \\ 
	&+\frac{153}{20} g_{1}^{2} g_{2}^{2} \lambda_4 -\frac{191}{8} g_{2}^{4} \lambda_4 +\frac{12}{5} g_{1}^{2} \lambda_1 \lambda_4 -7 \lambda_{1}^{2} \lambda_4 +\frac{12}{5} g_{1}^{2} \lambda_2 \lambda_4 -7 \lambda_{2}^{2} \lambda_4 +\frac{12}{5} g_{1}^{2} \lambda_3 \lambda_4 \nonumber \\ 
	&+36 g_{2}^{2} \lambda_3 \lambda_4 -40 \lambda_1 \lambda_3 \lambda_4 -40 \lambda_2 \lambda_3 \lambda_4 -28 \lambda_{3}^{2} \lambda_4 +\frac{24}{5} g_{1}^{2} \lambda_{4}^{2} +18 g_{2}^{2} \lambda_{4}^{2} -20 \lambda_1 \lambda_{4}^{2} \nonumber \\ 
	&-20 \lambda_2 \lambda_{4}^{2} -28 \lambda_3 \lambda_{4}^{2} -\frac{27}{5} g_{1}^{2} g_{2}^{2} \mbox{Tr}\Big({Y_d  Y_{d}^{\dagger}}\Big) \nonumber \\ 
	&+\frac{5}{4} g_{1}^{2} \lambda_4 \mbox{Tr}\Big({Y_d  Y_{d}^{\dagger}}\Big) +\frac{45}{4} g_{2}^{2} \lambda_4 \mbox{Tr}\Big({Y_d  Y_{d}^{\dagger}}\Big) +40 g_{3}^{2} \lambda_4 \mbox{Tr}\Big({Y_d  Y_{d}^{\dagger}}\Big) -12 \lambda_1 \lambda_4 \mbox{Tr}\Big({Y_d  Y_{d}^{\dagger}}\Big) \nonumber \\ 
	&-24 \lambda_3 \lambda_4 \mbox{Tr}\Big({Y_d  Y_{d}^{\dagger}}\Big) -12 \lambda_{4}^{2} \mbox{Tr}\Big({Y_d  Y_{d}^{\dagger}}\Big) -\frac{33}{5} g_{1}^{2} g_{2}^{2} \mbox{Tr}\Big({Y_e  Y_{e}^{\dagger}}\Big) +\frac{15}{4} g_{1}^{2} \lambda_4 \mbox{Tr}\Big({Y_e  Y_{e}^{\dagger}}\Big) \nonumber \\ 
	&+\frac{15}{4} g_{2}^{2} \lambda_4 \mbox{Tr}\Big({Y_e  Y_{e}^{\dagger}}\Big) -4 \lambda_1 \lambda_4 \mbox{Tr}\Big({Y_e  Y_{e}^{\dagger}}\Big) -8 \lambda_3 \lambda_4 \mbox{Tr}\Big({Y_e  Y_{e}^{\dagger}}\Big) -4 \lambda_{4}^{2} \mbox{Tr}\Big({Y_e  Y_{e}^{\dagger}}\Big) \nonumber \\ 
	&-\frac{63}{5} g_{1}^{2} g_{2}^{2} \mbox{Tr}\Big({Y_u  Y_{u}^{\dagger}}\Big) +\frac{17}{4} g_{1}^{2} \lambda_4 \mbox{Tr}\Big({Y_u  Y_{u}^{\dagger}}\Big) +\frac{45}{4} g_{2}^{2} \lambda_4 \mbox{Tr}\Big({Y_u  Y_{u}^{\dagger}}\Big) +40 g_{3}^{2} \lambda_4 \mbox{Tr}\Big({Y_u  Y_{u}^{\dagger}}\Big) \nonumber \\ 
	&-12 \lambda_2 \lambda_4 \mbox{Tr}\Big({Y_u  Y_{u}^{\dagger}}\Big) -24 \lambda_3 \lambda_4 \mbox{Tr}\Big({Y_u  Y_{u}^{\dagger}}\Big) -12 \lambda_{4}^{2} \mbox{Tr}\Big({Y_u  Y_{u}^{\dagger}}\Big) -\frac{27}{2} \lambda_4 \mbox{Tr}\Big({Y_d  Y_{d}^{\dagger}  Y_d  Y_{d}^{\dagger}}\Big) \nonumber \\ 
	&+\frac{4}{5} g_{1}^{2} \mbox{Tr}\Big({Y_d  Y_{u}^{\dagger}  Y_u  Y_{d}^{\dagger}}\Big) +64 g_{3}^{2} \mbox{Tr}\Big({Y_d  Y_{u}^{\dagger}  Y_u  Y_{d}^{\dagger}}\Big) -24 \lambda_3 \mbox{Tr}\Big({Y_d  Y_{u}^{\dagger}  Y_u  Y_{d}^{\dagger}}\Big) \nonumber \\ 
	&-33 \lambda_4 \mbox{Tr}\Big({Y_d  Y_{u}^{\dagger}  Y_u  Y_{d}^{\dagger}}\Big) -\frac{9}{2} \lambda_4 \mbox{Tr}\Big({Y_e  Y_{e}^{\dagger}  Y_e  Y_{e}^{\dagger}}\Big) -\frac{27}{2} \lambda_4 \mbox{Tr}\Big({Y_u  Y_{u}^{\dagger}  Y_u  Y_{u}^{\dagger}}\Big)\nonumber \\ 
	& -12 \mbox{Tr}\Big({Y_d  Y_{d}^{\dagger}  Y_d  Y_{u}^{\dagger}  Y_u  Y_{d}^{\dagger}}\Big) -12 \mbox{Tr}\Big({Y_d  Y_{u}^{\dagger}  Y_u  Y_{d}^{\dagger}  Y_d  Y_{d}^{\dagger}}\Big) -24 \mbox{Tr}\Big({Y_d  Y_{u}^{\dagger}  Y_u  Y_{u}^{\dagger}  Y_u  Y_{d}^{\dagger}}\Big), \nonumber\\
	\beta_{\lambda_4}^{(2,1)} & =0,\nonumber \\ 
	\beta_{\lambda_4}^{(2,2)} & =+\frac{3}{5} g_{1}^{2} g_{2}^{2} |y_3|^2 +\frac{3}{4} g_{1}^{2} \lambda_4 |y_3|^2 +\frac{15}{4} g_{2}^{2} \lambda_4 |y_3|^2 -4 \lambda_2 \lambda_4 |y_3|^2 -8 \lambda_3 \lambda_4 |y_3|^2 -4 \lambda_{4}^{2} |y_3|^2\nonumber\\
	& +\frac{3}{5} g_{1}^{2} g_{2}^{2} |y_4|^2 +\frac{3}{4} g_{1}^{2} \lambda_4 |y_4|^2 +\frac{15}{4} g_{2}^{2} \lambda_4 |y_4|^2 -4 \lambda_1 \lambda_4 |y_4|^2 -8 \lambda_3 \lambda_4 |y_4|^2 -4 \lambda_{4}^{2} |y_4|^2 \nonumber\\
	&-\frac{9}{2} \lambda_4 |y_3|^4 -\frac{9}{2} \lambda_4 |y_4|^4 +10 y_1 y_{3}^{2} y_{1}^{*3} +10 y_2 y_{4}^{2} y_{2}^{*3} +2 \lambda_1 y_{1}^{2} y_{3}^{*2} +2 \lambda_2 y_{1}^{2} y_{3}^{*2} \nonumber\\
	&+12 y_{1}^{2} |y_4|^2 y_{3}^{*2} +10 y_{1}^{2} y_3 y_{3}^{*3} +2 \lambda_4 y_4 |y_3|^2 y_4^* +8 \lambda_3 y_1 y_2 y_3^* y_4^* +8 \lambda_4 y_1 y_2 y_3^* y_4^* \nonumber\\
	&+22 y_1 y_2 y_3 y_{3}^{*2} y_4^* +2 \lambda_1 y_{2}^{2} y_{4}^{*2} +2 \lambda_2 y_{2}^{2} y_{4}^{*2} +12 y_{2}^{2} |y_3|^2 y_{4}^{*2} +22 y_1 y_2 y_4 y_3^* y_{4}^{*2} +10 y_{2}^{2} y_4 y_{4}^{*3} \nonumber\\
	&+y_{1}^{*2} \Big(-\frac{9}{2} \lambda_4 y_{1}^{2} +2 \lambda_1 y_{3}^{2} +2 \lambda_2 y_{3}^{2} +12 y_{3}^{2} |y_4|^2 +2 y_3 \Big(11 y_1 y_4  + 6 y_2 y_3 \Big)y_2^* +10 \Big(2 y_{1}^{2} y_3 \nonumber\\
	& + y_{3}^{3}\Big)y_3^* +22 y_1 y_2 y_3 y_4^* \Big)+y_{2}^{*2} \Big(10 \Big(2 y_{2}^{2} y_4  + y_{4}^{3}\Big)y_4^*  + 2 \lambda_1 y_{4}^{2}  + 2 \lambda_2 y_{4}^{2}  + 2 y_4 \Big(11 y_1 y_2 \nonumber\\
	& + 6 y_3 y_4 \Big)y_3^*  -\frac{9}{2} \lambda_4 y_{2}^{2} \Big)+y_1^* \Big(\frac{3}{5} g_{1}^{2} g_{2}^{2} y_1 +\frac{3}{4} g_{1}^{2} \lambda_4 y_1 +\frac{15}{4} g_{2}^{2} \lambda_4 y_1 -4 \lambda_1 \lambda_4 y_1 -8 \lambda_3 \lambda_4 y_1 \nonumber\\
	&-4 \lambda_{4}^{2} y_1 -2 \lambda_4 y_1 |y_4|^2 +2 y_4 \Big(11 y_2 y_3  + 6 y_1 y_4 \Big)y_{2}^{*2} +10 \Big(2 y_1 y_{3}^{2}  + y_{1}^{3}\Big)y_{3}^{*2} +4 \lambda_1 y_2 y_3 y_4^* \nonumber\\
	&+4 \lambda_2 y_2 y_3 y_4^* -4 \lambda_4 y_2 y_3 y_4^* +12 y_1 y_{2}^{2} y_{4}^{*2} +22 y_2 y_3 y_4 y_{4}^{*2} +y_3^* \Big(\Big(22 y_{1}^{2} y_2  + 22 y_2 y_{3}^{2}  \nonumber\\
	&+ 24 y_1 y_3 y_4 \Big)y_4^*  + \Big(8 \lambda_3  - \lambda_4 \Big)y_1 y_3 \Big)+2 y_2^* \Big(\lambda_4 y_1 y_2 +4 \lambda_3 y_3 y_4 +4 \lambda_4 y_3 y_4 +\Big(11 y_{1}^{2} y_4 \nonumber\\
	& + 11 y_{3}^{2} y_4  + 12 y_1 y_2 y_3 \Big)y_3^* +\Big(11 y_{2}^{2} y_3  + 11 y_3 y_{4}^{2}  + 12 y_1 y_2 y_4 \Big)y_4^* \Big)\Big)+\frac{1}{20} y_2^* \Big(40 y_1 \Big(11 y_3 y_4  \nonumber\\
	&+ 6 y_1 y_2 \Big)y_{3}^{*2} -40 y_3^* \Big(- \Big(11 y_1 \Big(y_{2}^{2} + y_{4}^{2}\Big) + 12 y_2 y_3 y_4 \Big)y_4^*  -2 \Big(\lambda_1 + \lambda_2\Big)y_1 y_4  + \lambda_4 \Big(2 y_1 y_4  \nonumber\\
	&+ y_2 y_3 \Big)\Big)+y_2 \Big(200 \Big(2 y_{4}^{2}  + y_{2}^{2}\Big)y_{4}^{*2}  + 20 \Big(8 \lambda_3  - \lambda_4 \Big)|y_4|^2  + 3 g_{1}^{2} \Big(4 g_{2}^{2}  + 5 \lambda_4 \Big) + 75 g_{2}^{2} \lambda_4 \nonumber\\
	& -80 \lambda_4 \Big(2 \lambda_3  + \lambda_2 + \lambda_4\Big)\Big)\Big),\nonumber \\ 
	\beta_{\lambda_3}^{(1,0)} & =  
	+\frac{27}{100} g_{1}^{4} -\frac{9}{10} g_{1}^{2} g_{2}^{2} +\frac{9}{4} g_{2}^{4} -\frac{9}{5} g_{1}^{2} \lambda_3 -9 g_{2}^{2} \lambda_3 +6 \lambda_1 \lambda_3 +6 \lambda_2 \lambda_3 +4 \lambda_{3}^{2} +2 \lambda_1 \lambda_4 \nonumber \\ 
	&+2 \lambda_2 \lambda_4 +2 \lambda_{4}^{2} +6 \lambda_3 \mbox{Tr}\Big({Y_d  Y_{d}^{\dagger}}\Big) +2 \lambda_3 \mbox{Tr}\Big({Y_e  Y_{e}^{\dagger}}\Big) +6 \lambda_3 \mbox{Tr}\Big({Y_u  Y_{u}^{\dagger}}\Big) -12 \mbox{Tr}\Big({Y_d  Y_{u}^{\dagger}  Y_u  Y_{d}^{\dagger}}\Big),\nonumber \\ 
	\beta_{\lambda_3}^{(1,1)} & = 0,\nonumber\\
	\beta_{\lambda_3}^{(1,2)} & =+2 \lambda_3 |y_3|^2 +2 \lambda_3 |y_4|^2 +2 |y_1|^2 \Big(-2 y_2 y_2^*  -2 y_3 y_3^*  + \lambda_3\Big)-4 y_4 |y_3|^2 y_4^* +2 |y_2|^2 \Big(-2 y_4 y_4^*  + \lambda_3\Big),\nonumber\\
	\beta_{\lambda_3}^{(2,0)} & =  
	-\frac{153}{40} g_{1}^{6} +\frac{201}{40} g_{1}^{4} g_{2}^{2} +\frac{13}{8} g_{1}^{2} g_{2}^{4} +\frac{259}{8} g_{2}^{6} +\frac{27}{20} g_{1}^{4} \lambda_1 -\frac{3}{2} g_{1}^{2} g_{2}^{2} \lambda_1 +\frac{45}{4} g_{2}^{4} \lambda_1 +\frac{27}{20} g_{1}^{4} \lambda_2 \nonumber \\ 
	&-\frac{3}{2} g_{1}^{2} g_{2}^{2} \lambda_2 +\frac{45}{4} g_{2}^{4} \lambda_2 +\frac{1893}{200} g_{1}^{4} \lambda_3 +\frac{33}{20} g_{1}^{2} g_{2}^{2} \lambda_3 -\frac{71}{8} g_{2}^{4} \lambda_3 +\frac{36}{5} g_{1}^{2} \lambda_1 \lambda_3 +36 g_{2}^{2} \lambda_1 \lambda_3 \nonumber \\ 
	&-15 \lambda_{1}^{2} \lambda_3 +\frac{36}{5} g_{1}^{2} \lambda_2 \lambda_3 +36 g_{2}^{2} \lambda_2 \lambda_3 -15 \lambda_{2}^{2} \lambda_3 +\frac{6}{5} g_{1}^{2} \lambda_{3}^{2} +6 g_{2}^{2} \lambda_{3}^{2} -36 \lambda_1 \lambda_{3}^{2} -36 \lambda_2 \lambda_{3}^{2} \nonumber \\ 
	&-12 \lambda_{3}^{3} +\frac{9}{10} g_{1}^{4} \lambda_4 -\frac{9}{5} g_{1}^{2} g_{2}^{2} \lambda_4 +\frac{15}{2} g_{2}^{4} \lambda_4 +\frac{12}{5} g_{1}^{2} \lambda_1 \lambda_4 +18 g_{2}^{2} \lambda_1 \lambda_4 -4 \lambda_{1}^{2} \lambda_4 +\frac{12}{5} g_{1}^{2} \lambda_2 \lambda_4 \nonumber \\ 
	&+18 g_{2}^{2} \lambda_2 \lambda_4 -4 \lambda_{2}^{2} \lambda_4 -12 g_{2}^{2} \lambda_3 \lambda_4 -16 \lambda_1 \lambda_3 \lambda_4 -16 \lambda_2 \lambda_3 \lambda_4 -4 \lambda_{3}^{2} \lambda_4 -\frac{6}{5} g_{1}^{2} \lambda_{4}^{2} \nonumber \\ 
	&+6 g_{2}^{2} \lambda_{4}^{2} -14 \lambda_1 \lambda_{4}^{2} -14 \lambda_2 \lambda_{4}^{2} -16 \lambda_3 \lambda_{4}^{2} -12 \lambda_{4}^{3}+\frac{9}{20} g_{1}^{4} \mbox{Tr}\Big({Y_d  Y_{d}^{\dagger}}\Big) -\frac{27}{10} g_{1}^{2} g_{2}^{2} \mbox{Tr}\Big({Y_d  Y_{d}^{\dagger}}\Big)  \nonumber \\ 
	&-\frac{9}{4} g_{2}^{4} \mbox{Tr}\Big({Y_d  Y_{d}^{\dagger}}\Big) +\frac{5}{4} g_{1}^{2} \lambda_3 \mbox{Tr}\Big({Y_d  Y_{d}^{\dagger}}\Big)+\frac{45}{4} g_{2}^{2} \lambda_3 \mbox{Tr}\Big({Y_d  Y_{d}^{\dagger}}\Big) +40 g_{3}^{2} \lambda_3 \mbox{Tr}\Big({Y_d  Y_{d}^{\dagger}}\Big) \nonumber \\ 
	& -36 \lambda_1 \lambda_3 \mbox{Tr}\Big({Y_d  Y_{d}^{\dagger}}\Big) -12 \lambda_{3}^{2} \mbox{Tr}\Big({Y_d  Y_{d}^{\dagger}}\Big) -12 \lambda_1 \lambda_4 \mbox{Tr}\Big({Y_d  Y_{d}^{\dagger}}\Big) -6 \lambda_{4}^{2} \mbox{Tr}\Big({Y_d  Y_{d}^{\dagger}}\Big) \nonumber \\ 
	&-\frac{9}{4} g_{1}^{4} \mbox{Tr}\Big({Y_e  Y_{e}^{\dagger}}\Big) -\frac{33}{10} g_{1}^{2} g_{2}^{2} \mbox{Tr}\Big({Y_e  Y_{e}^{\dagger}}\Big) -\frac{3}{4} g_{2}^{4} \mbox{Tr}\Big({Y_e  Y_{e}^{\dagger}}\Big) +\frac{15}{4} g_{1}^{2} \lambda_3 \mbox{Tr}\Big({Y_e  Y_{e}^{\dagger}}\Big) \nonumber\\
	&+\frac{15}{4} g_{2}^{2} \lambda_3 \mbox{Tr}\Big({Y_e  Y_{e}^{\dagger}}\Big) -12 \lambda_1 \lambda_3 \mbox{Tr}\Big({Y_e  Y_{e}^{\dagger}}\Big)-4 \lambda_{3}^{2} \mbox{Tr}\Big({Y_e  Y_{e}^{\dagger}}\Big) -4 \lambda_1 \lambda_4 \mbox{Tr}\Big({Y_e  Y_{e}^{\dagger}}\Big)  \nonumber \\ 
	&-2 \lambda_{4}^{2} \mbox{Tr}\Big({Y_e  Y_{e}^{\dagger}}\Big) -\frac{171}{100} g_{1}^{4} \mbox{Tr}\Big({Y_u  Y_{u}^{\dagger}}\Big) -\frac{63}{10} g_{1}^{2} g_{2}^{2} \mbox{Tr}\Big({Y_u  Y_{u}^{\dagger}}\Big) -\frac{9}{4} g_{2}^{4} \mbox{Tr}\Big({Y_u  Y_{u}^{\dagger}}\Big) \nonumber \\ &
	+\frac{17}{4} g_{1}^{2} \lambda_3 \mbox{Tr}\Big({Y_u  Y_{u}^{\dagger}}\Big) +\frac{45}{4} g_{2}^{2} \lambda_3 \mbox{Tr}\Big({Y_u  Y_{u}^{\dagger}}\Big) +40 g_{3}^{2} \lambda_3 \mbox{Tr}\Big({Y_u  Y_{u}^{\dagger}}\Big) \nonumber\\&-36 \lambda_2 \lambda_3 \mbox{Tr}\Big({Y_u  Y_{u}^{\dagger}}\Big) -12 \lambda_{3}^{2} \mbox{Tr}\Big({Y_u  Y_{u}^{\dagger}}\Big) -12 \lambda_2 \lambda_4 \mbox{Tr}\Big({Y_u  Y_{u}^{\dagger}}\Big) \nonumber \\ 
	&-6 \lambda_{4}^{2} \mbox{Tr}\Big({Y_u  Y_{u}^{\dagger}}\Big) -\frac{27}{2} \lambda_3 \mbox{Tr}\Big({Y_d  Y_{d}^{\dagger}  Y_d  Y_{d}^{\dagger}}\Big) -\frac{4}{5} g_{1}^{2} \mbox{Tr}\Big({Y_d  Y_{u}^{\dagger}  Y_u  Y_{d}^{\dagger}}\Big) \nonumber \\ 
	&-64 g_{3}^{2} \mbox{Tr}\Big({Y_d  Y_{u}^{\dagger}  Y_u  Y_{d}^{\dagger}}\Big) +15 \lambda_3 \mbox{Tr}\Big({Y_d  Y_{u}^{\dagger}  Y_u  Y_{d}^{\dagger}}\Big) -\frac{9}{2} \lambda_3 \mbox{Tr}\Big({Y_e  Y_{e}^{\dagger}  Y_e  Y_{e}^{\dagger}}\Big) \nonumber \\ 
	&-\frac{27}{2} \lambda_3 \mbox{Tr}\Big({Y_u  Y_{u}^{\dagger}  Y_u  Y_{u}^{\dagger}}\Big) +12 \mbox{Tr}\Big({Y_d  Y_{d}^{\dagger}  Y_d  Y_{u}^{\dagger}  Y_u  Y_{d}^{\dagger}}\Big) +24 \mbox{Tr}\Big({Y_d  Y_{u}^{\dagger}  Y_u  Y_{d}^{\dagger}  Y_d  Y_{d}^{\dagger}}\Big) \nonumber \\ 
	&+36 \mbox{Tr}\Big({Y_d  Y_{u}^{\dagger}  Y_u  Y_{u}^{\dagger}  Y_u  Y_{d}^{\dagger}}\Big), \nonumber\\ 
		\beta_{\lambda_3}^{(2,1)} & =  0,\nonumber\\
	\beta_{\lambda_3}^{(2,2)} & = -\frac{9}{100} g_{1}^{4} |y_3|^2 +\frac{3}{10} g_{1}^{2} g_{2}^{2} |y_3|^2 -\frac{3}{4} g_{2}^{4} |y_3|^2 +\frac{3}{4} g_{1}^{2} \lambda_3 |y_3|^2 +\frac{15}{4} g_{2}^{2} \lambda_3 |y_3|^2 -12 \lambda_2 \lambda_3 |y_3|^2 \nonumber\\
	&-4 \lambda_{3}^{2} |y_3|^2 -4 \lambda_2 \lambda_4 |y_3|^2 -2 \lambda_{4}^{2} |y_3|^2 -\frac{9}{100} g_{1}^{4} |y_4|^2 +\frac{3}{10} g_{1}^{2} g_{2}^{2} |y_4|^2 -\frac{3}{4} g_{2}^{4} |y_4|^2 +\frac{3}{4} g_{1}^{2} \lambda_3 |y_4|^2 \nonumber\\
	&+\frac{15}{4} g_{2}^{2} \lambda_3 |y_4|^2 -12 \lambda_1 \lambda_3 |y_4|^2 -4 \lambda_{3}^{2} |y_4|^2 -4 \lambda_1 \lambda_4 |y_4|^2 -2 \lambda_{4}^{2} |y_4|^2-\frac{9}{2} \lambda_3 |y_3|^4 +14 |y_4|^2 |y_3|^4 \nonumber\\
	& -\frac{9}{2} \lambda_3 |y_4|^4 +14 |y_3|^2 |y_4|^4 -2 \Big(- \Big(19 y_2 y_3  + 4 y_1 y_4 \Big)|y_4|^2  -2 \Big(\lambda_1 + \lambda_2\Big)y_1 y_4  + \lambda_3 \Big(2 y_1 y_4  \nonumber\\
	&+ y_2 y_3 \Big)\Big)y_2^* y_3^* +8 y_1 y_3 y_4 y_2^* y_{3}^{*2} +\frac{1}{2} y_2 y_{2}^{*2} \Big(16 y_1 y_4 y_3^*  + 40 y_2 |y_4|^2  -9 \lambda_3 y_2 \Big)+2 \lambda_3 y_4 |y_3|^2 y_4^*\nonumber\\
	& -\frac{1}{2} y_1 y_{1}^{*2} \Big(-16 y_2 y_3 y_4^*  -28 y_1 |y_2|^2  -40 y_1 |y_3|^2  + 9 \lambda_3 y_1 \Big)-\frac{1}{100} |y_2|^2 \Big(9 g_{1}^{4} -15 g_{1}^{2} \Big(2 g_{2}^{2}  + 5 \lambda_3 \Big)\nonumber\\
	&+25 \Big(-15 g_{2}^{2} \lambda_3  + 3 g_{2}^{4}  + 8 \Big(2 \lambda_2 \lambda_4  + 2 \lambda_{3}^{2}  + 6 \lambda_2 \lambda_3  + \lambda_{4}^{2}\Big)\Big)-2000 |y_4|^4 +100 \Big(-8 \lambda_4  \nonumber\\
	&+ \lambda_3\Big)y_4 y_4^* \Big)+y_1^* \Big(-\frac{9}{100} g_{1}^{4} y_1 +\frac{3}{10} g_{1}^{2} g_{2}^{2} y_1 -\frac{3}{4} g_{2}^{4} y_1 +\frac{3}{4} g_{1}^{2} \lambda_3 y_1 +\frac{15}{4} g_{2}^{2} \lambda_3 y_1\nonumber\\
	& -12 \lambda_1 \lambda_3 y_1 -4 \lambda_{3}^{2} y_1 -4 \lambda_1 \lambda_4 y_1 -2 \lambda_{4}^{2} y_1 -2 \lambda_3 y_1 |y_4|^2 +14 y_1 |y_2|^4 +20 y_1 |y_3|^4 \nonumber\\
	&+4 \lambda_1 y_2 y_3 y_4^* +4 \lambda_2 y_2 y_3 y_4^* -4 \lambda_3 y_2 y_3 y_4^* +8 y_2 y_3 y_4 y_{4}^{*2} +|y_3|^2 \Big(38 y_1 y_4 y_4^*  + 8 \lambda_4 y_1  + 8 y_2 y_3 y_4^* \nonumber\\
	& - \lambda_3 y_1 \Big)+2 y_2^* \Big(\lambda_3 y_1 y_2  + y_1 \Big(19 y_2 y_3  + 4 y_1 y_4 \Big)y_3^*  + y_2 \Big(19 y_1 y_4  + 4 y_2 y_3 \Big)y_4^* \Big)\Big)\nonumber \\  
	&+ 9 g_{1}^{4} \Big)|y_2|^2 -\frac{1}{2} \Big(-28 |y_2|^2  + 9 \lambda_3 \Big)|y_1|^4 -\frac{9}{2} \lambda_3 |y_2|^4 -\frac{1}{100} |y_1|^2 \Big(9 g_{1}^{4} -15 g_{1}^{2} \Big(2 g_{2}^{2}  + 5 \lambda_3 \Big)\nonumber \\ 
	&+25 \Big(-15 g_{2}^{2} \lambda_3  + 3 g_{2}^{4}  + 8 \Big(2 \lambda_1 \lambda_4  + 2 \lambda_{3}^{2}  + 6 \lambda_1 \lambda_3  + \lambda_{4}^{2}\Big)\Big)-1400 |y_2|^4 -200 \lambda_3 y_2 y_2^* \Big),\nonumber   \\
	\beta_{\lambda_2}^{(1,0)} & =  
	+\frac{27}{100} g_{1}^{4} +\frac{9}{10} g_{1}^{2} g_{2}^{2} +\frac{9}{4} g_{2}^{4} -\frac{9}{5} g_{1}^{2} \lambda_2 -9 g_{2}^{2} \lambda_2 +12 \lambda_{2}^{2} +4 \lambda_{3}^{2} +4 \lambda_3 \lambda_4 +2 \lambda_{4}^{2} \nonumber \\ 
	&+12 \lambda_2 \mbox{Tr}\Big({Y_u  Y_{u}^{\dagger}}\Big) -12 \mbox{Tr}\Big({Y_u  Y_{u}^{\dagger}  Y_u  Y_{u}^{\dagger}}\Big),\nonumber \\ 
	\beta_{\lambda_2}^{(1,1)} & = 0,\nonumber\\
	\beta_{\lambda_2}^{(1,2)} & = +4 \lambda_2 |y_3|^2 -4 |y_2|^4 -4 |y_3|^4 +4 |y_2|^2 \Big(-2 y_3 y_3^*  + \lambda_2\Big),\nonumber\\
	\beta_{\lambda_2}^{(2,0)} & =  
	-\frac{153}{40} g_{1}^{6} -\frac{363}{40} g_{1}^{4} g_{2}^{2} -\frac{67}{8} g_{1}^{2} g_{2}^{4} +\frac{259}{8} g_{2}^{6} +\frac{2073}{200} g_{1}^{4} \lambda_2 +\frac{117}{20} g_{1}^{2} g_{2}^{2} \lambda_2 -\frac{11}{8} g_{2}^{4} \lambda_2 \nonumber \\ &+\frac{54}{5} g_{1}^{2} \lambda_{2}^{2} +54 g_{2}^{2} \lambda_{2}^{2} -78 \lambda_{2}^{3} +\frac{9}{5} g_{1}^{4} \lambda_3 +15 g_{2}^{4} \lambda_3 +\frac{24}{5} g_{1}^{2} \lambda_{3}^{2} +24 g_{2}^{2} \lambda_{3}^{2} -20 \lambda_2 \lambda_{3}^{2} -16 \lambda_{3}^{3} \nonumber \\ 
	&+\frac{9}{10} g_{1}^{4} \lambda_4 +3 g_{1}^{2} g_{2}^{2} \lambda_4 +\frac{15}{2} g_{2}^{4} \lambda_4 +\frac{24}{5} g_{1}^{2} \lambda_3 \lambda_4 +24 g_{2}^{2} \lambda_3 \lambda_4 -20 \lambda_2 \lambda_3 \lambda_4 -24 \lambda_{3}^{2} \lambda_4 \nonumber \\ 
	&+\frac{12}{5} g_{1}^{2} \lambda_{4}^{2} +6 g_{2}^{2} \lambda_{4}^{2} -12 \lambda_2 \lambda_{4}^{2} -32 \lambda_3 \lambda_{4}^{2} -12 \lambda_{4}^{3}-24 \lambda_{3}^{2} \mbox{Tr}\Big({Y_d  Y_{d}^{\dagger}}\Big) -24 \lambda_3 \lambda_4 \mbox{Tr}\Big({Y_d  Y_{d}^{\dagger}}\Big) \nonumber \\ 
	&-12 \lambda_{4}^{2} \mbox{Tr}\Big({Y_d  Y_{d}^{\dagger}}\Big) -8 \lambda_{3}^{2} \mbox{Tr}\Big({Y_e  Y_{e}^{\dagger}}\Big) -8 \lambda_3 \lambda_4 \mbox{Tr}\Big({Y_e  Y_{e}^{\dagger}}\Big) -4 \lambda_{4}^{2} \mbox{Tr}\Big({Y_e  Y_{e}^{\dagger}}\Big) \nonumber \\ 
	&-\frac{171}{50} g_{1}^{4} \mbox{Tr}\Big({Y_u  Y_{u}^{\dagger}}\Big) +\frac{63}{5} g_{1}^{2} g_{2}^{2} \mbox{Tr}\Big({Y_u  Y_{u}^{\dagger}}\Big) -\frac{9}{2} g_{2}^{4} \mbox{Tr}\Big({Y_u  Y_{u}^{\dagger}}\Big) +\frac{17}{2} g_{1}^{2} \lambda_2 \mbox{Tr}\Big({Y_u  Y_{u}^{\dagger}}\Big) \nonumber \\ 
	&+\frac{45}{2} g_{2}^{2} \lambda_2 \mbox{Tr}\Big({Y_u  Y_{u}^{\dagger}}\Big) +80 g_{3}^{2} \lambda_2 \mbox{Tr}\Big({Y_u  Y_{u}^{\dagger}}\Big) -72 \lambda_{2}^{2} \mbox{Tr}\Big({Y_u  Y_{u}^{\dagger}}\Big) -9 \lambda_2 \mbox{Tr}\Big({Y_d  Y_{u}^{\dagger}  Y_u  Y_{d}^{\dagger}}\Big) \nonumber \\ 
	&-\frac{16}{5} g_{1}^{2} \mbox{Tr}\Big({Y_u  Y_{u}^{\dagger}  Y_u  Y_{u}^{\dagger}}\Big) -64 g_{3}^{2} \mbox{Tr}\Big({Y_u  Y_{u}^{\dagger}  Y_u  Y_{u}^{\dagger}}\Big) -3 \lambda_2 \mbox{Tr}\Big({Y_u  Y_{u}^{\dagger}  Y_u  Y_{u}^{\dagger}}\Big) \nonumber \\ 
	&+12 \mbox{Tr}\Big({Y_d  Y_{u}^{\dagger}  Y_u  Y_{u}^{\dagger}  Y_u  Y_{d}^{\dagger}}\Big) +60 \mbox{Tr}\Big({Y_u  Y_{u}^{\dagger}  Y_u  Y_{u}^{\dagger}  Y_u  Y_{u}^{\dagger}}\Big), \nonumber\\
\beta_{\lambda_2}^{(2,1)} & = 0,\nonumber\\
\beta_{\lambda_2}^{(2,2)} & = -\frac{9}{50} g_{1}^{4} |y_3|^2 -\frac{3}{5} g_{1}^{2} g_{2}^{2} |y_3|^2 -\frac{3}{2} g_{2}^{4} |y_3|^2 +\frac{3}{2} g_{1}^{2} \lambda_2 |y_3|^2 +\frac{15}{2} g_{2}^{2} \lambda_2 |y_3|^2 -24 \lambda_{2}^{2} |y_3|^2\nonumber\\
& -8 \lambda_{3}^{2} |y_4|^2 -8 \lambda_3 \lambda_4 |y_4|^2 -4 \lambda_{4}^{2} |y_4|^2 - \lambda_2 |y_3|^4 +8 |y_4|^2 |y_3|^4 +20 y_{2}^{3} y_{2}^{*3} +4 \Big(2 \Big(\lambda_3 \nonumber\\
&+ \lambda_4\Big)y_1 y_4  + 7 y_2 y_3 |y_4|^2  + \lambda_2 \Big(3 y_2 y_3  - y_1 y_4 \Big)\Big)y_2^* y_3^* +4 y_3 \Big(17 y_2 y_3  + 4 y_1 y_4 \Big)y_2^* y_{3}^{*2} +20 y_{3}^{3} y_{3}^{*3} \nonumber \\ 
&+y_2 y_{2}^{*2} \Big(16 y_1 y_4 y_3^*  + 20 y_2 |y_4|^2  + 68 y_2 |y_3|^2  - \lambda_2 y_2 \Big)-6 \lambda_2 y_4 |y_3|^2 y_4^* -\frac{3}{50} |y_2|^2 \Big(150 \lambda_2 y_4 y_4^*  \nonumber\\
&+ 25 \Big(16 \lambda_{2}^{2}  -5 g_{2}^{2} \lambda_2  + g_{2}^{4}\Big) + 3 g_{1}^{4}  + 5 g_{1}^{2} \Big(2 g_{2}^{2}  -5 \lambda_2 \Big)\Big)+y_1^* \Big(8 y_1 |y_2|^4 +20 y_1 |y_3|^4 \nonumber\\
&+2 |y_2|^2 \Big(14 y_1 y_3 y_3^*  -3 \lambda_2 y_1  + 8 y_2 y_3 y_4^* \Big)+|y_3|^2 \Big(16 y_2 y_3 y_4^*  -9 \lambda_2 y_1 \Big)\nonumber \\ 
&-4 \Big(\Big(2 \lambda_{3}^{2}  + 2 \lambda_3 \lambda_4  + \lambda_{4}^{2}\Big)y_1  + \Big(-2 \Big(\lambda_3 + \lambda_4\Big) + \lambda_2\Big)y_2 y_3 y_4^* \Big)\Big).\nonumber \
	\end{align}} 

\section{Results for different values of top pole mass}
\label{app:top_mass}
\begin{figure}[t]
\centering
\subfigure{\includegraphics[width=0.24\textwidth]{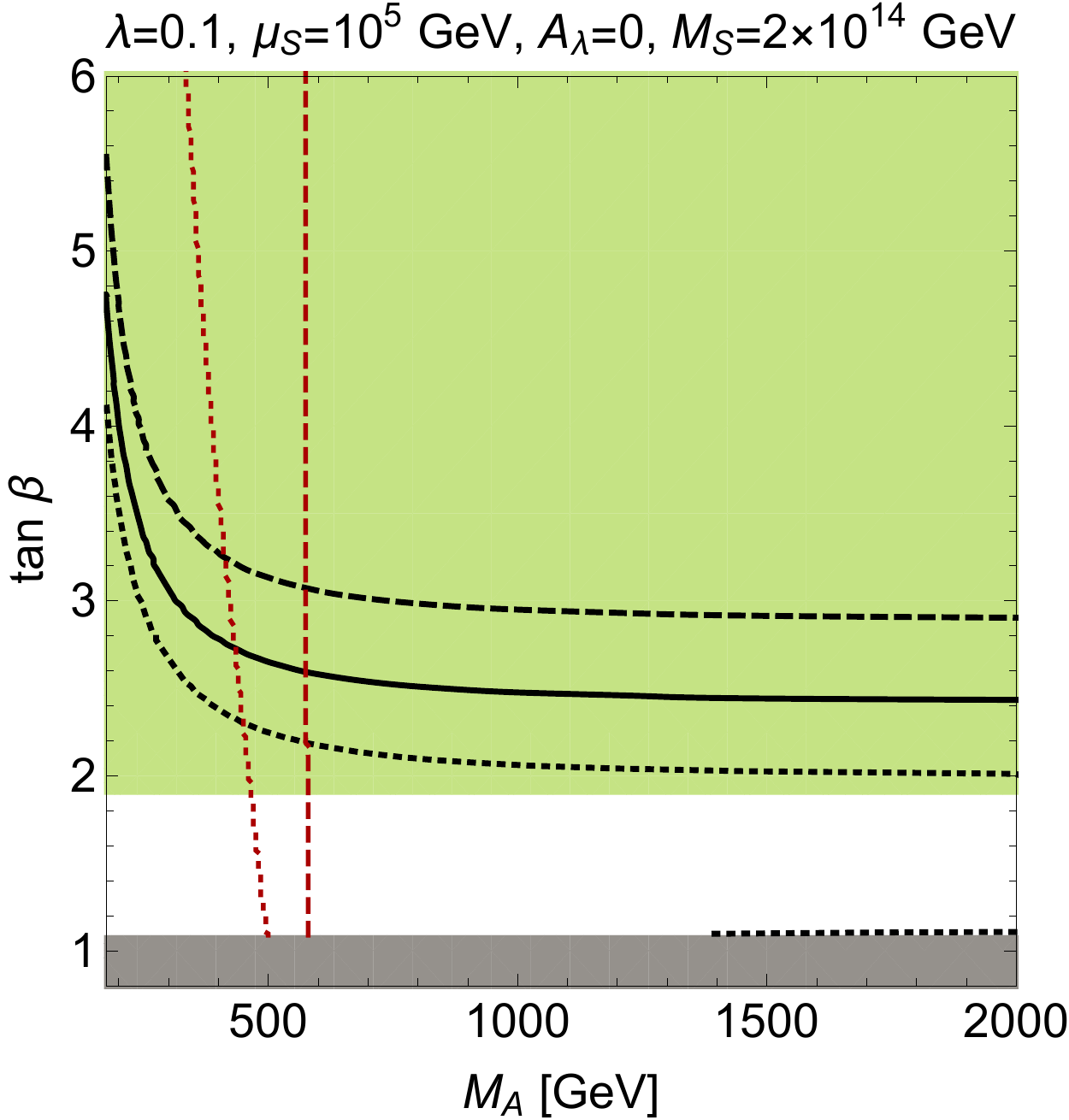}}
\subfigure{\includegraphics[width=0.24\textwidth]{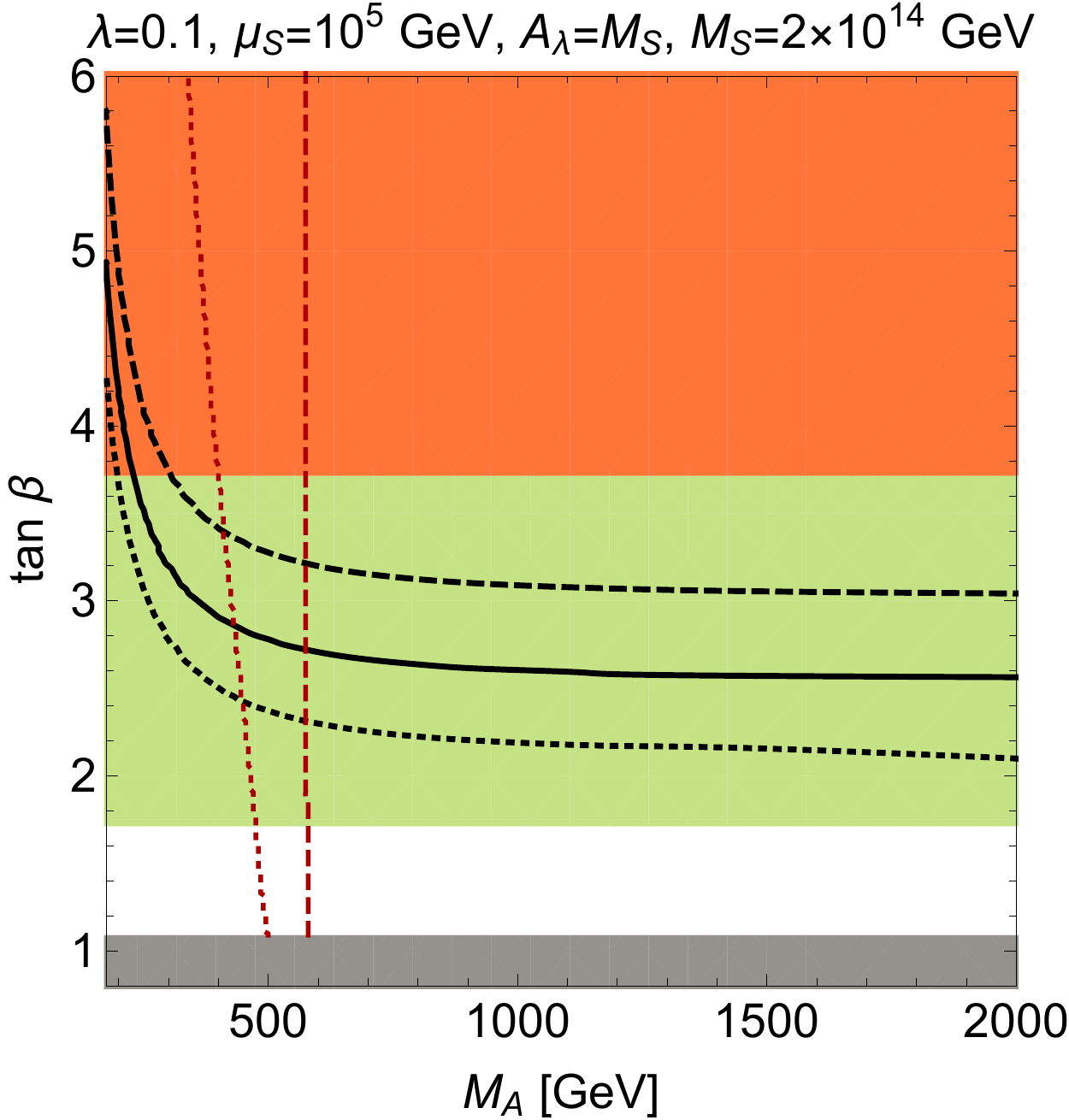}}
\subfigure{\includegraphics[width=0.24\textwidth]{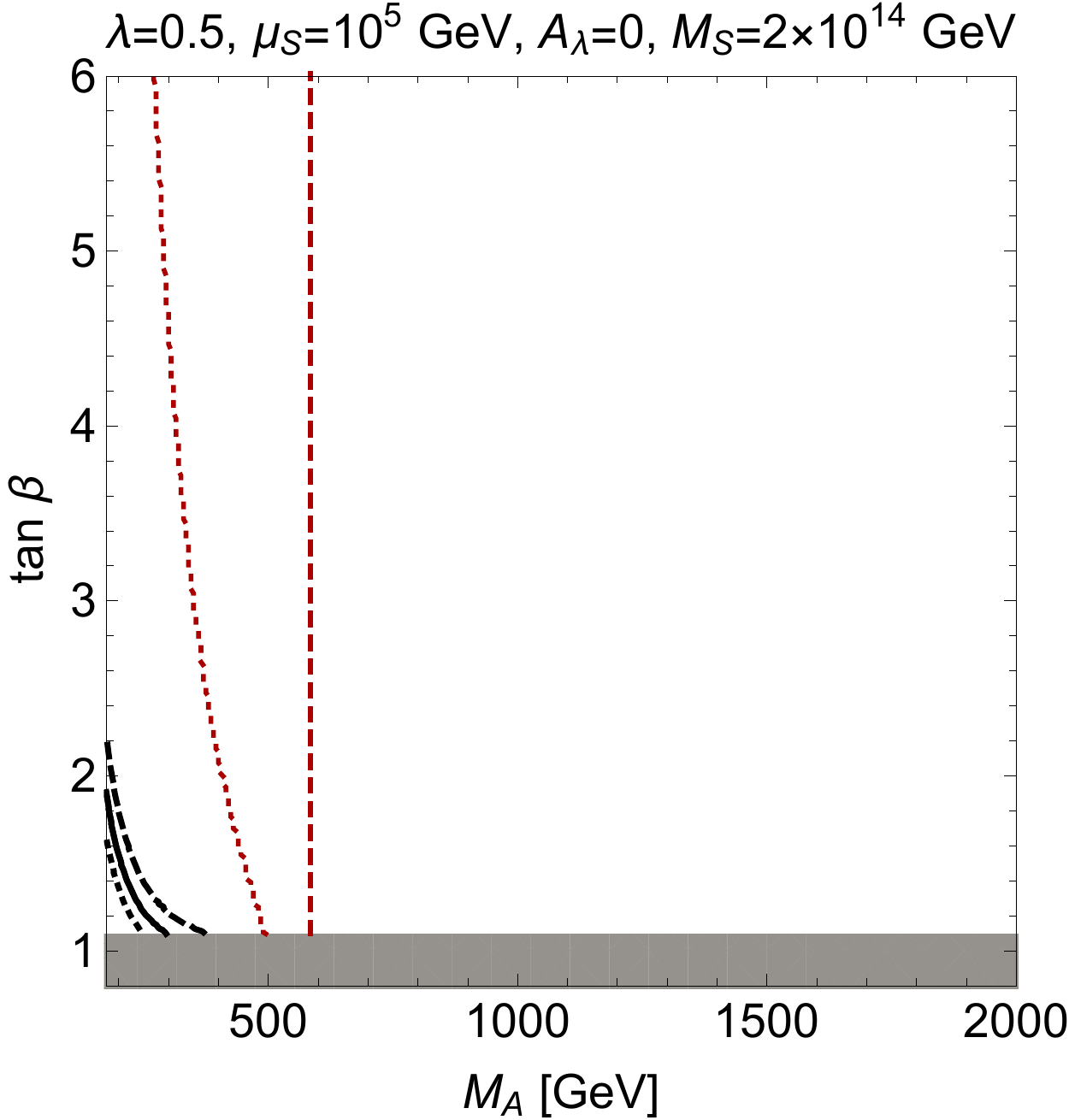}}
\subfigure{\includegraphics[width=0.24\textwidth]{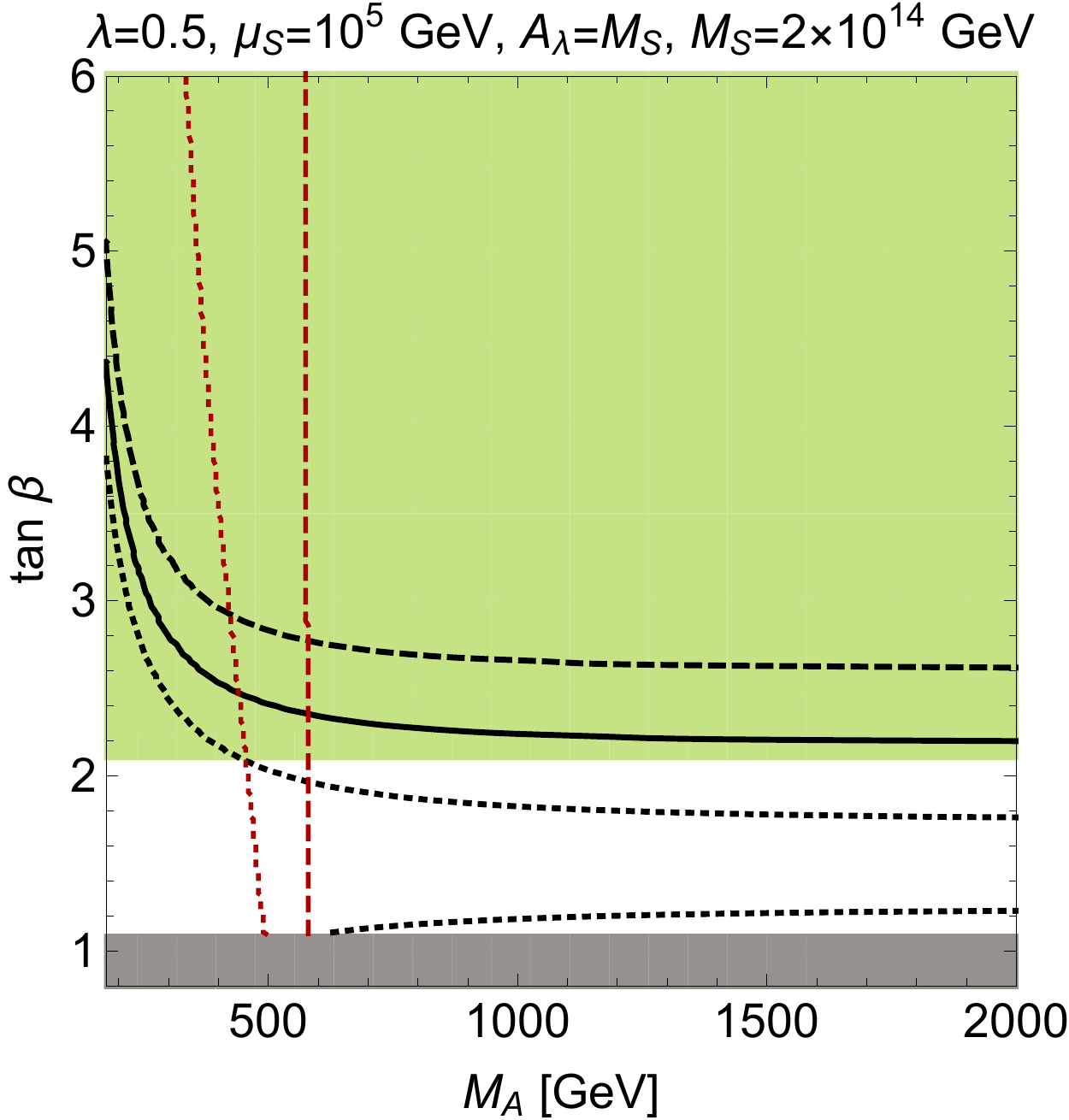}}\\
\subfigure{\includegraphics[width=0.24\textwidth]{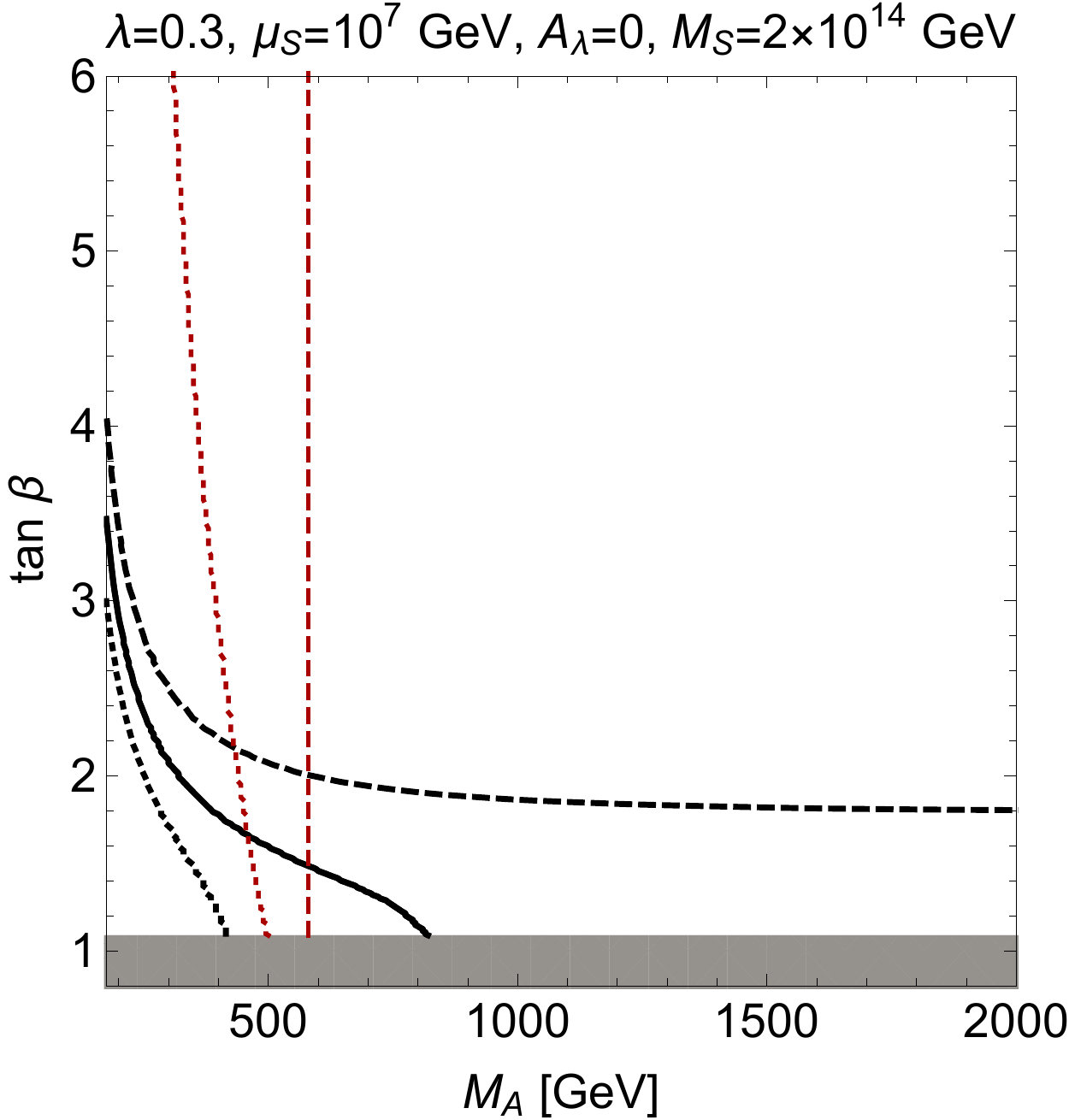}}
\subfigure{\includegraphics[width=0.24\textwidth]{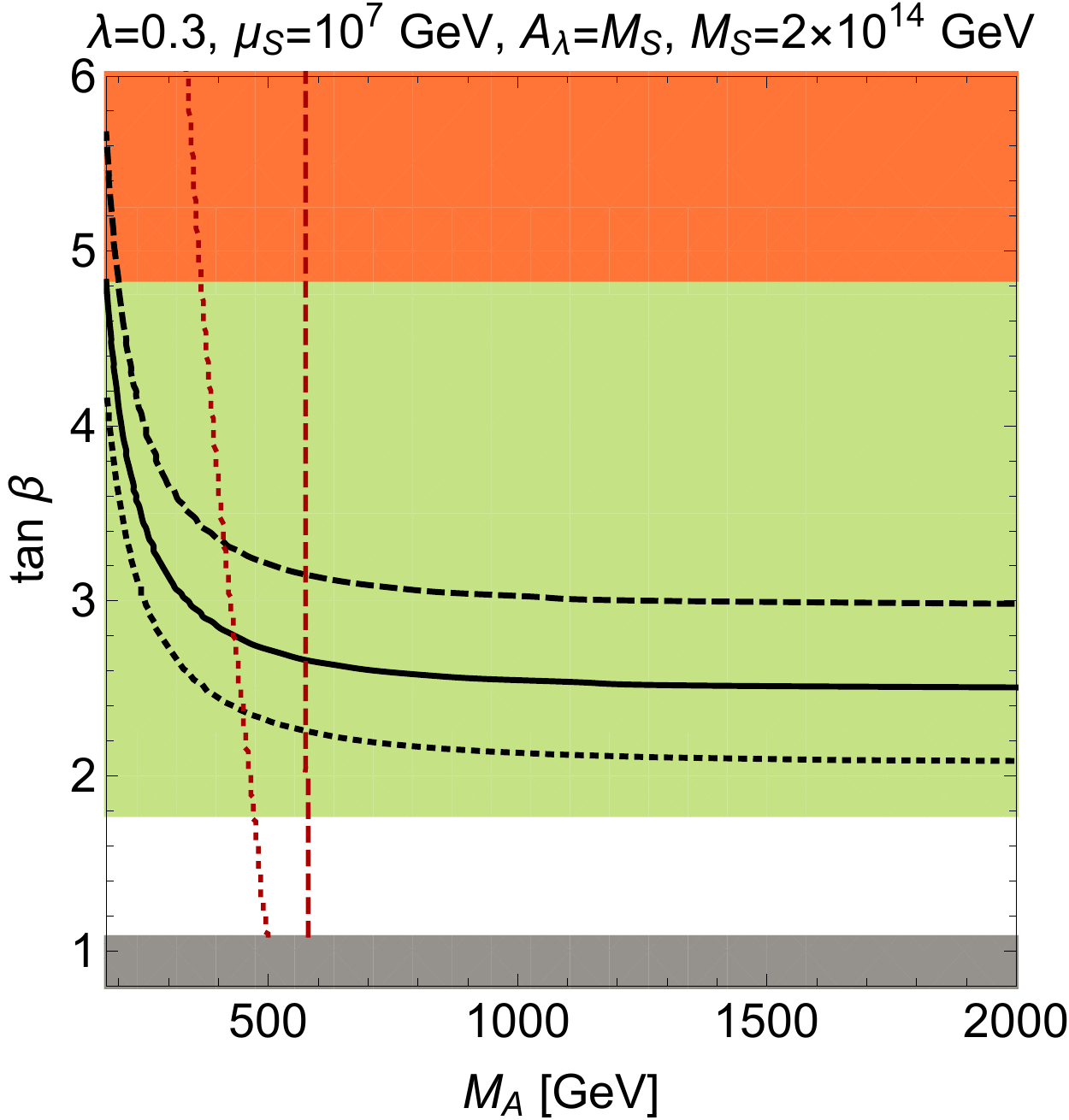}}
\subfigure{\includegraphics[width=0.24\textwidth]{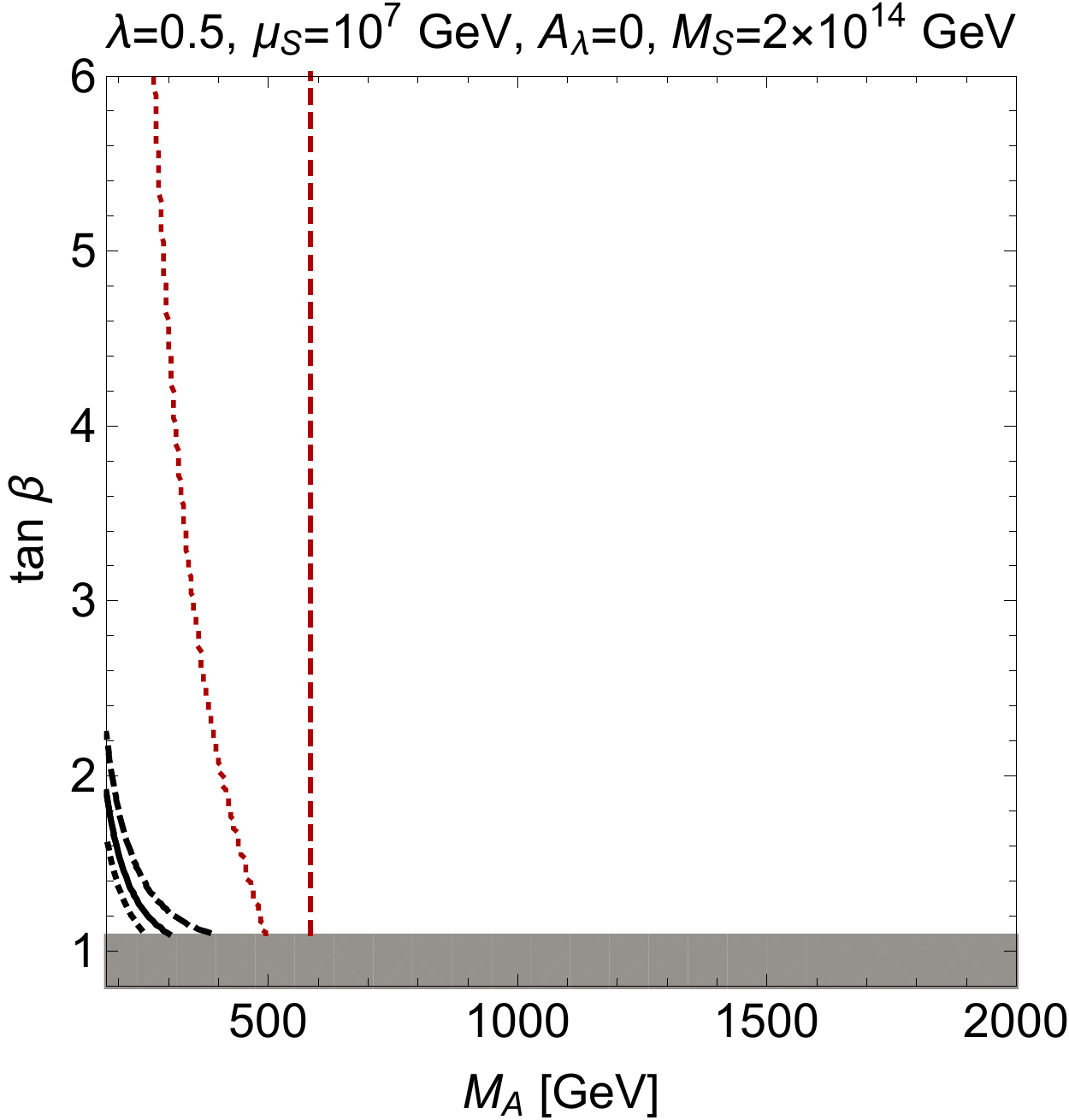}}
\subfigure{\includegraphics[width=0.24\textwidth]{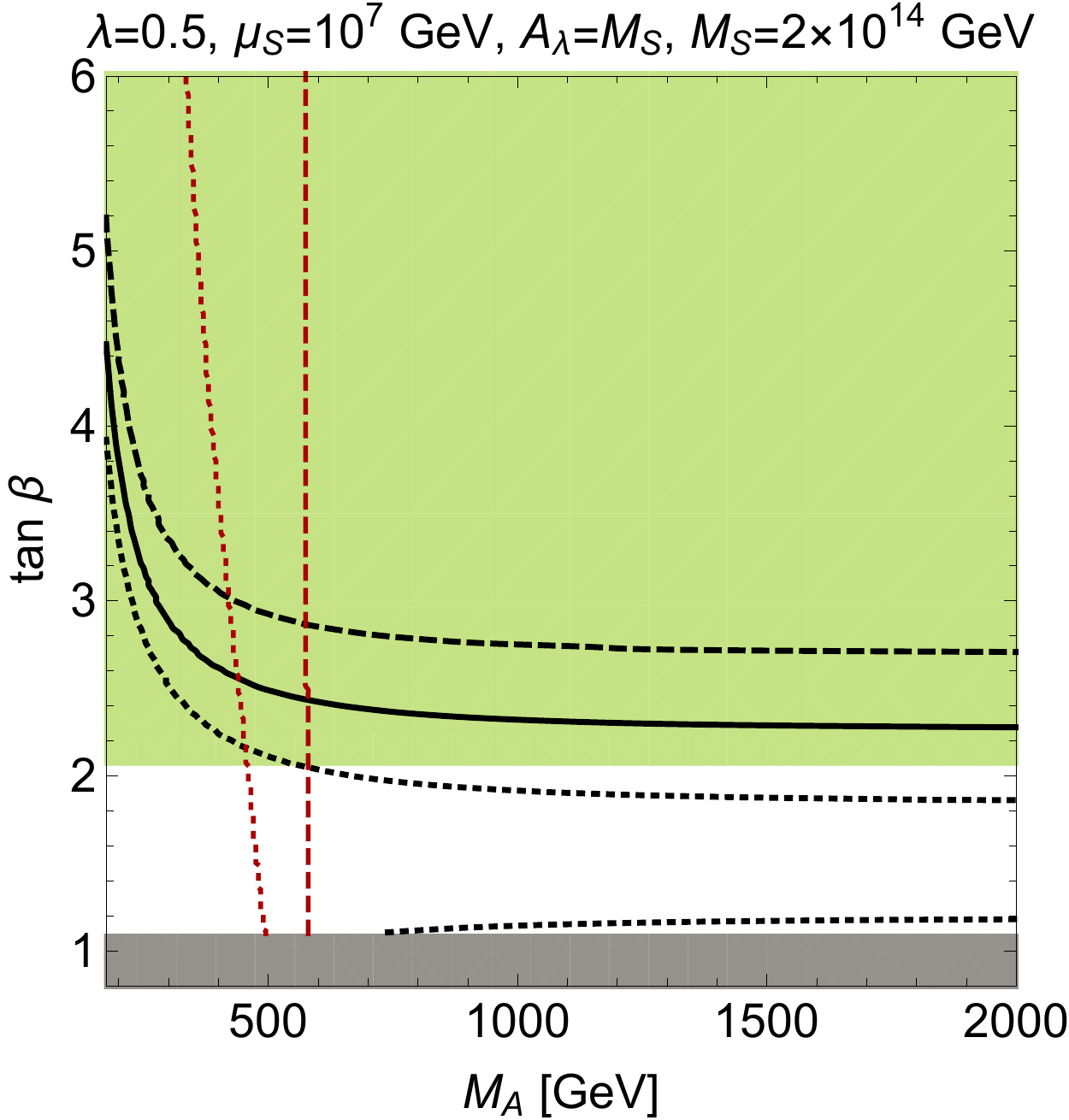}}\\
\caption{Same as Figs. \ref{fig:Higgs_mus5}, \ref{fig:Higgs_mus7} but for $M_t = 172.2$ GeV.}
\label{fig:app_mtm}
\end{figure}

\begin{figure}[t]
\centering
\subfigure{\includegraphics[width=0.24\textwidth]{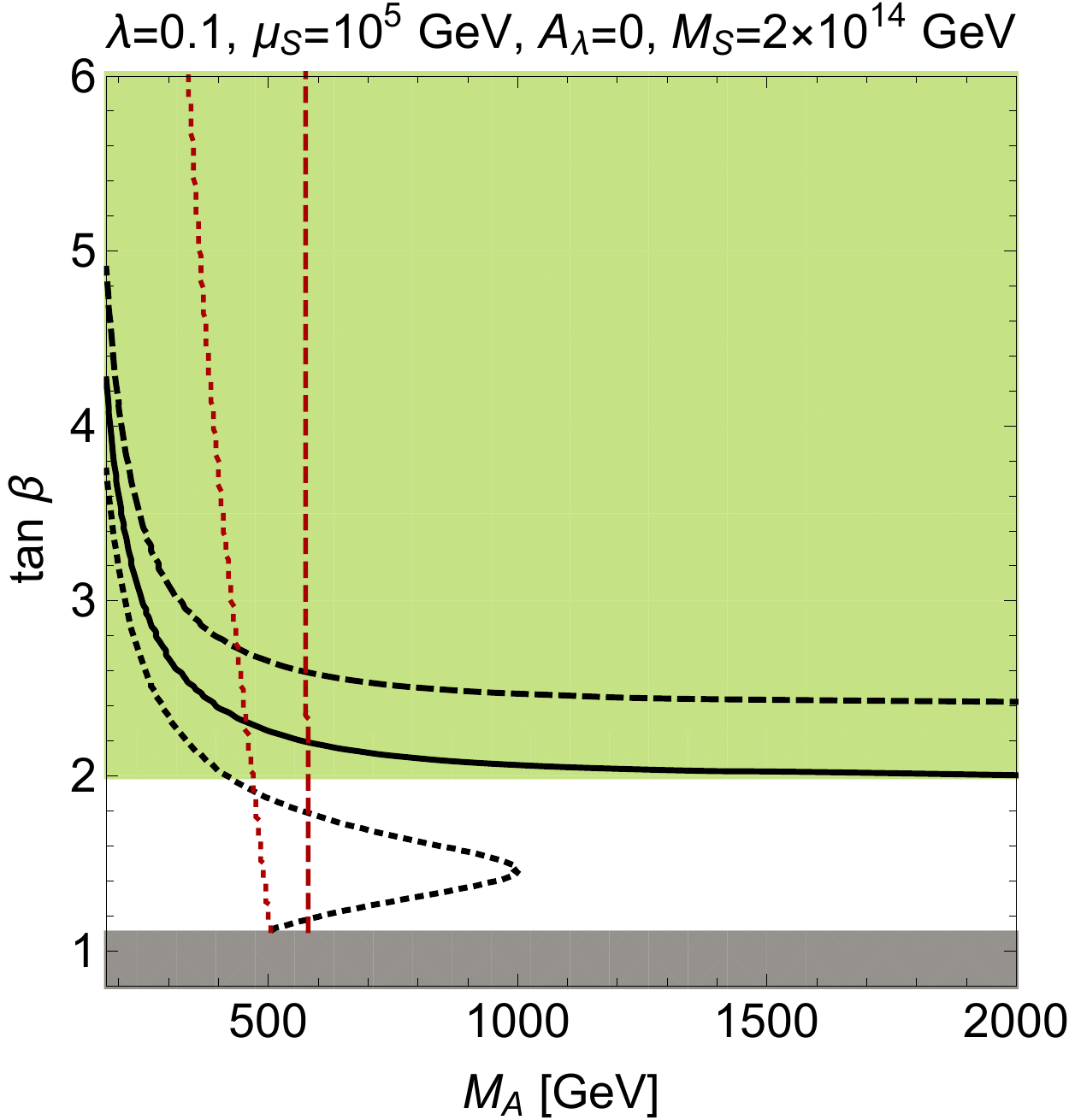}}
\subfigure{\includegraphics[width=0.24\textwidth]{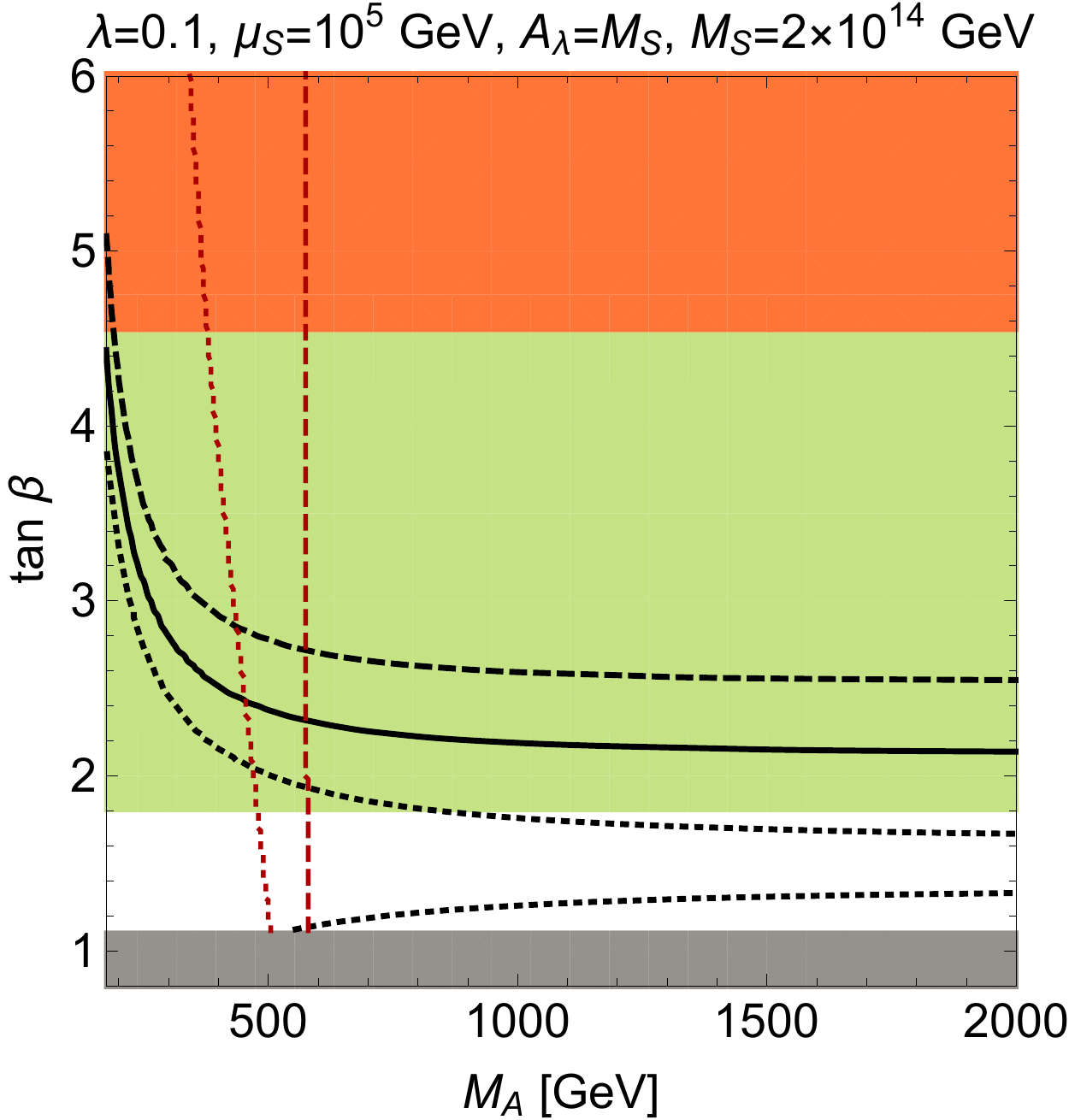}}
\subfigure{\includegraphics[width=0.24\textwidth]{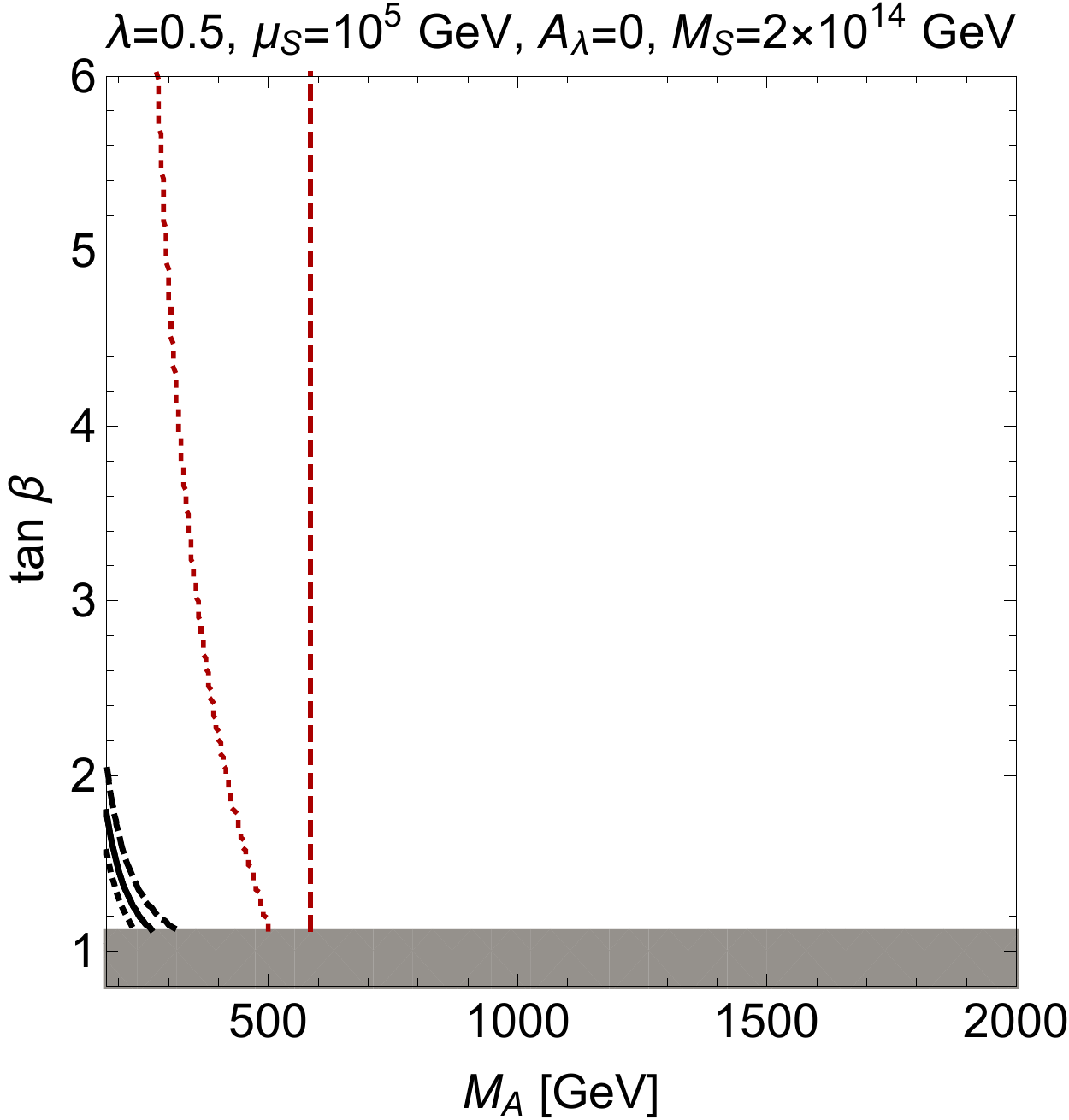}}
\subfigure{\includegraphics[width=0.24\textwidth]{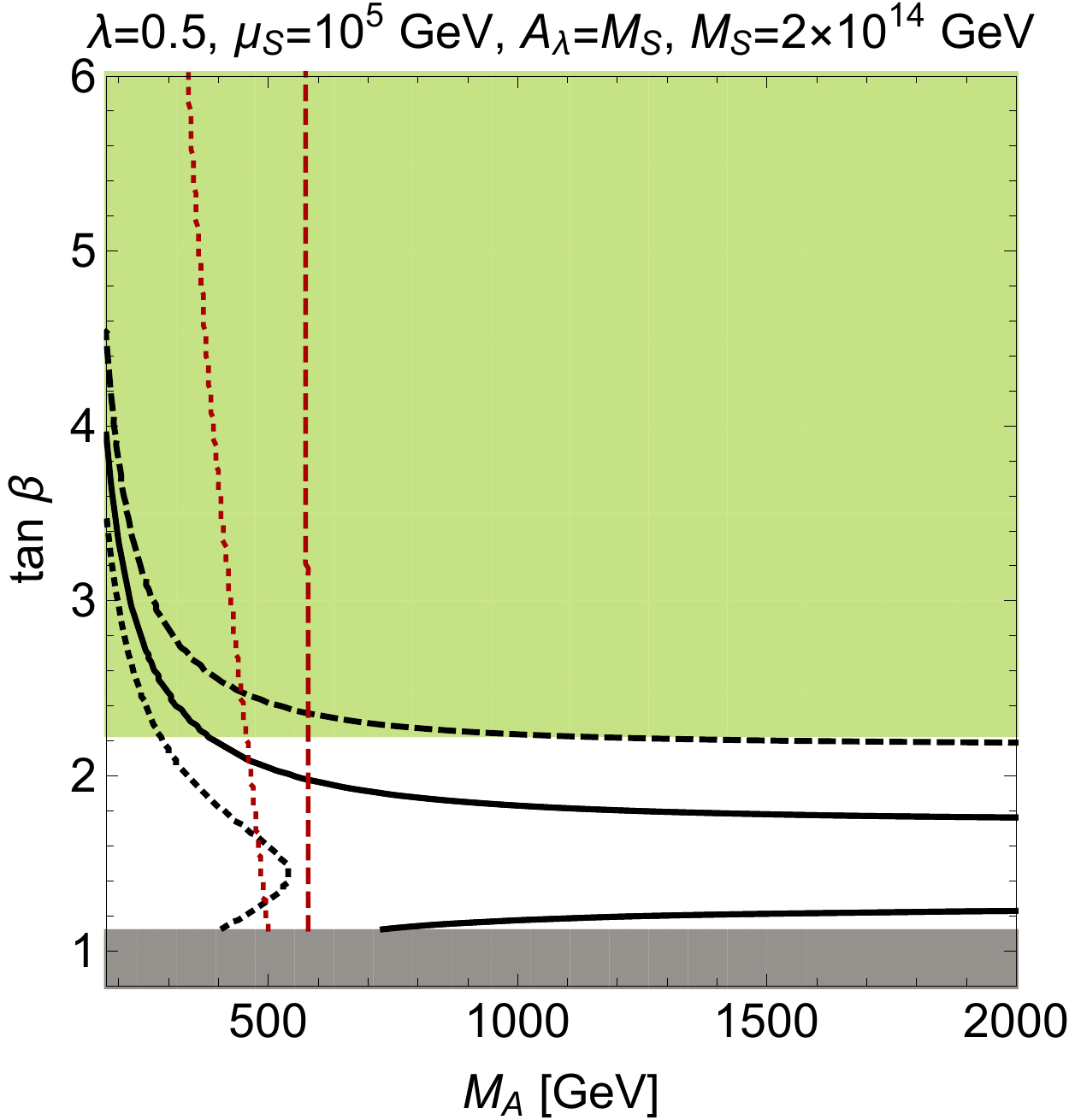}}\\
\subfigure{\includegraphics[width=0.24\textwidth]{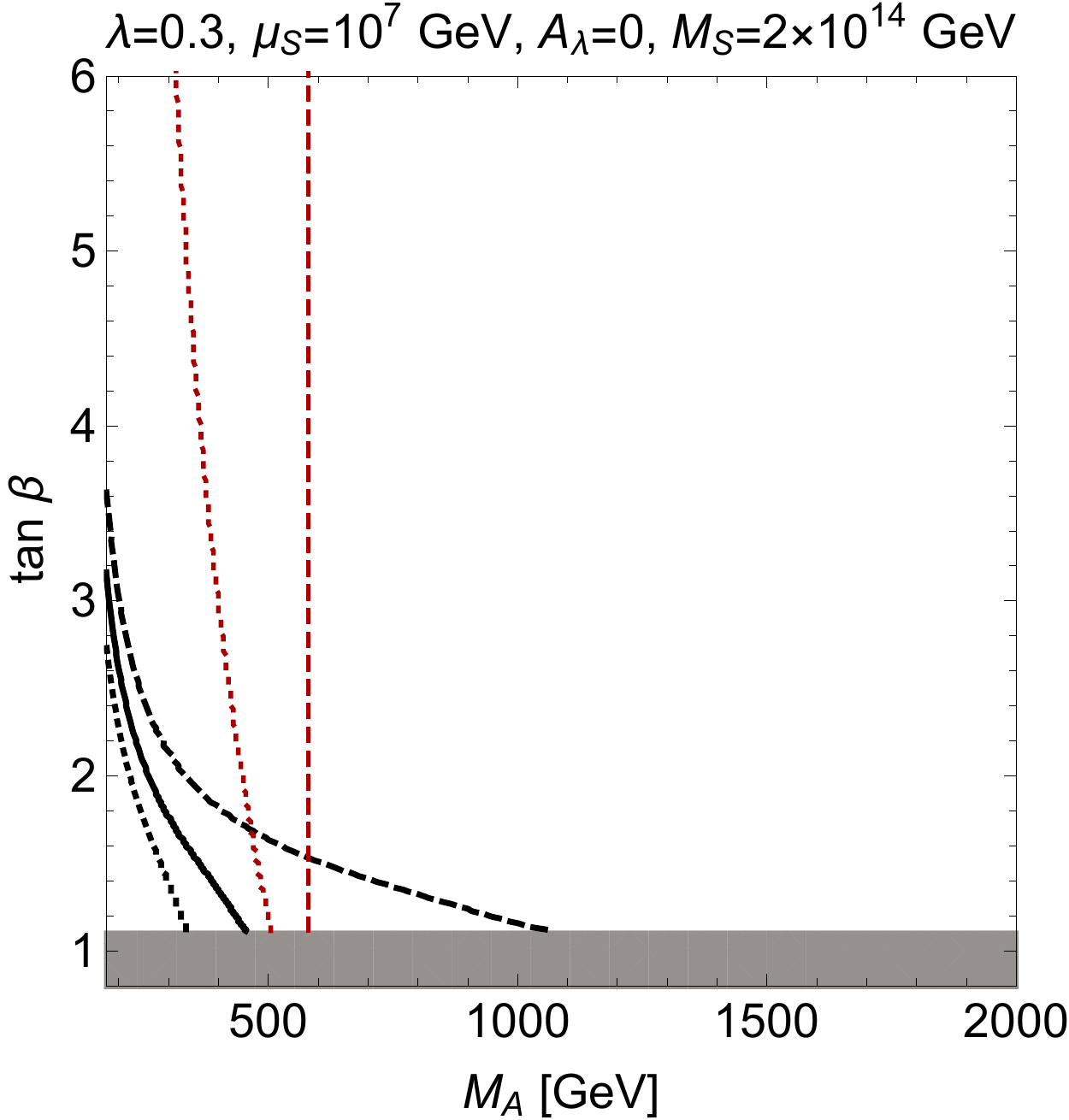}}
\subfigure{\includegraphics[width=0.24\textwidth]{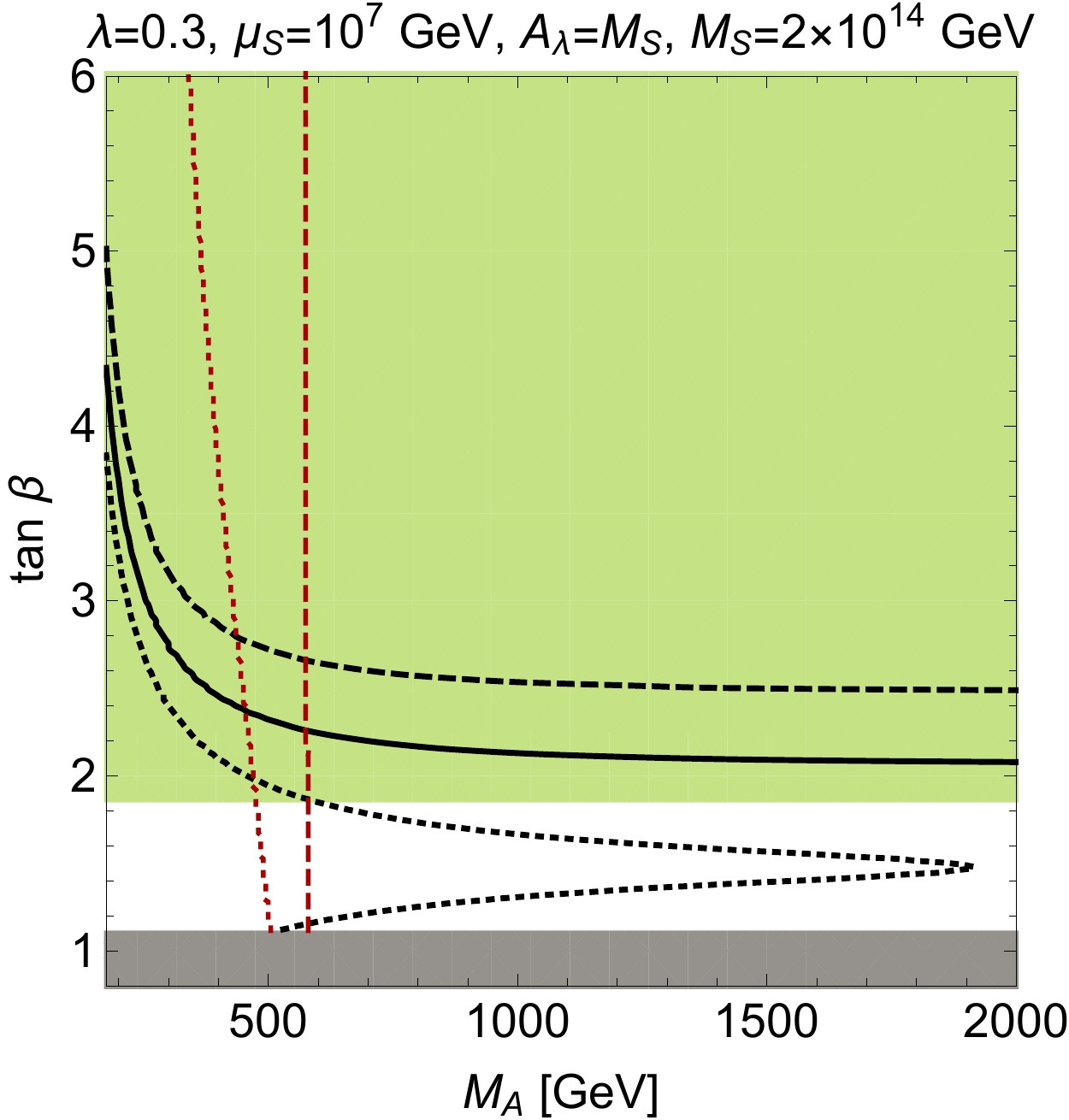}}
\subfigure{\includegraphics[width=0.24\textwidth]{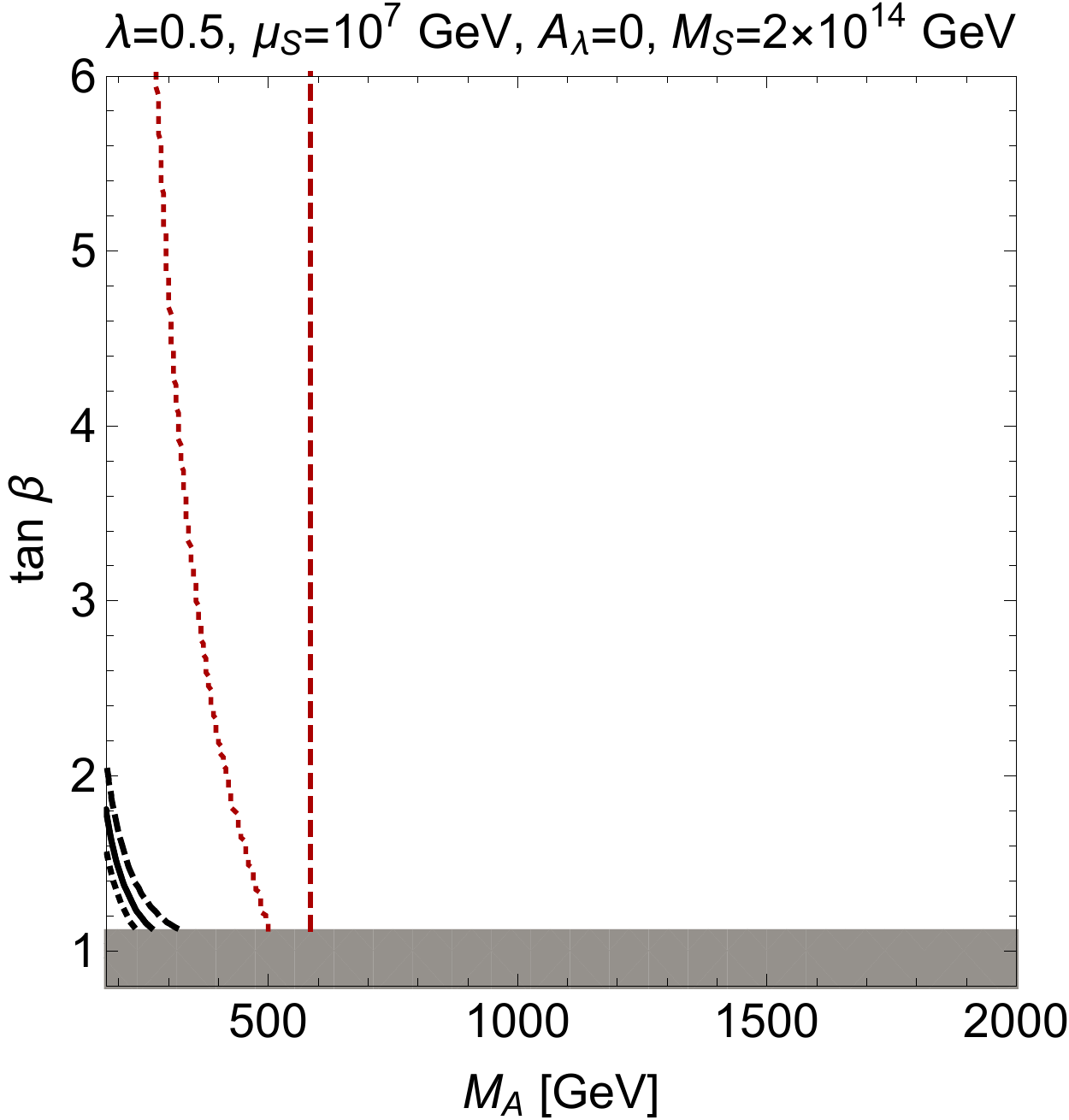}}
\subfigure{\includegraphics[width=0.24\textwidth]{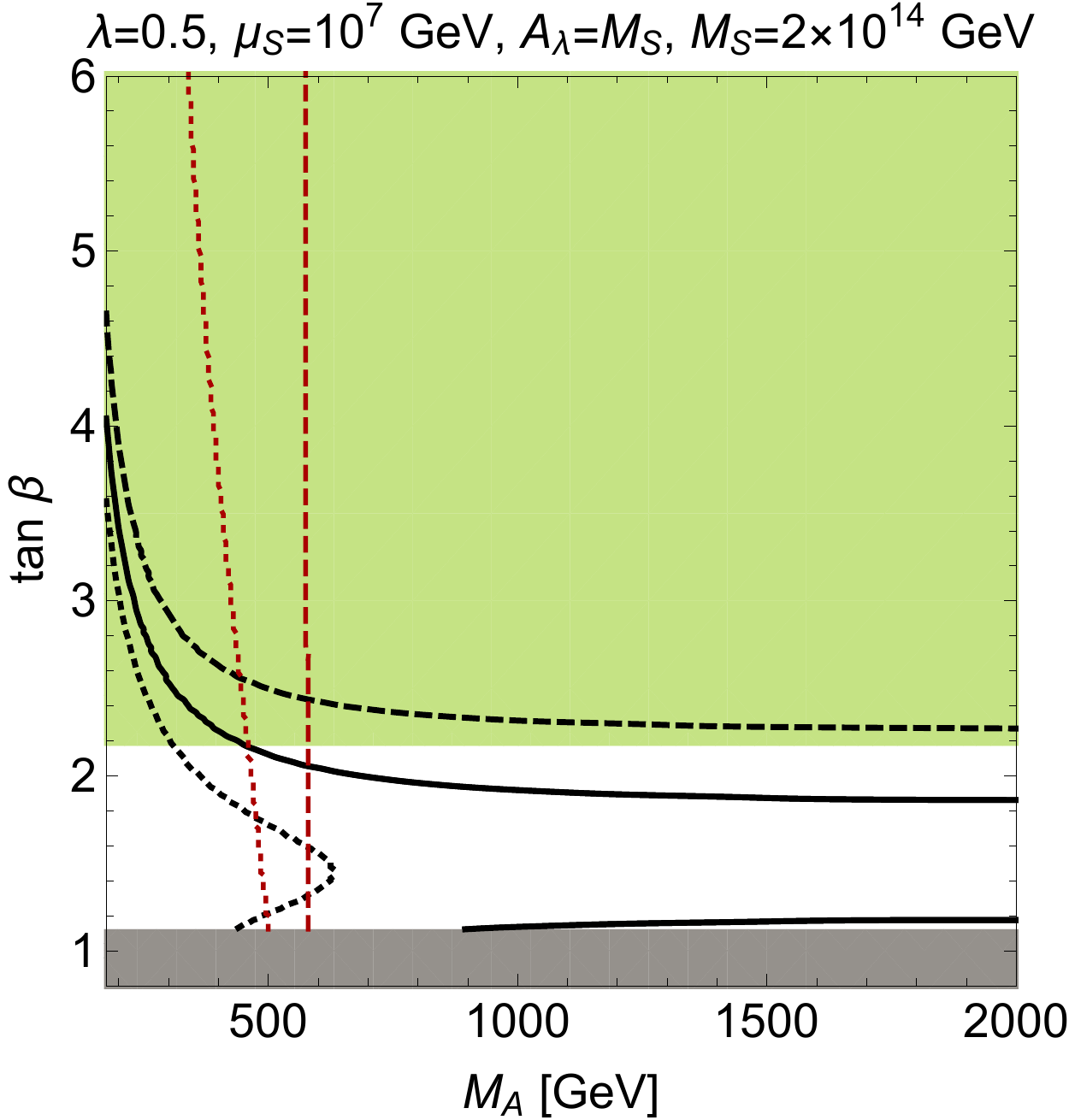}}\\
\caption{Same as  Figs. \ref{fig:Higgs_mus5}, \ref{fig:Higgs_mus7} but for $M_t = 174$ GeV.}
\label{fig:app_mtp}
\end{figure}

In this Appendix, we show the effects of uncertainty in the measurement of top pole mass on our  results. The results shown in Figs. \ref{fig:Higgs_mus5} and \ref{fig:Higgs_mus7}, obtained for $M_t=173.1$ GeV, are regenerated for $M_t=172.2$ GeV and $M_t=174$ GeV and are displayed as Figs. \ref{fig:app_mtm} and \ref{fig:app_mtp}, respectively.

\bibliography{references}

\begin{thebibliography}{68}%
\makeatletter
\providecommand \@ifxundefined [1]{%
 \@ifx{#1\undefined}
}%
\providecommand \@ifnum [1]{%
 \ifnum #1\expandafter \@firstoftwo
 \else \expandafter \@secondoftwo
 \fi
}%
\providecommand \@ifx [1]{%
 \ifx #1\expandafter \@firstoftwo
 \else \expandafter \@secondoftwo
 \fi
}%
\providecommand \natexlab [1]{#1}%
\providecommand \enquote  [1]{``#1''}%
\providecommand \bibnamefont  [1]{#1}%
\providecommand \bibfnamefont [1]{#1}%
\providecommand \citenamefont [1]{#1}%
\providecommand \href@noop [0]{\@secondoftwo}%
\providecommand \href [0]{\begingroup \@sanitize@url \@href}%
\providecommand \@href[1]{\@@startlink{#1}\@@href}%
\providecommand \@@href[1]{\endgroup#1\@@endlink}%
\providecommand \@sanitize@url [0]{\catcode `\\12\catcode `\$12\catcode
  `\&12\catcode `\#12\catcode `\^12\catcode `\_12\catcode `\%12\relax}%
\providecommand \@@startlink[1]{}%
\providecommand \@@endlink[0]{}%
\providecommand \url  [0]{\begingroup\@sanitize@url \@url }%
\providecommand \@url [1]{\endgroup\@href {#1}{\urlprefix }}%
\providecommand \urlprefix  [0]{URL }%
\providecommand \Eprint [0]{\href }%
\providecommand \doibase [0]{http://dx.doi.org/}%
\providecommand \selectlanguage [0]{\@gobble}%
\providecommand \bibinfo  [0]{\@secondoftwo}%
\providecommand \bibfield  [0]{\@secondoftwo}%
\providecommand \translation [1]{[#1]}%
\providecommand \BibitemOpen [0]{}%
\providecommand \bibitemStop [0]{}%
\providecommand \bibitemNoStop [0]{.\EOS\space}%
\providecommand \EOS [0]{\spacefactor3000\relax}%
\providecommand \BibitemShut  [1]{\csname bibitem#1\endcsname}%
\let\auto@bib@innerbib\@empty
\bibitem [{\citenamefont {Giudice}\ and\ \citenamefont
  {Romanino}(2004)}]{Giudice:2004tc}%
  \BibitemOpen
  \bibfield  {author} {\bibinfo {author} {\bibfnamefont {G.~F.}\ \bibnamefont
  {Giudice}}\ and\ \bibinfo {author} {\bibfnamefont {A.}~\bibnamefont
  {Romanino}},\ }\bibfield  {title} {\enquote {\bibinfo {title} {{Split
  supersymmetry}},}\ }\href {\doibase 10.1016/j.nuclphysb.2004.11.048,
  10.1016/j.nuclphysb.2004.08.001} {\bibfield  {journal} {\bibinfo  {journal}
  {Nucl. Phys.}\ }\textbf {\bibinfo {volume} {B699}},\ \bibinfo {pages}
  {65--89} (\bibinfo {year} {2004})},\ \bibinfo {note} {[Erratum: Nucl.
  Phys.B706,487(2005)]},\ \Eprint {http://arxiv.org/abs/hep-ph/0406088}
  {arXiv:hep-ph/0406088 [hep-ph]} \BibitemShut {NoStop}%
\bibitem [{\citenamefont {Arkani-Hamed}\ and\ \citenamefont
  {Dimopoulos}(2005)}]{ArkaniHamed:2004fb}%
  \BibitemOpen
  \bibfield  {author} {\bibinfo {author} {\bibfnamefont {Nima}\ \bibnamefont
  {Arkani-Hamed}}\ and\ \bibinfo {author} {\bibfnamefont {Savas}\ \bibnamefont
  {Dimopoulos}},\ }\bibfield  {title} {\enquote {\bibinfo {title}
  {{Supersymmetric unification without low energy supersymmetry and signatures
  for fine-tuning at the LHC}},}\ }\href {\doibase
  10.1088/1126-6708/2005/06/073} {\bibfield  {journal} {\bibinfo  {journal}
  {JHEP}\ }\textbf {\bibinfo {volume} {06}},\ \bibinfo {pages} {073} (\bibinfo
  {year} {2005})},\ \Eprint {http://arxiv.org/abs/hep-th/0405159}
  {arXiv:hep-th/0405159 [hep-th]} \BibitemShut {NoStop}%
\bibitem [{\citenamefont {Green}\ \emph {et~al.}(1988)\citenamefont {Green},
  \citenamefont {Schwarz},\ and\ \citenamefont {Witten}}]{Green:1987sp}%
  \BibitemOpen
  \bibfield  {author} {\bibinfo {author} {\bibfnamefont {Michael~B.}\
  \bibnamefont {Green}}, \bibinfo {author} {\bibfnamefont {J.~H.}\ \bibnamefont
  {Schwarz}}, \ and\ \bibinfo {author} {\bibfnamefont {Edward}\ \bibnamefont
  {Witten}},\ }\href
  {http://www.cambridge.org/us/academic/subjects/physics/theoretical-physics-and-mathematical-physics/superstring-theory-volume-1}
  {\emph {\bibinfo {title} {{SUPERSTRING THEORY. VOL. 1: INTRODUCTION}}}},\
  Cambridge Monographs on Mathematical Physics\ (\bibinfo {year}
  {1988})\BibitemShut {NoStop}%
\bibitem [{\citenamefont {Asaka}\ \emph
  {et~al.}(2003{\natexlab{a}})\citenamefont {Asaka}, \citenamefont
  {Buchmuller},\ and\ \citenamefont {Covi}}]{Asaka:2002my}%
  \BibitemOpen
  \bibfield  {author} {\bibinfo {author} {\bibfnamefont {T.}~\bibnamefont
  {Asaka}}, \bibinfo {author} {\bibfnamefont {W.}~\bibnamefont {Buchmuller}}, \
  and\ \bibinfo {author} {\bibfnamefont {L.}~\bibnamefont {Covi}},\ }\bibfield
  {title} {\enquote {\bibinfo {title} {{Bulk and brane anomalies in
  six-dimensions}},}\ }\href {\doibase 10.1016/S0550-3213(02)00976-8}
  {\bibfield  {journal} {\bibinfo  {journal} {Nucl. Phys.}\ }\textbf {\bibinfo
  {volume} {B648}},\ \bibinfo {pages} {231--253} (\bibinfo {year}
  {2003}{\natexlab{a}})},\ \Eprint {http://arxiv.org/abs/hep-ph/0209144}
  {arXiv:hep-ph/0209144 [hep-ph]} \BibitemShut {NoStop}%
\bibitem [{\citenamefont {Kim}\ and\ \citenamefont {Raby}(2003)}]{Kim:2002im}%
  \BibitemOpen
  \bibfield  {author} {\bibinfo {author} {\bibfnamefont {Hyung~Do}\
  \bibnamefont {Kim}}\ and\ \bibinfo {author} {\bibfnamefont {Stuart}\
  \bibnamefont {Raby}},\ }\bibfield  {title} {\enquote {\bibinfo {title}
  {{Unification in 5D SO(10)}},}\ }\href {\doibase
  10.1088/1126-6708/2003/01/056} {\bibfield  {journal} {\bibinfo  {journal}
  {JHEP}\ }\textbf {\bibinfo {volume} {01}},\ \bibinfo {pages} {056} (\bibinfo
  {year} {2003})},\ \Eprint {http://arxiv.org/abs/hep-ph/0212348}
  {arXiv:hep-ph/0212348 [hep-ph]} \BibitemShut {NoStop}%
\bibitem [{\citenamefont {Kitano}\ and\ \citenamefont
  {Li}(2003)}]{Kitano:2003cn}%
  \BibitemOpen
  \bibfield  {author} {\bibinfo {author} {\bibfnamefont {Ryuichiro}\
  \bibnamefont {Kitano}}\ and\ \bibinfo {author} {\bibfnamefont {Tian-jun}\
  \bibnamefont {Li}},\ }\bibfield  {title} {\enquote {\bibinfo {title} {{Flavor
  hierarchy in SO(10) grand unified theories via five-dimensional wave function
  localization}},}\ }\href {\doibase 10.1103/PhysRevD.67.116004} {\bibfield
  {journal} {\bibinfo  {journal} {Phys. Rev.}\ }\textbf {\bibinfo {volume}
  {D67}},\ \bibinfo {pages} {116004} (\bibinfo {year} {2003})},\ \Eprint
  {http://arxiv.org/abs/hep-ph/0302073} {arXiv:hep-ph/0302073 [hep-ph]}
  \BibitemShut {NoStop}%
\bibitem [{\citenamefont {Asaka}\ \emph
  {et~al.}(2003{\natexlab{b}})\citenamefont {Asaka}, \citenamefont
  {Buchmuller},\ and\ \citenamefont {Covi}}]{Asaka:2003iy}%
  \BibitemOpen
  \bibfield  {author} {\bibinfo {author} {\bibfnamefont {T.}~\bibnamefont
  {Asaka}}, \bibinfo {author} {\bibfnamefont {W.}~\bibnamefont {Buchmuller}}, \
  and\ \bibinfo {author} {\bibfnamefont {L.}~\bibnamefont {Covi}},\ }\bibfield
  {title} {\enquote {\bibinfo {title} {{Quarks and leptons between branes and
  bulk}},}\ }\href {\doibase 10.1016/S0370-2693(03)00644-0} {\bibfield
  {journal} {\bibinfo  {journal} {Phys. Lett.}\ }\textbf {\bibinfo {volume}
  {B563}},\ \bibinfo {pages} {209--216} (\bibinfo {year}
  {2003}{\natexlab{b}})},\ \Eprint {http://arxiv.org/abs/hep-ph/0304142}
  {arXiv:hep-ph/0304142 [hep-ph]} \BibitemShut {NoStop}%
\bibitem [{\citenamefont {Kobayashi}\ \emph {et~al.}(2004)\citenamefont
  {Kobayashi}, \citenamefont {Raby},\ and\ \citenamefont
  {Zhang}}]{Kobayashi:2004ud}%
  \BibitemOpen
  \bibfield  {author} {\bibinfo {author} {\bibfnamefont {Tatsuo}\ \bibnamefont
  {Kobayashi}}, \bibinfo {author} {\bibfnamefont {Stuart}\ \bibnamefont
  {Raby}}, \ and\ \bibinfo {author} {\bibfnamefont {Ren-Jie}\ \bibnamefont
  {Zhang}},\ }\bibfield  {title} {\enquote {\bibinfo {title} {{Constructing 5-D
  orbifold grand unified theories from heterotic strings}},}\ }\href {\doibase
  10.1016/j.physletb.2004.04.058} {\bibfield  {journal} {\bibinfo  {journal}
  {Phys. Lett.}\ }\textbf {\bibinfo {volume} {B593}},\ \bibinfo {pages}
  {262--270} (\bibinfo {year} {2004})},\ \Eprint
  {http://arxiv.org/abs/hep-ph/0403065} {arXiv:hep-ph/0403065 [hep-ph]}
  \BibitemShut {NoStop}%
\bibitem [{\citenamefont {Feruglio}\ \emph {et~al.}(2014)\citenamefont
  {Feruglio}, \citenamefont {Patel},\ and\ \citenamefont
  {Vicino}}]{Feruglio:2014jla}%
  \BibitemOpen
  \bibfield  {author} {\bibinfo {author} {\bibfnamefont {Ferruccio}\
  \bibnamefont {Feruglio}}, \bibinfo {author} {\bibfnamefont {Ketan~M.}\
  \bibnamefont {Patel}}, \ and\ \bibinfo {author} {\bibfnamefont {Denise}\
  \bibnamefont {Vicino}},\ }\bibfield  {title} {\enquote {\bibinfo {title}
  {{Order and Anarchy hand in hand in 5D SO(10)}},}\ }\href {\doibase
  10.1007/JHEP09(2014)095} {\bibfield  {journal} {\bibinfo  {journal} {JHEP}\
  }\textbf {\bibinfo {volume} {09}},\ \bibinfo {pages} {095} (\bibinfo {year}
  {2014})},\ \Eprint {http://arxiv.org/abs/1407.2913} {arXiv:1407.2913
  [hep-ph]} \BibitemShut {NoStop}%
\bibitem [{\citenamefont {Feruglio}\ \emph {et~al.}(2015)\citenamefont
  {Feruglio}, \citenamefont {Patel},\ and\ \citenamefont
  {Vicino}}]{Feruglio:2015iua}%
  \BibitemOpen
  \bibfield  {author} {\bibinfo {author} {\bibfnamefont {Ferruccio}\
  \bibnamefont {Feruglio}}, \bibinfo {author} {\bibfnamefont {Ketan~M.}\
  \bibnamefont {Patel}}, \ and\ \bibinfo {author} {\bibfnamefont {Denise}\
  \bibnamefont {Vicino}},\ }\bibfield  {title} {\enquote {\bibinfo {title} {{A
  realistic pattern of fermion masses from a five-dimensional SO(10) model}},}\
  }\href {\doibase 10.1007/JHEP09(2015)040} {\bibfield  {journal} {\bibinfo
  {journal} {JHEP}\ }\textbf {\bibinfo {volume} {09}},\ \bibinfo {pages} {040}
  (\bibinfo {year} {2015})},\ \Eprint {http://arxiv.org/abs/1507.00669}
  {arXiv:1507.00669 [hep-ph]} \BibitemShut {NoStop}%
\bibitem [{\citenamefont {Buchmuller}\ \emph {et~al.}(2015)\citenamefont
  {Buchmuller}, \citenamefont {Dierigl}, \citenamefont {Ruehle},\ and\
  \citenamefont {Schweizer}}]{Buchmuller:2015jna}%
  \BibitemOpen
  \bibfield  {author} {\bibinfo {author} {\bibfnamefont {Wilfried}\
  \bibnamefont {Buchmuller}}, \bibinfo {author} {\bibfnamefont {Markus}\
  \bibnamefont {Dierigl}}, \bibinfo {author} {\bibfnamefont {Fabian}\
  \bibnamefont {Ruehle}}, \ and\ \bibinfo {author} {\bibfnamefont {Julian}\
  \bibnamefont {Schweizer}},\ }\bibfield  {title} {\enquote {\bibinfo {title}
  {{Split symmetries}},}\ }\href {\doibase 10.1016/j.physletb.2015.09.069}
  {\bibfield  {journal} {\bibinfo  {journal} {Phys. Lett.}\ }\textbf {\bibinfo
  {volume} {B750}},\ \bibinfo {pages} {615--619} (\bibinfo {year} {2015})},\
  \Eprint {http://arxiv.org/abs/1507.06819} {arXiv:1507.06819 [hep-th]}
  \BibitemShut {NoStop}%
\bibitem [{\citenamefont {Buchmuller}\ and\ \citenamefont
  {Schweizer}(2017)}]{Buchmuller:2017vho}%
  \BibitemOpen
  \bibfield  {author} {\bibinfo {author} {\bibfnamefont {Wilfried}\
  \bibnamefont {Buchmuller}}\ and\ \bibinfo {author} {\bibfnamefont {Julian}\
  \bibnamefont {Schweizer}},\ }\bibfield  {title} {\enquote {\bibinfo {title}
  {{Flavor mixings in flux compactifications}},}\ }\href {\doibase
  10.1103/PhysRevD.95.075024} {\bibfield  {journal} {\bibinfo  {journal} {Phys.
  Rev.}\ }\textbf {\bibinfo {volume} {D95}},\ \bibinfo {pages} {075024}
  (\bibinfo {year} {2017})},\ \Eprint {http://arxiv.org/abs/1701.06935}
  {arXiv:1701.06935 [hep-ph]} \BibitemShut {NoStop}%
\bibitem [{\citenamefont {Buchmuller}\ and\ \citenamefont
  {Patel}(2018)}]{Buchmuller:2017vut}%
  \BibitemOpen
  \bibfield  {author} {\bibinfo {author} {\bibfnamefont {Wilfried}\
  \bibnamefont {Buchmuller}}\ and\ \bibinfo {author} {\bibfnamefont {Ketan~M.}\
  \bibnamefont {Patel}},\ }\bibfield  {title} {\enquote {\bibinfo {title}
  {{Flavour physics without flavour symmetries}},}\ }\href {\doibase
  10.1103/PhysRevD.97.075019} {\bibfield  {journal} {\bibinfo  {journal} {Phys.
  Rev.}\ }\textbf {\bibinfo {volume} {D97}},\ \bibinfo {pages} {075019}
  (\bibinfo {year} {2018})},\ \Eprint {http://arxiv.org/abs/1712.06862}
  {arXiv:1712.06862 [hep-ph]} \BibitemShut {NoStop}%
\bibitem [{\citenamefont {Giudice}\ and\ \citenamefont
  {Strumia}(2012)}]{Giudice:2011cg}%
  \BibitemOpen
  \bibfield  {author} {\bibinfo {author} {\bibfnamefont {Gian~F.}\ \bibnamefont
  {Giudice}}\ and\ \bibinfo {author} {\bibfnamefont {Alessandro}\ \bibnamefont
  {Strumia}},\ }\bibfield  {title} {\enquote {\bibinfo {title} {{Probing
  High-Scale and Split Supersymmetry with Higgs Mass Measurements}},}\ }\href
  {\doibase 10.1016/j.nuclphysb.2012.01.001} {\bibfield  {journal} {\bibinfo
  {journal} {Nucl. Phys.}\ }\textbf {\bibinfo {volume} {B858}},\ \bibinfo
  {pages} {63--83} (\bibinfo {year} {2012})},\ \Eprint
  {http://arxiv.org/abs/1108.6077} {arXiv:1108.6077 [hep-ph]} \BibitemShut
  {NoStop}%
\bibitem [{\citenamefont {Elias-Miro}\ \emph {et~al.}(2012)\citenamefont
  {Elias-Miro}, \citenamefont {Espinosa}, \citenamefont {Giudice},
  \citenamefont {Isidori}, \citenamefont {Riotto},\ and\ \citenamefont
  {Strumia}}]{EliasMiro:2011aa}%
  \BibitemOpen
  \bibfield  {author} {\bibinfo {author} {\bibfnamefont {Joan}\ \bibnamefont
  {Elias-Miro}}, \bibinfo {author} {\bibfnamefont {Jose~R.}\ \bibnamefont
  {Espinosa}}, \bibinfo {author} {\bibfnamefont {Gian~F.}\ \bibnamefont
  {Giudice}}, \bibinfo {author} {\bibfnamefont {Gino}\ \bibnamefont {Isidori}},
  \bibinfo {author} {\bibfnamefont {Antonio}\ \bibnamefont {Riotto}}, \ and\
  \bibinfo {author} {\bibfnamefont {Alessandro}\ \bibnamefont {Strumia}},\
  }\bibfield  {title} {\enquote {\bibinfo {title} {{Higgs mass implications on
  the stability of the electroweak vacuum}},}\ }\href {\doibase
  10.1016/j.physletb.2012.02.013} {\bibfield  {journal} {\bibinfo  {journal}
  {Phys. Lett.}\ }\textbf {\bibinfo {volume} {B709}},\ \bibinfo {pages}
  {222--228} (\bibinfo {year} {2012})},\ \Eprint
  {http://arxiv.org/abs/1112.3022} {arXiv:1112.3022 [hep-ph]} \BibitemShut
  {NoStop}%
\bibitem [{\citenamefont {Draper}\ \emph {et~al.}(2014)\citenamefont {Draper},
  \citenamefont {Lee},\ and\ \citenamefont {Wagner}}]{Draper:2013oza}%
  \BibitemOpen
  \bibfield  {author} {\bibinfo {author} {\bibfnamefont {Patrick}\ \bibnamefont
  {Draper}}, \bibinfo {author} {\bibfnamefont {Gabriel}\ \bibnamefont {Lee}}, \
  and\ \bibinfo {author} {\bibfnamefont {Carlos E.~M.}\ \bibnamefont
  {Wagner}},\ }\bibfield  {title} {\enquote {\bibinfo {title} {{Precise
  estimates of the Higgs mass in heavy supersymmetry}},}\ }\href {\doibase
  10.1103/PhysRevD.89.055023} {\bibfield  {journal} {\bibinfo  {journal} {Phys.
  Rev.}\ }\textbf {\bibinfo {volume} {D89}},\ \bibinfo {pages} {055023}
  (\bibinfo {year} {2014})},\ \Eprint {http://arxiv.org/abs/1312.5743}
  {arXiv:1312.5743 [hep-ph]} \BibitemShut {NoStop}%
\bibitem [{\citenamefont {Ellis}\ and\ \citenamefont
  {Wells}(2017)}]{Ellis:2017erg}%
  \BibitemOpen
  \bibfield  {author} {\bibinfo {author} {\bibfnamefont {Sebastian A.~R.}\
  \bibnamefont {Ellis}}\ and\ \bibinfo {author} {\bibfnamefont {James~D.}\
  \bibnamefont {Wells}},\ }\bibfield  {title} {\enquote {\bibinfo {title}
  {{High-scale supersymmetry, the Higgs boson mass, and gauge unification}},}\
  }\href {\doibase 10.1103/PhysRevD.96.055024} {\bibfield  {journal} {\bibinfo
  {journal} {Phys. Rev.}\ }\textbf {\bibinfo {volume} {D96}},\ \bibinfo {pages}
  {055024} (\bibinfo {year} {2017})},\ \Eprint
  {http://arxiv.org/abs/1706.00013} {arXiv:1706.00013 [hep-ph]} \BibitemShut
  {NoStop}%
\bibitem [{\citenamefont {Bagnaschi}\ \emph {et~al.}(2016)\citenamefont
  {Bagnaschi}, \citenamefont {Brummer}, \citenamefont {Buchmuller},
  \citenamefont {Voigt},\ and\ \citenamefont {Weiglein}}]{Bagnaschi:2015pwa}%
  \BibitemOpen
  \bibfield  {author} {\bibinfo {author} {\bibfnamefont {Emanuele}\
  \bibnamefont {Bagnaschi}}, \bibinfo {author} {\bibfnamefont {Felix}\
  \bibnamefont {Brummer}}, \bibinfo {author} {\bibfnamefont {Wilfried}\
  \bibnamefont {Buchmuller}}, \bibinfo {author} {\bibfnamefont {Alexander}\
  \bibnamefont {Voigt}}, \ and\ \bibinfo {author} {\bibfnamefont {Georg}\
  \bibnamefont {Weiglein}},\ }\bibfield  {title} {\enquote {\bibinfo {title}
  {{Vacuum stability and supersymmetry at high scales with two Higgs
  doublets}},}\ }\href {\doibase 10.1007/JHEP03(2016)158} {\bibfield  {journal}
  {\bibinfo  {journal} {JHEP}\ }\textbf {\bibinfo {volume} {03}},\ \bibinfo
  {pages} {158} (\bibinfo {year} {2016})},\ \Eprint
  {http://arxiv.org/abs/1512.07761} {arXiv:1512.07761 [hep-ph]} \BibitemShut
  {NoStop}%
\bibitem [{\citenamefont {Mummidi}\ \emph {et~al.}(2018)\citenamefont
  {Mummidi}, \citenamefont {K.},\ and\ \citenamefont
  {Patel}}]{Mummidi:2018nph}%
  \BibitemOpen
  \bibfield  {author} {\bibinfo {author} {\bibfnamefont {V.~Suryanarayana}\
  \bibnamefont {Mummidi}}, \bibinfo {author} {\bibfnamefont {Vishnu~P.}\
  \bibnamefont {K.}}, \ and\ \bibinfo {author} {\bibfnamefont {Ketan~M.}\
  \bibnamefont {Patel}},\ }\bibfield  {title} {\enquote {\bibinfo {title}
  {{Effects of heavy neutrinos on vacuum stability in two-Higgs-doublet model
  with GUT scale supersymmetry}},}\ }\href {\doibase 10.1007/JHEP08(2018)134}
  {\bibfield  {journal} {\bibinfo  {journal} {JHEP}\ }\textbf {\bibinfo
  {volume} {08}},\ \bibinfo {pages} {134} (\bibinfo {year} {2018})},\ \Eprint
  {http://arxiv.org/abs/1805.08005} {arXiv:1805.08005 [hep-ph]} \BibitemShut
  {NoStop}%
\bibitem [{\citenamefont {Lee}\ and\ \citenamefont
  {Wagner}(2015)}]{Lee:2015uza}%
  \BibitemOpen
  \bibfield  {author} {\bibinfo {author} {\bibfnamefont {Gabriel}\ \bibnamefont
  {Lee}}\ and\ \bibinfo {author} {\bibfnamefont {Carlos E.~M.}\ \bibnamefont
  {Wagner}},\ }\bibfield  {title} {\enquote {\bibinfo {title} {{Higgs bosons in
  heavy supersymmetry with an intermediate m$_A$}},}\ }\href {\doibase
  10.1103/PhysRevD.92.075032} {\bibfield  {journal} {\bibinfo  {journal} {Phys.
  Rev.}\ }\textbf {\bibinfo {volume} {D92}},\ \bibinfo {pages} {075032}
  (\bibinfo {year} {2015})},\ \Eprint {http://arxiv.org/abs/1508.00576}
  {arXiv:1508.00576 [hep-ph]} \BibitemShut {NoStop}%
\bibitem [{\citenamefont {Servant}\ and\ \citenamefont
  {Tait}(2002)}]{Servant:2002hb}%
  \BibitemOpen
  \bibfield  {author} {\bibinfo {author} {\bibfnamefont {Geraldine}\
  \bibnamefont {Servant}}\ and\ \bibinfo {author} {\bibfnamefont {Timothy
  M.~P.}\ \bibnamefont {Tait}},\ }\bibfield  {title} {\enquote {\bibinfo
  {title} {{Elastic Scattering and Direct Detection of Kaluza-Klein Dark
  Matter}},}\ }\href {\doibase 10.1088/1367-2630/4/1/399} {\bibfield  {journal}
  {\bibinfo  {journal} {New J. Phys.}\ }\textbf {\bibinfo {volume} {4}},\
  \bibinfo {pages} {99} (\bibinfo {year} {2002})},\ \Eprint
  {http://arxiv.org/abs/hep-ph/0209262} {arXiv:hep-ph/0209262 [hep-ph]}
  \BibitemShut {NoStop}%
\bibitem [{\citenamefont {Nagata}\ and\ \citenamefont
  {Shirai}(2015)}]{Nagata:2014wma}%
  \BibitemOpen
  \bibfield  {author} {\bibinfo {author} {\bibfnamefont {Natsumi}\ \bibnamefont
  {Nagata}}\ and\ \bibinfo {author} {\bibfnamefont {Satoshi}\ \bibnamefont
  {Shirai}},\ }\bibfield  {title} {\enquote {\bibinfo {title} {{Higgsino Dark
  Matter in High-Scale Supersymmetry}},}\ }\href {\doibase
  10.1007/JHEP01(2015)029} {\bibfield  {journal} {\bibinfo  {journal} {JHEP}\
  }\textbf {\bibinfo {volume} {01}},\ \bibinfo {pages} {029} (\bibinfo {year}
  {2015})},\ \Eprint {http://arxiv.org/abs/1410.4549} {arXiv:1410.4549
  [hep-ph]} \BibitemShut {NoStop}%
\bibitem [{\citenamefont {Martin}(1997)}]{Martin:1997ns}%
  \BibitemOpen
  \bibfield  {author} {\bibinfo {author} {\bibfnamefont {Stephen~P.}\
  \bibnamefont {Martin}},\ }\bibfield  {title} {\enquote {\bibinfo {title} {{A
  Supersymmetry primer}},}\ }\href {\doibase 10.1142/9789812839657_0001,
  10.1142/9789814307505_0001} {\ ,\ \bibinfo {pages} {1--98} (\bibinfo {year}
  {1997})},\ \bibinfo {note} {[Adv. Ser. Direct. High Energy
  Phys.18,1(1998)]},\ \Eprint {http://arxiv.org/abs/hep-ph/9709356}
  {arXiv:hep-ph/9709356 [hep-ph]} \BibitemShut {NoStop}%
\bibitem [{\citenamefont {Ellwanger}\ \emph {et~al.}(2010)\citenamefont
  {Ellwanger}, \citenamefont {Hugonie},\ and\ \citenamefont
  {Teixeira}}]{Ellwanger:2009dp}%
  \BibitemOpen
  \bibfield  {author} {\bibinfo {author} {\bibfnamefont {Ulrich}\ \bibnamefont
  {Ellwanger}}, \bibinfo {author} {\bibfnamefont {Cyril}\ \bibnamefont
  {Hugonie}}, \ and\ \bibinfo {author} {\bibfnamefont {Ana~M.}\ \bibnamefont
  {Teixeira}},\ }\bibfield  {title} {\enquote {\bibinfo {title} {{The
  Next-to-Minimal Supersymmetric Standard Model}},}\ }\href {\doibase
  10.1016/j.physrep.2010.07.001} {\bibfield  {journal} {\bibinfo  {journal}
  {Phys. Rept.}\ }\textbf {\bibinfo {volume} {496}},\ \bibinfo {pages} {1--77}
  (\bibinfo {year} {2010})},\ \Eprint {http://arxiv.org/abs/0910.1785}
  {arXiv:0910.1785 [hep-ph]} \BibitemShut {NoStop}%
\bibitem [{\citenamefont {Bagger}\ and\ \citenamefont
  {Poppitz}(1993)}]{Bagger:1993ji}%
  \BibitemOpen
  \bibfield  {author} {\bibinfo {author} {\bibfnamefont {Jonathan}\
  \bibnamefont {Bagger}}\ and\ \bibinfo {author} {\bibfnamefont {Erich}\
  \bibnamefont {Poppitz}},\ }\bibfield  {title} {\enquote {\bibinfo {title}
  {{Destabilizing divergences in supergravity coupled supersymmetric
  theories}},}\ }\href {\doibase 10.1103/PhysRevLett.71.2380} {\bibfield
  {journal} {\bibinfo  {journal} {Phys. Rev. Lett.}\ }\textbf {\bibinfo
  {volume} {71}},\ \bibinfo {pages} {2380--2382} (\bibinfo {year} {1993})},\
  \Eprint {http://arxiv.org/abs/hep-ph/9307317} {arXiv:hep-ph/9307317 [hep-ph]}
  \BibitemShut {NoStop}%
\bibitem [{\citenamefont {Kim}\ and\ \citenamefont
  {Nilles}(1984)}]{Kim:1983dt}%
  \BibitemOpen
  \bibfield  {author} {\bibinfo {author} {\bibfnamefont {Jihn~E.}\ \bibnamefont
  {Kim}}\ and\ \bibinfo {author} {\bibfnamefont {Hans~Peter}\ \bibnamefont
  {Nilles}},\ }\bibfield  {title} {\enquote {\bibinfo {title} {{The mu Problem
  and the Strong CP Problem}},}\ }\href {\doibase 10.1016/0370-2693(84)91890-2}
  {\bibfield  {journal} {\bibinfo  {journal} {Phys. Lett.}\ }\textbf {\bibinfo
  {volume} {138B}},\ \bibinfo {pages} {150--154} (\bibinfo {year}
  {1984})}\BibitemShut {NoStop}%
\bibitem [{\citenamefont {Branco}\ \emph {et~al.}(2012)\citenamefont {Branco},
  \citenamefont {Ferreira}, \citenamefont {Lavoura}, \citenamefont {Rebelo},
  \citenamefont {Sher},\ and\ \citenamefont {Silva}}]{Branco:2011iw}%
  \BibitemOpen
  \bibfield  {author} {\bibinfo {author} {\bibfnamefont {G.~C.}\ \bibnamefont
  {Branco}}, \bibinfo {author} {\bibfnamefont {P.~M.}\ \bibnamefont
  {Ferreira}}, \bibinfo {author} {\bibfnamefont {L.}~\bibnamefont {Lavoura}},
  \bibinfo {author} {\bibfnamefont {M.~N.}\ \bibnamefont {Rebelo}}, \bibinfo
  {author} {\bibfnamefont {Marc}\ \bibnamefont {Sher}}, \ and\ \bibinfo
  {author} {\bibfnamefont {Joao~P.}\ \bibnamefont {Silva}},\ }\bibfield
  {title} {\enquote {\bibinfo {title} {{Theory and phenomenology of
  two-Higgs-doublet models}},}\ }\href {\doibase 10.1016/j.physrep.2012.02.002}
  {\bibfield  {journal} {\bibinfo  {journal} {Phys. Rept.}\ }\textbf {\bibinfo
  {volume} {516}},\ \bibinfo {pages} {1--102} (\bibinfo {year} {2012})},\
  \Eprint {http://arxiv.org/abs/1106.0034} {arXiv:1106.0034 [hep-ph]}
  \BibitemShut {NoStop}%
\bibitem [{\citenamefont {Cirelli}\ \emph {et~al.}(2006)\citenamefont
  {Cirelli}, \citenamefont {Fornengo},\ and\ \citenamefont
  {Strumia}}]{Cirelli:2005uq}%
  \BibitemOpen
  \bibfield  {author} {\bibinfo {author} {\bibfnamefont {Marco}\ \bibnamefont
  {Cirelli}}, \bibinfo {author} {\bibfnamefont {Nicolao}\ \bibnamefont
  {Fornengo}}, \ and\ \bibinfo {author} {\bibfnamefont {Alessandro}\
  \bibnamefont {Strumia}},\ }\bibfield  {title} {\enquote {\bibinfo {title}
  {{Minimal dark matter}},}\ }\href {\doibase 10.1016/j.nuclphysb.2006.07.012}
  {\bibfield  {journal} {\bibinfo  {journal} {Nucl. Phys.}\ }\textbf {\bibinfo
  {volume} {B753}},\ \bibinfo {pages} {178--194} (\bibinfo {year} {2006})},\
  \Eprint {http://arxiv.org/abs/hep-ph/0512090} {arXiv:hep-ph/0512090 [hep-ph]}
  \BibitemShut {NoStop}%
\bibitem [{\citenamefont {Hisano}\ \emph {et~al.}(2007)\citenamefont {Hisano},
  \citenamefont {Matsumoto}, \citenamefont {Nagai}, \citenamefont {Saito},\
  and\ \citenamefont {Senami}}]{Hisano:2006nn}%
  \BibitemOpen
  \bibfield  {author} {\bibinfo {author} {\bibfnamefont {Junji}\ \bibnamefont
  {Hisano}}, \bibinfo {author} {\bibfnamefont {Shigeki}\ \bibnamefont
  {Matsumoto}}, \bibinfo {author} {\bibfnamefont {Minoru}\ \bibnamefont
  {Nagai}}, \bibinfo {author} {\bibfnamefont {Osamu}\ \bibnamefont {Saito}}, \
  and\ \bibinfo {author} {\bibfnamefont {Masato}\ \bibnamefont {Senami}},\
  }\bibfield  {title} {\enquote {\bibinfo {title} {{Non-perturbative effect on
  thermal relic abundance of dark matter}},}\ }\href {\doibase
  10.1016/j.physletb.2007.01.012} {\bibfield  {journal} {\bibinfo  {journal}
  {Phys. Lett.}\ }\textbf {\bibinfo {volume} {B646}},\ \bibinfo {pages}
  {34--38} (\bibinfo {year} {2007})},\ \Eprint
  {http://arxiv.org/abs/hep-ph/0610249} {arXiv:hep-ph/0610249 [hep-ph]}
  \BibitemShut {NoStop}%
\bibitem [{\citenamefont {Cirelli}\ \emph {et~al.}(2007)\citenamefont
  {Cirelli}, \citenamefont {Strumia},\ and\ \citenamefont
  {Tamburini}}]{Cirelli:2007xd}%
  \BibitemOpen
  \bibfield  {author} {\bibinfo {author} {\bibfnamefont {Marco}\ \bibnamefont
  {Cirelli}}, \bibinfo {author} {\bibfnamefont {Alessandro}\ \bibnamefont
  {Strumia}}, \ and\ \bibinfo {author} {\bibfnamefont {Matteo}\ \bibnamefont
  {Tamburini}},\ }\bibfield  {title} {\enquote {\bibinfo {title} {{Cosmology
  and Astrophysics of Minimal Dark Matter}},}\ }\href {\doibase
  10.1016/j.nuclphysb.2007.07.023} {\bibfield  {journal} {\bibinfo  {journal}
  {Nucl. Phys.}\ }\textbf {\bibinfo {volume} {B787}},\ \bibinfo {pages}
  {152--175} (\bibinfo {year} {2007})},\ \Eprint
  {http://arxiv.org/abs/0706.4071} {arXiv:0706.4071 [hep-ph]} \BibitemShut
  {NoStop}%
\bibitem [{\citenamefont {Fox}\ \emph {et~al.}(2014)\citenamefont {Fox},
  \citenamefont {Kribs},\ and\ \citenamefont {Martin}}]{Fox:2014moa}%
  \BibitemOpen
  \bibfield  {author} {\bibinfo {author} {\bibfnamefont {Patrick~J.}\
  \bibnamefont {Fox}}, \bibinfo {author} {\bibfnamefont {Graham~D.}\
  \bibnamefont {Kribs}}, \ and\ \bibinfo {author} {\bibfnamefont {Adam}\
  \bibnamefont {Martin}},\ }\bibfield  {title} {\enquote {\bibinfo {title}
  {{Split Dirac Supersymmetry: An Ultraviolet Completion of Higgsino Dark
  Matter}},}\ }\href {\doibase 10.1103/PhysRevD.90.075006} {\bibfield
  {journal} {\bibinfo  {journal} {Phys. Rev.}\ }\textbf {\bibinfo {volume}
  {D90}},\ \bibinfo {pages} {075006} (\bibinfo {year} {2014})},\ \Eprint
  {http://arxiv.org/abs/1405.3692} {arXiv:1405.3692 [hep-ph]} \BibitemShut
  {NoStop}%
\bibitem [{\citenamefont {Giudice}\ and\ \citenamefont
  {Pomarol}(1996)}]{Giudice:1995np}%
  \BibitemOpen
  \bibfield  {author} {\bibinfo {author} {\bibfnamefont {Gian~F.}\ \bibnamefont
  {Giudice}}\ and\ \bibinfo {author} {\bibfnamefont {Alex}\ \bibnamefont
  {Pomarol}},\ }\bibfield  {title} {\enquote {\bibinfo {title} {{Mass
  degeneracy of the Higgsinos}},}\ }\href {\doibase
  10.1016/0370-2693(96)00060-3} {\bibfield  {journal} {\bibinfo  {journal}
  {Phys. Lett.}\ }\textbf {\bibinfo {volume} {B372}},\ \bibinfo {pages}
  {253--258} (\bibinfo {year} {1996})},\ \Eprint
  {http://arxiv.org/abs/hep-ph/9512337} {arXiv:hep-ph/9512337 [hep-ph]}
  \BibitemShut {NoStop}%
\bibitem [{\citenamefont {Krall}\ and\ \citenamefont
  {Reece}(2018)}]{Krall:2017xij}%
  \BibitemOpen
  \bibfield  {author} {\bibinfo {author} {\bibfnamefont {Rebecca}\ \bibnamefont
  {Krall}}\ and\ \bibinfo {author} {\bibfnamefont {Matthew}\ \bibnamefont
  {Reece}},\ }\bibfield  {title} {\enquote {\bibinfo {title} {{Last Electroweak
  WIMP Standing: Pseudo-Dirac Higgsino Status and Compact Stars as Future
  Probes}},}\ }\href {\doibase 10.1088/1674-1137/42/4/043105} {\bibfield
  {journal} {\bibinfo  {journal} {Chin. Phys.}\ }\textbf {\bibinfo {volume}
  {C42}},\ \bibinfo {pages} {043105} (\bibinfo {year} {2018})},\ \Eprint
  {http://arxiv.org/abs/1705.04843} {arXiv:1705.04843 [hep-ph]} \BibitemShut
  {NoStop}%
\bibitem [{\citenamefont {Kowalska}\ and\ \citenamefont
  {Sessolo}(2018)}]{Kowalska:2018toh}%
  \BibitemOpen
  \bibfield  {author} {\bibinfo {author} {\bibfnamefont {Kamila}\ \bibnamefont
  {Kowalska}}\ and\ \bibinfo {author} {\bibfnamefont {Enrico~Maria}\
  \bibnamefont {Sessolo}},\ }\bibfield  {title} {\enquote {\bibinfo {title}
  {{The discreet charm of higgsino dark matter - a pocket review}},}\ }\href
  {\doibase 10.1155/2018/6828560} {\bibfield  {journal} {\bibinfo  {journal}
  {Adv. High Energy Phys.}\ }\textbf {\bibinfo {volume} {2018}},\ \bibinfo
  {pages} {6828560} (\bibinfo {year} {2018})},\ \Eprint
  {http://arxiv.org/abs/1802.04097} {arXiv:1802.04097 [hep-ph]} \BibitemShut
  {NoStop}%
\bibitem [{\citenamefont {Ackermann}\ \emph {et~al.}(2015)\citenamefont
  {Ackermann} \emph {et~al.}}]{Ackermann:2015zua}%
  \BibitemOpen
  \bibfield  {author} {\bibinfo {author} {\bibfnamefont {M.}~\bibnamefont
  {Ackermann}} \emph {et~al.} (\bibinfo {collaboration} {Fermi-LAT}),\
  }\bibfield  {title} {\enquote {\bibinfo {title} {{Searching for Dark Matter
  Annihilation from Milky Way Dwarf Spheroidal Galaxies with Six Years of Fermi
  Large Area Telescope Data}},}\ }\href {\doibase
  10.1103/PhysRevLett.115.231301} {\bibfield  {journal} {\bibinfo  {journal}
  {Phys. Rev. Lett.}\ }\textbf {\bibinfo {volume} {115}},\ \bibinfo {pages}
  {231301} (\bibinfo {year} {2015})},\ \Eprint
  {http://arxiv.org/abs/1503.02641} {arXiv:1503.02641 [astro-ph.HE]}
  \BibitemShut {NoStop}%
\bibitem [{\citenamefont {Abdallah}\ \emph {et~al.}(2016)\citenamefont
  {Abdallah} \emph {et~al.}}]{Abdallah:2016ygi}%
  \BibitemOpen
  \bibfield  {author} {\bibinfo {author} {\bibfnamefont {H.}~\bibnamefont
  {Abdallah}} \emph {et~al.} (\bibinfo {collaboration} {H.E.S.S.}),\ }\bibfield
   {title} {\enquote {\bibinfo {title} {{Search for dark matter annihilations
  towards the inner Galactic halo from 10 years of observations with
  H.E.S.S}},}\ }\href {\doibase 10.1103/PhysRevLett.117.111301} {\bibfield
  {journal} {\bibinfo  {journal} {Phys. Rev. Lett.}\ }\textbf {\bibinfo
  {volume} {117}},\ \bibinfo {pages} {111301} (\bibinfo {year} {2016})},\
  \Eprint {http://arxiv.org/abs/1607.08142} {arXiv:1607.08142 [astro-ph.HE]}
  \BibitemShut {NoStop}%
\bibitem [{\citenamefont {Aguilar}\ \emph {et~al.}(2016)\citenamefont {Aguilar}
  \emph {et~al.}}]{Aguilar:2016kjl}%
  \BibitemOpen
  \bibfield  {author} {\bibinfo {author} {\bibfnamefont {M.}~\bibnamefont
  {Aguilar}} \emph {et~al.} (\bibinfo {collaboration} {AMS}),\ }\bibfield
  {title} {\enquote {\bibinfo {title} {{Antiproton Flux, Antiproton-to-Proton
  Flux Ratio, and Properties of Elementary Particle Fluxes in Primary Cosmic
  Rays Measured with the Alpha Magnetic Spectrometer on the International Space
  Station}},}\ }\href {\doibase 10.1103/PhysRevLett.117.091103} {\bibfield
  {journal} {\bibinfo  {journal} {Phys. Rev. Lett.}\ }\textbf {\bibinfo
  {volume} {117}},\ \bibinfo {pages} {091103} (\bibinfo {year}
  {2016})}\BibitemShut {NoStop}%
\bibitem [{\citenamefont {Gunion}\ and\ \citenamefont
  {Haber}(2003)}]{Gunion:2002zf}%
  \BibitemOpen
  \bibfield  {author} {\bibinfo {author} {\bibfnamefont {John~F.}\ \bibnamefont
  {Gunion}}\ and\ \bibinfo {author} {\bibfnamefont {Howard~E.}\ \bibnamefont
  {Haber}},\ }\bibfield  {title} {\enquote {\bibinfo {title} {{The CP
  conserving two Higgs doublet model: The Approach to the decoupling limit}},}\
  }\href {\doibase 10.1103/PhysRevD.67.075019} {\bibfield  {journal} {\bibinfo
  {journal} {Phys. Rev.}\ }\textbf {\bibinfo {volume} {D67}},\ \bibinfo {pages}
  {075019} (\bibinfo {year} {2003})},\ \Eprint
  {http://arxiv.org/abs/hep-ph/0207010} {arXiv:hep-ph/0207010 [hep-ph]}
  \BibitemShut {NoStop}%
\bibitem [{\citenamefont {Isidori}\ \emph {et~al.}(2001)\citenamefont
  {Isidori}, \citenamefont {Ridolfi},\ and\ \citenamefont
  {Strumia}}]{Isidori:2001bm}%
  \BibitemOpen
  \bibfield  {author} {\bibinfo {author} {\bibfnamefont {Gino}\ \bibnamefont
  {Isidori}}, \bibinfo {author} {\bibfnamefont {Giovanni}\ \bibnamefont
  {Ridolfi}}, \ and\ \bibinfo {author} {\bibfnamefont {Alessandro}\
  \bibnamefont {Strumia}},\ }\bibfield  {title} {\enquote {\bibinfo {title}
  {{On the metastability of the standard model vacuum}},}\ }\href {\doibase
  10.1016/S0550-3213(01)00302-9} {\bibfield  {journal} {\bibinfo  {journal}
  {Nucl. Phys.}\ }\textbf {\bibinfo {volume} {B609}},\ \bibinfo {pages}
  {387--409} (\bibinfo {year} {2001})},\ \Eprint
  {http://arxiv.org/abs/hep-ph/0104016} {arXiv:hep-ph/0104016 [hep-ph]}
  \BibitemShut {NoStop}%
\bibitem [{\citenamefont {Aad}\ \emph {et~al.}(2015)\citenamefont {Aad} \emph
  {et~al.}}]{Aad:2015zhl}%
  \BibitemOpen
  \bibfield  {author} {\bibinfo {author} {\bibfnamefont {Georges}\ \bibnamefont
  {Aad}} \emph {et~al.} (\bibinfo {collaboration} {ATLAS, CMS}),\ }\bibfield
  {title} {\enquote {\bibinfo {title} {{Combined Measurement of the Higgs Boson
  Mass in $pp$ Collisions at $\sqrt{s}=7$ and 8 TeV with the ATLAS and CMS
  Experiments}},}\ }\href {\doibase 10.1103/PhysRevLett.114.191803} {\bibfield
  {journal} {\bibinfo  {journal} {Phys. Rev. Lett.}\ }\textbf {\bibinfo
  {volume} {114}},\ \bibinfo {pages} {191803} (\bibinfo {year} {2015})},\
  \Eprint {http://arxiv.org/abs/1503.07589} {arXiv:1503.07589 [hep-ex]}
  \BibitemShut {NoStop}%
\bibitem [{\citenamefont {Chowdhury}\ and\ \citenamefont
  {Eberhardt}(2017)}]{Chowdhury:2017aav}%
  \BibitemOpen
  \bibfield  {author} {\bibinfo {author} {\bibfnamefont {Debtosh}\ \bibnamefont
  {Chowdhury}}\ and\ \bibinfo {author} {\bibfnamefont {Otto}\ \bibnamefont
  {Eberhardt}},\ }\bibfield  {title} {\enquote {\bibinfo {title} {{Update of
  Global Two-Higgs-Doublet Model Fits}},}\ }\href@noop {} {\  (\bibinfo {year}
  {2017})},\ \Eprint {http://arxiv.org/abs/1711.02095} {arXiv:1711.02095
  [hep-ph]} \BibitemShut {NoStop}%
\bibitem [{\citenamefont {Misiak}\ and\ \citenamefont
  {Steinhauser}(2017)}]{Misiak:2017bgg}%
  \BibitemOpen
  \bibfield  {author} {\bibinfo {author} {\bibfnamefont {Mikolaj}\ \bibnamefont
  {Misiak}}\ and\ \bibinfo {author} {\bibfnamefont {Matthias}\ \bibnamefont
  {Steinhauser}},\ }\bibfield  {title} {\enquote {\bibinfo {title} {{Weak
  radiative decays of the B meson and bounds on $M_{H^\pm }$ in the
  Two-Higgs-Doublet Model}},}\ }\href {\doibase 10.1140/epjc/s10052-017-4776-y}
  {\bibfield  {journal} {\bibinfo  {journal} {Eur. Phys. J.}\ }\textbf
  {\bibinfo {volume} {C77}},\ \bibinfo {pages} {201} (\bibinfo {year}
  {2017})},\ \Eprint {http://arxiv.org/abs/1702.04571} {arXiv:1702.04571
  [hep-ph]} \BibitemShut {NoStop}%
\bibitem [{\citenamefont {Broggio}\ \emph {et~al.}(2014)\citenamefont
  {Broggio}, \citenamefont {Chun}, \citenamefont {Passera}, \citenamefont
  {Patel},\ and\ \citenamefont {Vempati}}]{Broggio:2014mna}%
  \BibitemOpen
  \bibfield  {author} {\bibinfo {author} {\bibfnamefont {Alessandro}\
  \bibnamefont {Broggio}}, \bibinfo {author} {\bibfnamefont {Eung~Jin}\
  \bibnamefont {Chun}}, \bibinfo {author} {\bibfnamefont {Massimo}\
  \bibnamefont {Passera}}, \bibinfo {author} {\bibfnamefont {Ketan~M.}\
  \bibnamefont {Patel}}, \ and\ \bibinfo {author} {\bibfnamefont {Sudhir~K.}\
  \bibnamefont {Vempati}},\ }\bibfield  {title} {\enquote {\bibinfo {title}
  {{Limiting two-Higgs-doublet models}},}\ }\href {\doibase
  10.1007/JHEP11(2014)058} {\bibfield  {journal} {\bibinfo  {journal} {JHEP}\
  }\textbf {\bibinfo {volume} {11}},\ \bibinfo {pages} {058} (\bibinfo {year}
  {2014})},\ \Eprint {http://arxiv.org/abs/1409.3199} {arXiv:1409.3199
  [hep-ph]} \BibitemShut {NoStop}%
\bibitem [{\citenamefont {Staub}(2014)}]{Staub:2013tta}%
  \BibitemOpen
  \bibfield  {author} {\bibinfo {author} {\bibfnamefont {Florian}\ \bibnamefont
  {Staub}},\ }\bibfield  {title} {\enquote {\bibinfo {title} {{SARAH 4 : A tool
  for (not only SUSY) model builders}},}\ }\href {\doibase
  10.1016/j.cpc.2014.02.018} {\bibfield  {journal} {\bibinfo  {journal}
  {Comput. Phys. Commun.}\ }\textbf {\bibinfo {volume} {185}},\ \bibinfo
  {pages} {1773--1790} (\bibinfo {year} {2014})},\ \Eprint
  {http://arxiv.org/abs/1309.7223} {arXiv:1309.7223 [hep-ph]} \BibitemShut
  {NoStop}%
\bibitem [{\citenamefont {Tanabashi}\ \emph {et~al.}(2018)\citenamefont
  {Tanabashi} \emph {et~al.}}]{Tanabashi:2018oca}%
  \BibitemOpen
  \bibfield  {author} {\bibinfo {author} {\bibfnamefont {M.}~\bibnamefont
  {Tanabashi}} \emph {et~al.} (\bibinfo {collaboration} {Particle Data
  Group}),\ }\bibfield  {title} {\enquote {\bibinfo {title} {{Review of
  Particle Physics}},}\ }\href {\doibase 10.1103/PhysRevD.98.030001} {\bibfield
   {journal} {\bibinfo  {journal} {Phys. Rev.}\ }\textbf {\bibinfo {volume}
  {D98}},\ \bibinfo {pages} {030001} (\bibinfo {year} {2018})}\BibitemShut
  {NoStop}%
\bibitem [{\citenamefont {Haber}\ and\ \citenamefont
  {Hempfling}(1993)}]{Haber:1993an}%
  \BibitemOpen
  \bibfield  {author} {\bibinfo {author} {\bibfnamefont {Howard~E.}\
  \bibnamefont {Haber}}\ and\ \bibinfo {author} {\bibfnamefont {Ralf}\
  \bibnamefont {Hempfling}},\ }\bibfield  {title} {\enquote {\bibinfo {title}
  {{The Renormalization group improved Higgs sector of the minimal
  supersymmetric model}},}\ }\href {\doibase 10.1103/PhysRevD.48.4280}
  {\bibfield  {journal} {\bibinfo  {journal} {Phys. Rev.}\ }\textbf {\bibinfo
  {volume} {D48}},\ \bibinfo {pages} {4280--4309} (\bibinfo {year} {1993})},\
  \Eprint {http://arxiv.org/abs/hep-ph/9307201} {arXiv:hep-ph/9307201 [hep-ph]}
  \BibitemShut {NoStop}%
\bibitem [{\citenamefont {Antusch}\ \emph {et~al.}(2002)\citenamefont
  {Antusch}, \citenamefont {Drees}, \citenamefont {Kersten}, \citenamefont
  {Lindner},\ and\ \citenamefont {Ratz}}]{Antusch:2001vn}%
  \BibitemOpen
  \bibfield  {author} {\bibinfo {author} {\bibfnamefont {Stefan}\ \bibnamefont
  {Antusch}}, \bibinfo {author} {\bibfnamefont {Manuel}\ \bibnamefont {Drees}},
  \bibinfo {author} {\bibfnamefont {Jörn}\ \bibnamefont {Kersten}}, \bibinfo
  {author} {\bibfnamefont {Manfred}\ \bibnamefont {Lindner}}, \ and\ \bibinfo
  {author} {\bibfnamefont {Michael}\ \bibnamefont {Ratz}},\ }\bibfield  {title}
  {\enquote {\bibinfo {title} {{Neutrino mass operator renormalization in two
  Higgs doublet models and the MSSM}},}\ }\href {\doibase
  10.1016/S0370-2693(01)01414-9} {\bibfield  {journal} {\bibinfo  {journal}
  {Phys. Lett.}\ }\textbf {\bibinfo {volume} {B525}},\ \bibinfo {pages}
  {130--134} (\bibinfo {year} {2002})},\ \Eprint
  {http://arxiv.org/abs/hep-ph/0110366} {arXiv:hep-ph/0110366 [hep-ph]}
  \BibitemShut {NoStop}%
\bibitem [{\citenamefont {Antusch}\ \emph {et~al.}(2001)\citenamefont
  {Antusch}, \citenamefont {Drees}, \citenamefont {Kersten}, \citenamefont
  {Lindner},\ and\ \citenamefont {Ratz}}]{Antusch:2001ck}%
  \BibitemOpen
  \bibfield  {author} {\bibinfo {author} {\bibfnamefont {Stefan}\ \bibnamefont
  {Antusch}}, \bibinfo {author} {\bibfnamefont {Manuel}\ \bibnamefont {Drees}},
  \bibinfo {author} {\bibfnamefont {Jörn}\ \bibnamefont {Kersten}}, \bibinfo
  {author} {\bibfnamefont {Manfred}\ \bibnamefont {Lindner}}, \ and\ \bibinfo
  {author} {\bibfnamefont {Michael}\ \bibnamefont {Ratz}},\ }\bibfield  {title}
  {\enquote {\bibinfo {title} {{Neutrino mass operator renormalization
  revisited}},}\ }\href {\doibase 10.1016/S0370-2693(01)01127-3} {\bibfield
  {journal} {\bibinfo  {journal} {Phys. Lett.}\ }\textbf {\bibinfo {volume}
  {B519}},\ \bibinfo {pages} {238--242} (\bibinfo {year} {2001})},\ \Eprint
  {http://arxiv.org/abs/hep-ph/0108005} {arXiv:hep-ph/0108005 [hep-ph]}
  \BibitemShut {NoStop}%
\bibitem [{\citenamefont {Berggren}\ \emph {et~al.}(2013)\citenamefont
  {Berggren}, \citenamefont {Brümmer}, \citenamefont {List}, \citenamefont
  {Moortgat-Pick}, \citenamefont {Robens}, \citenamefont {Rolbiecki},\ and\
  \citenamefont {Sert}}]{Berggren:2013vfa}%
  \BibitemOpen
  \bibfield  {author} {\bibinfo {author} {\bibfnamefont {Mikael}\ \bibnamefont
  {Berggren}}, \bibinfo {author} {\bibfnamefont {Felix}\ \bibnamefont
  {Brümmer}}, \bibinfo {author} {\bibfnamefont {Jenny}\ \bibnamefont {List}},
  \bibinfo {author} {\bibfnamefont {Gudrid}\ \bibnamefont {Moortgat-Pick}},
  \bibinfo {author} {\bibfnamefont {Tania}\ \bibnamefont {Robens}}, \bibinfo
  {author} {\bibfnamefont {Krzysztof}\ \bibnamefont {Rolbiecki}}, \ and\
  \bibinfo {author} {\bibfnamefont {Hale}\ \bibnamefont {Sert}},\ }\bibfield
  {title} {\enquote {\bibinfo {title} {{Tackling light higgsinos at the
  ILC}},}\ }\href {\doibase 10.1140/epjc/s10052-013-2660-y} {\bibfield
  {journal} {\bibinfo  {journal} {Eur. Phys. J.}\ }\textbf {\bibinfo {volume}
  {C73}},\ \bibinfo {pages} {2660} (\bibinfo {year} {2013})},\ \Eprint
  {http://arxiv.org/abs/1307.3566} {arXiv:1307.3566 [hep-ph]} \BibitemShut
  {NoStop}%
\bibitem [{\citenamefont {Schwaller}\ and\ \citenamefont
  {Zurita}(2014)}]{Schwaller:2013baa}%
  \BibitemOpen
  \bibfield  {author} {\bibinfo {author} {\bibfnamefont {Pedro}\ \bibnamefont
  {Schwaller}}\ and\ \bibinfo {author} {\bibfnamefont {Jose}\ \bibnamefont
  {Zurita}},\ }\bibfield  {title} {\enquote {\bibinfo {title} {{Compressed
  electroweakino spectra at the LHC}},}\ }\href {\doibase
  10.1007/JHEP03(2014)060} {\bibfield  {journal} {\bibinfo  {journal} {JHEP}\
  }\textbf {\bibinfo {volume} {03}},\ \bibinfo {pages} {060} (\bibinfo {year}
  {2014})},\ \Eprint {http://arxiv.org/abs/1312.7350} {arXiv:1312.7350
  [hep-ph]} \BibitemShut {NoStop}%
\bibitem [{\citenamefont {Low}\ and\ \citenamefont {Wang}(2014)}]{Low:2014cba}%
  \BibitemOpen
  \bibfield  {author} {\bibinfo {author} {\bibfnamefont {Matthew}\ \bibnamefont
  {Low}}\ and\ \bibinfo {author} {\bibfnamefont {Lian-Tao}\ \bibnamefont
  {Wang}},\ }\bibfield  {title} {\enquote {\bibinfo {title} {{Neutralino dark
  matter at 14 TeV and 100 TeV}},}\ }\href {\doibase 10.1007/JHEP08(2014)161}
  {\bibfield  {journal} {\bibinfo  {journal} {JHEP}\ }\textbf {\bibinfo
  {volume} {08}},\ \bibinfo {pages} {161} (\bibinfo {year} {2014})},\ \Eprint
  {http://arxiv.org/abs/1404.0682} {arXiv:1404.0682 [hep-ph]} \BibitemShut
  {NoStop}%
\bibitem [{\citenamefont {Han}\ \emph {et~al.}(2014)\citenamefont {Han},
  \citenamefont {Kribs}, \citenamefont {Martin},\ and\ \citenamefont
  {Menon}}]{Han:2014kaa}%
  \BibitemOpen
  \bibfield  {author} {\bibinfo {author} {\bibfnamefont {Zhenyu}\ \bibnamefont
  {Han}}, \bibinfo {author} {\bibfnamefont {Graham~D.}\ \bibnamefont {Kribs}},
  \bibinfo {author} {\bibfnamefont {Adam}\ \bibnamefont {Martin}}, \ and\
  \bibinfo {author} {\bibfnamefont {Arjun}\ \bibnamefont {Menon}},\ }\bibfield
  {title} {\enquote {\bibinfo {title} {{Hunting quasidegenerate Higgsinos}},}\
  }\href {\doibase 10.1103/PhysRevD.89.075007} {\bibfield  {journal} {\bibinfo
  {journal} {Phys. Rev.}\ }\textbf {\bibinfo {volume} {D89}},\ \bibinfo {pages}
  {075007} (\bibinfo {year} {2014})},\ \Eprint {http://arxiv.org/abs/1401.1235}
  {arXiv:1401.1235 [hep-ph]} \BibitemShut {NoStop}%
\bibitem [{\citenamefont {Bobrovskyi}\ \emph {et~al.}(2012)\citenamefont
  {Bobrovskyi}, \citenamefont {Brummer}, \citenamefont {Buchmuller},\ and\
  \citenamefont {Hajer}}]{Bobrovskyi:2011jj}%
  \BibitemOpen
  \bibfield  {author} {\bibinfo {author} {\bibfnamefont {S.}~\bibnamefont
  {Bobrovskyi}}, \bibinfo {author} {\bibfnamefont {F.}~\bibnamefont {Brummer}},
  \bibinfo {author} {\bibfnamefont {W.}~\bibnamefont {Buchmuller}}, \ and\
  \bibinfo {author} {\bibfnamefont {J.}~\bibnamefont {Hajer}},\ }\bibfield
  {title} {\enquote {\bibinfo {title} {{Searching for light higgsinos with
  b-jets and missing leptons}},}\ }\href {\doibase 10.1007/JHEP01(2012)122}
  {\bibfield  {journal} {\bibinfo  {journal} {JHEP}\ }\textbf {\bibinfo
  {volume} {01}},\ \bibinfo {pages} {122} (\bibinfo {year} {2012})},\ \Eprint
  {http://arxiv.org/abs/1111.6005} {arXiv:1111.6005 [hep-ph]} \BibitemShut
  {NoStop}%
\bibitem [{\citenamefont {Xiang}\ \emph {et~al.}(2016)\citenamefont {Xiang},
  \citenamefont {Bi}, \citenamefont {Yin},\ and\ \citenamefont
  {Yu}}]{Xiang:2016ndq}%
  \BibitemOpen
  \bibfield  {author} {\bibinfo {author} {\bibfnamefont {Qian-Fei}\
  \bibnamefont {Xiang}}, \bibinfo {author} {\bibfnamefont {Xiao-Jun}\
  \bibnamefont {Bi}}, \bibinfo {author} {\bibfnamefont {Peng-Fei}\ \bibnamefont
  {Yin}}, \ and\ \bibinfo {author} {\bibfnamefont {Zhao-Huan}\ \bibnamefont
  {Yu}},\ }\bibfield  {title} {\enquote {\bibinfo {title} {{Searching for
  Singlino-Higgsino Dark Matter in the NMSSM}},}\ }\href {\doibase
  10.1103/PhysRevD.94.055031} {\bibfield  {journal} {\bibinfo  {journal} {Phys.
  Rev.}\ }\textbf {\bibinfo {volume} {D94}},\ \bibinfo {pages} {055031}
  (\bibinfo {year} {2016})},\ \Eprint {http://arxiv.org/abs/1606.02149}
  {arXiv:1606.02149 [hep-ph]} \BibitemShut {NoStop}%
\bibitem [{\citenamefont {Mahbubani}\ \emph {et~al.}(2017)\citenamefont
  {Mahbubani}, \citenamefont {Schwaller},\ and\ \citenamefont
  {Zurita}}]{Mahbubani:2017gjh}%
  \BibitemOpen
  \bibfield  {author} {\bibinfo {author} {\bibfnamefont {Rakhi}\ \bibnamefont
  {Mahbubani}}, \bibinfo {author} {\bibfnamefont {Pedro}\ \bibnamefont
  {Schwaller}}, \ and\ \bibinfo {author} {\bibfnamefont {Jose}\ \bibnamefont
  {Zurita}},\ }\bibfield  {title} {\enquote {\bibinfo {title} {{Closing the
  window for compressed Dark Sectors with disappearing charged tracks}},}\
  }\href {\doibase 10.1007/JHEP06(2017)119, 10.1007/JHEP10(2017)061} {\bibfield
   {journal} {\bibinfo  {journal} {JHEP}\ }\textbf {\bibinfo {volume} {06}},\
  \bibinfo {pages} {119} (\bibinfo {year} {2017})},\ \bibinfo {note} {[Erratum:
  JHEP10,061(2017)]},\ \Eprint {http://arxiv.org/abs/1703.05327}
  {arXiv:1703.05327 [hep-ph]} \BibitemShut {NoStop}%
\bibitem [{\citenamefont {Thomas}\ and\ \citenamefont
  {Wells}(1998)}]{Thomas:1998wy}%
  \BibitemOpen
  \bibfield  {author} {\bibinfo {author} {\bibfnamefont {Scott~D.}\
  \bibnamefont {Thomas}}\ and\ \bibinfo {author} {\bibfnamefont {James~D.}\
  \bibnamefont {Wells}},\ }\bibfield  {title} {\enquote {\bibinfo {title}
  {{Phenomenology of Massive Vectorlike Doublet Leptons}},}\ }\href {\doibase
  10.1103/PhysRevLett.81.34} {\bibfield  {journal} {\bibinfo  {journal} {Phys.
  Rev. Lett.}\ }\textbf {\bibinfo {volume} {81}},\ \bibinfo {pages} {34--37}
  (\bibinfo {year} {1998})},\ \Eprint {http://arxiv.org/abs/hep-ph/9804359}
  {arXiv:hep-ph/9804359 [hep-ph]} \BibitemShut {NoStop}%
\bibitem [{\citenamefont {Curtin}\ \emph {et~al.}(2018)\citenamefont {Curtin},
  \citenamefont {Deshpande}, \citenamefont {Fischer},\ and\ \citenamefont
  {Zurita}}]{Curtin:2017bxr}%
  \BibitemOpen
  \bibfield  {author} {\bibinfo {author} {\bibfnamefont {David}\ \bibnamefont
  {Curtin}}, \bibinfo {author} {\bibfnamefont {Kaustubh}\ \bibnamefont
  {Deshpande}}, \bibinfo {author} {\bibfnamefont {Oliver}\ \bibnamefont
  {Fischer}}, \ and\ \bibinfo {author} {\bibfnamefont {José}\ \bibnamefont
  {Zurita}},\ }\bibfield  {title} {\enquote {\bibinfo {title} {{New Physics
  Opportunities for Long-Lived Particles at Electron-Proton Colliders}},}\
  }\href {\doibase 10.1007/JHEP07(2018)024} {\bibfield  {journal} {\bibinfo
  {journal} {JHEP}\ }\textbf {\bibinfo {volume} {07}},\ \bibinfo {pages} {024}
  (\bibinfo {year} {2018})},\ \Eprint {http://arxiv.org/abs/1712.07135}
  {arXiv:1712.07135 [hep-ph]} \BibitemShut {NoStop}%
\bibitem [{\citenamefont {Giveon}\ \emph {et~al.}(1991)\citenamefont {Giveon},
  \citenamefont {Hall},\ and\ \citenamefont {Sarid}}]{Giveon:1991zm}%
  \BibitemOpen
  \bibfield  {author} {\bibinfo {author} {\bibfnamefont {Amit}\ \bibnamefont
  {Giveon}}, \bibinfo {author} {\bibfnamefont {Lawrence~J.}\ \bibnamefont
  {Hall}}, \ and\ \bibinfo {author} {\bibfnamefont {Uri}\ \bibnamefont
  {Sarid}},\ }\bibfield  {title} {\enquote {\bibinfo {title} {{SU(5)
  unification revisited}},}\ }\href {\doibase 10.1016/0370-2693(91)91289-8}
  {\bibfield  {journal} {\bibinfo  {journal} {Phys. Lett.}\ }\textbf {\bibinfo
  {volume} {B271}},\ \bibinfo {pages} {138--144} (\bibinfo {year}
  {1991})}\BibitemShut {NoStop}%
\bibitem [{\citenamefont {Patel}\ and\ \citenamefont
  {Sharma}(2011)}]{Patel:2011eh}%
  \BibitemOpen
  \bibfield  {author} {\bibinfo {author} {\bibfnamefont {Ketan~M.}\
  \bibnamefont {Patel}}\ and\ \bibinfo {author} {\bibfnamefont {Pankaj}\
  \bibnamefont {Sharma}},\ }\bibfield  {title} {\enquote {\bibinfo {title}
  {{Forward-backward asymmetry in top quark production from light colored
  scalars in SO(10) model}},}\ }\href {\doibase 10.1007/JHEP04(2011)085}
  {\bibfield  {journal} {\bibinfo  {journal} {JHEP}\ }\textbf {\bibinfo
  {volume} {04}},\ \bibinfo {pages} {085} (\bibinfo {year} {2011})},\ \Eprint
  {http://arxiv.org/abs/1102.4736} {arXiv:1102.4736 [hep-ph]} \BibitemShut
  {NoStop}%
\bibitem [{\citenamefont {Abe}\ \emph {et~al.}(2017)\citenamefont {Abe} \emph
  {et~al.}}]{Miura:2016krn}%
  \BibitemOpen
  \bibfield  {author} {\bibinfo {author} {\bibfnamefont {K.}~\bibnamefont
  {Abe}} \emph {et~al.} (\bibinfo {collaboration} {Super-Kamiokande}),\
  }\bibfield  {title} {\enquote {\bibinfo {title} {{Search for proton decay via
  $p \to e^+\pi^0$ and $p \to \mu^+\pi^0$ in 0.31  megaton·years exposure
  of the Super-Kamiokande water Cherenkov detector}},}\ }\href {\doibase
  10.1103/PhysRevD.95.012004} {\bibfield  {journal} {\bibinfo  {journal} {Phys.
  Rev.}\ }\textbf {\bibinfo {volume} {D95}},\ \bibinfo {pages} {012004}
  (\bibinfo {year} {2017})},\ \Eprint {http://arxiv.org/abs/1610.03597}
  {arXiv:1610.03597 [hep-ex]} \BibitemShut {NoStop}%
\bibitem [{\citenamefont {Akerib}\ \emph {et~al.}(2017)\citenamefont {Akerib}
  \emph {et~al.}}]{Akerib:2016vxi}%
  \BibitemOpen
  \bibfield  {author} {\bibinfo {author} {\bibfnamefont {D.~S.}\ \bibnamefont
  {Akerib}} \emph {et~al.} (\bibinfo {collaboration} {LUX}),\ }\bibfield
  {title} {\enquote {\bibinfo {title} {{Results from a search for dark matter
  in the complete LUX exposure}},}\ }\href {\doibase
  10.1103/PhysRevLett.118.021303} {\bibfield  {journal} {\bibinfo  {journal}
  {Phys. Rev. Lett.}\ }\textbf {\bibinfo {volume} {118}},\ \bibinfo {pages}
  {021303} (\bibinfo {year} {2017})},\ \Eprint
  {http://arxiv.org/abs/1608.07648} {arXiv:1608.07648 [astro-ph.CO]}
  \BibitemShut {NoStop}%
\bibitem [{\citenamefont {Angle}\ \emph {et~al.}(2009)\citenamefont {Angle}
  \emph {et~al.}}]{Angle:2009xb}%
  \BibitemOpen
  \bibfield  {author} {\bibinfo {author} {\bibfnamefont {J.}~\bibnamefont
  {Angle}} \emph {et~al.} (\bibinfo {collaboration} {XENON10}),\ }\bibfield
  {title} {\enquote {\bibinfo {title} {{Constraints on inelastic dark matter
  from XENON10}},}\ }\href {\doibase 10.1103/PhysRevD.80.115005} {\bibfield
  {journal} {\bibinfo  {journal} {Phys. Rev.}\ }\textbf {\bibinfo {volume}
  {D80}},\ \bibinfo {pages} {115005} (\bibinfo {year} {2009})},\ \Eprint
  {http://arxiv.org/abs/0910.3698} {arXiv:0910.3698 [astro-ph.CO]} \BibitemShut
  {NoStop}%
\bibitem [{\citenamefont {Aprile}\ \emph {et~al.}(2012)\citenamefont {Aprile}
  \emph {et~al.}}]{Aprile:2012nq}%
  \BibitemOpen
  \bibfield  {author} {\bibinfo {author} {\bibfnamefont {E.}~\bibnamefont
  {Aprile}} \emph {et~al.} (\bibinfo {collaboration} {XENON100}),\ }\bibfield
  {title} {\enquote {\bibinfo {title} {{Dark Matter Results from 225 Live Days
  of XENON100 Data}},}\ }\href {\doibase 10.1103/PhysRevLett.109.181301}
  {\bibfield  {journal} {\bibinfo  {journal} {Phys. Rev. Lett.}\ }\textbf
  {\bibinfo {volume} {109}},\ \bibinfo {pages} {181301} (\bibinfo {year}
  {2012})},\ \Eprint {http://arxiv.org/abs/1207.5988} {arXiv:1207.5988
  [astro-ph.CO]} \BibitemShut {NoStop}%
\bibitem [{\citenamefont {Aprile}\ \emph {et~al.}(2018)\citenamefont {Aprile}
  \emph {et~al.}}]{Aprile:2018dbl}%
  \BibitemOpen
  \bibfield  {author} {\bibinfo {author} {\bibfnamefont {E.}~\bibnamefont
  {Aprile}} \emph {et~al.} (\bibinfo {collaboration} {XENON}),\ }\bibfield
  {title} {\enquote {\bibinfo {title} {{Dark Matter Search Results from a One
  Ton-Year Exposure of XENON1T}},}\ }\href {\doibase
  10.1103/PhysRevLett.121.111302} {\bibfield  {journal} {\bibinfo  {journal}
  {Phys. Rev. Lett.}\ }\textbf {\bibinfo {volume} {121}},\ \bibinfo {pages}
  {111302} (\bibinfo {year} {2018})},\ \Eprint
  {http://arxiv.org/abs/1805.12562} {arXiv:1805.12562 [astro-ph.CO]}
  \BibitemShut {NoStop}%
\bibitem [{\citenamefont {Savage}\ \emph {et~al.}(2006)\citenamefont {Savage},
  \citenamefont {Freese},\ and\ \citenamefont {Gondolo}}]{Savage:2006qr}%
  \BibitemOpen
  \bibfield  {author} {\bibinfo {author} {\bibfnamefont {Christopher}\
  \bibnamefont {Savage}}, \bibinfo {author} {\bibfnamefont {Katherine}\
  \bibnamefont {Freese}}, \ and\ \bibinfo {author} {\bibfnamefont {Paolo}\
  \bibnamefont {Gondolo}},\ }\bibfield  {title} {\enquote {\bibinfo {title}
  {{Annual Modulation of Dark Matter in the Presence of Streams}},}\ }\href
  {\doibase 10.1103/PhysRevD.74.043531} {\bibfield  {journal} {\bibinfo
  {journal} {Phys. Rev.}\ }\textbf {\bibinfo {volume} {D74}},\ \bibinfo {pages}
  {043531} (\bibinfo {year} {2006})},\ \Eprint
  {http://arxiv.org/abs/astro-ph/0607121} {arXiv:astro-ph/0607121 [astro-ph]}
  \BibitemShut {NoStop}%
\bibitem [{\citenamefont {Duda}\ \emph {et~al.}(2007)\citenamefont {Duda},
  \citenamefont {Kemper},\ and\ \citenamefont {Gondolo}}]{Duda:2006uk}%
  \BibitemOpen
  \bibfield  {author} {\bibinfo {author} {\bibfnamefont {Gintaras}\
  \bibnamefont {Duda}}, \bibinfo {author} {\bibfnamefont {Ann}\ \bibnamefont
  {Kemper}}, \ and\ \bibinfo {author} {\bibfnamefont {Paolo}\ \bibnamefont
  {Gondolo}},\ }\bibfield  {title} {\enquote {\bibinfo {title} {{Model
  Independent Form Factors for Spin Independent Neutralino-Nucleon Scattering
  from Elastic Electron Scattering Data}},}\ }\href {\doibase
  10.1088/1475-7516/2007/04/012} {\bibfield  {journal} {\bibinfo  {journal}
  {JCAP}\ }\textbf {\bibinfo {volume} {0704}},\ \bibinfo {pages} {012}
  (\bibinfo {year} {2007})},\ \Eprint {http://arxiv.org/abs/hep-ph/0608035}
  {arXiv:hep-ph/0608035 [hep-ph]} \BibitemShut {NoStop}%
\bibitem [{\citenamefont {Ohki}\ \emph {et~al.}(2013)\citenamefont {Ohki},
  \citenamefont {Takeda}, \citenamefont {Aoki}, \citenamefont {Hashimoto},
  \citenamefont {Kaneko}, \citenamefont {Matsufuru}, \citenamefont {Noaki},\
  and\ \citenamefont {Onogi}}]{Oksuzian:2012rzb}%
  \BibitemOpen
  \bibfield  {author} {\bibinfo {author} {\bibfnamefont {H.}~\bibnamefont
  {Ohki}}, \bibinfo {author} {\bibfnamefont {K.}~\bibnamefont {Takeda}},
  \bibinfo {author} {\bibfnamefont {S.}~\bibnamefont {Aoki}}, \bibinfo {author}
  {\bibfnamefont {S.}~\bibnamefont {Hashimoto}}, \bibinfo {author}
  {\bibfnamefont {T.}~\bibnamefont {Kaneko}}, \bibinfo {author} {\bibfnamefont
  {H.}~\bibnamefont {Matsufuru}}, \bibinfo {author} {\bibfnamefont
  {J.}~\bibnamefont {Noaki}}, \ and\ \bibinfo {author} {\bibfnamefont
  {T.}~\bibnamefont {Onogi}} (\bibinfo {collaboration} {JLQCD}),\ }\bibfield
  {title} {\enquote {\bibinfo {title} {{Nucleon strange quark content from
  $N_f=2+1$ lattice QCD with exact chiral symmetry}},}\ }\href {\doibase
  10.1103/PhysRevD.87.034509} {\bibfield  {journal} {\bibinfo  {journal} {Phys.
  Rev.}\ }\textbf {\bibinfo {volume} {D87}},\ \bibinfo {pages} {034509}
  (\bibinfo {year} {2013})},\ \Eprint {http://arxiv.org/abs/1208.4185}
  {arXiv:1208.4185 [hep-lat]} \BibitemShut {NoStop}%
\bibitem [{\citenamefont {Billard}\ \emph {et~al.}(2014)\citenamefont
  {Billard}, \citenamefont {Strigari},\ and\ \citenamefont
  {Figueroa-Feliciano}}]{Billard:2013qya}%
  \BibitemOpen
  \bibfield  {author} {\bibinfo {author} {\bibfnamefont {J.}~\bibnamefont
  {Billard}}, \bibinfo {author} {\bibfnamefont {L.}~\bibnamefont {Strigari}}, \
  and\ \bibinfo {author} {\bibfnamefont {E.}~\bibnamefont
  {Figueroa-Feliciano}},\ }\bibfield  {title} {\enquote {\bibinfo {title}
  {{Implication of neutrino backgrounds on the reach of next generation dark
  matter direct detection experiments}},}\ }\href {\doibase
  10.1103/PhysRevD.89.023524} {\bibfield  {journal} {\bibinfo  {journal} {Phys.
  Rev.}\ }\textbf {\bibinfo {volume} {D89}},\ \bibinfo {pages} {023524}
  (\bibinfo {year} {2014})},\ \Eprint {http://arxiv.org/abs/1307.5458}
  {arXiv:1307.5458 [hep-ph]} \BibitemShut {NoStop}%
\end{thebibliography}%
\end{document}